\documentclass[twocolumn]{aastex6}

\usepackage{graphicx}
\usepackage{amsmath}
\usepackage{psfig}
\usepackage{color}

\def\sec{\textrm{s}}

\def\gram{\textrm{g}}
\def\erg{\textrm{erg}}
\def\dyn{\textrm{dyn}}

\def\fm{\textrm{fm}}
\def\cm{\textrm{cm}}
\def\km{\textrm{km}}

\def\Mpc{\textrm{Mpc}}

\def\yr{\textrm{yr}}
\def\Myr{\textrm{Myr}}
\def\Gyr{\textrm{Gyr}}

\def\Kelv{\textrm{K}}

\def\Msun{\textrm{M}_{\odot}}

\def\Joule{\textrm{J}}
\def\kg{\textrm{kg}}
\def\mole{\textrm{mol}}

\def\amu{\textrm{amu}}

\def\PETI{{\cal P}_{\rm ETI}}
\def\PMin{{\cal P}_{\rm min}}
\def\PLife{{\cal P}_{\rm life}}
\def\PCrowded{P_{\rm crowded}}

\def\GHAT{\^G}

\def\HChem{\textrm{H}}
\def\HeChem{\textrm{He}}
\def\CChem{\textrm{C}}
\def\NChem{\textrm{N}}
\def\OChem{\textrm{O}}

\def\NaChem{\textrm{Na}}
\def\CaChem{\textrm{Ca}}

\def\ClChem{\textrm{Cl}}

\def\ArChem{\textrm{Ar}}

\def\ga{\gtrsim}
\def\la{\lesssim}

\shorttitle{The Log Log SETI Prior}
\shortauthors{Lacki}

\begin{document}

\title{The Log Log Prior for the Frequency of Extraterrestrial Intelligences}
\author{Brian C. Lacki$^\varnothing$}
\email{astrobrianlacki@gmail.com}
\noaffiliation

\begin{abstract}
It is unclear how frequently life and intelligence arise on planets.  I consider a Bayesian prior for the probability $\PETI$ that intelligence evolves at a suitable site, with weight distributed evenly over $\ln(1 - \ln \PETI)$.  This log log prior can handle a very wide range of $\PETI$ values, from $1$ to $10^{-10^{122}}$, while remaining responsive to evidence about extraterrestrial societies.  It is motivated by our uncertainty in the number of conditions that must be fulfilled for intelligence to arise, and it is related to considerations of information, entropy, and state space dimensionality.  After setting a lower limit to $\PETI$ from the number of possible genome sequences, I calculate a Bayesian confidence of $18\%$ that aliens exist within the observable Universe.  With different assumptions about the minimum $\PETI$ and the number of times intelligence can appear on a planet, this value falls between $1.4\%$ and $47\%$.  Overall, the prior leans towards our being isolated from extraterrestrial intelligences, but indicates that we should not be confident of this conclusion.  I discuss the implications of the prior for the Search for Extraterrestrial Intelligence, concluding that searches for interstellar probes from nearby societies seem relatively effective.  I also discuss the possibility of very small probabilities allowed by the prior for the origin of life and the Fermi Paradox, and note that similar priors might be constructed for interesting complex phenomena in general.
\end{abstract}


\section{Introduction}
\label{sec:Intro}

\emph{Of course} we are not alone.  We now know the Earth is but one of billions of planets in the Galaxy, and the Milky Way is but one of billions of galaxies \citep{Johnson10,Cassan12,Petigura13,Zackrisson16}.  Intelligent life arose on Earth through natural processes.  Since the laws of physics and astrophysical environments of galaxies are basically uniform through the observable Universe, there's no reason why analogous processes could not happen on other planets.  Self-organization is a very general phenomenon capable of generating the complexity required by life \citep{Kauffman95}, and life appeared early in Earth's history, suggesting its ubiquity \citep{Lineweaver02,Ward04}.  Intelligence may also be common: our biosphere is filled with examples of convergent evolution \citep{ConwayMorris03}, and several clades of animals demonstrate cognition and even tool use \citep[e.g.,][]{Marino02,Emery06,Hochner06}.  Even if the odds of it happening on a particular planet are one in a trillion, billions of other societies have evolved over the Universe's history \citep{Frank16}.

\emph{Of course} we are alone.  With a benign astronomical environment and rare geological processes, the Earth is a far better long-term home for life than a typical terrestrial planet \citep{Gonzalez01,ConwayMorris03,Ward04}.  Even simple proteins or self-replicating nucleic acids are enormously complex; the odds of a planet generating functional molecules that assemble themselves into a working cell may be beyond astronomically tiny \citep{Yockey00,ConwayMorris03}.  Even if life did appear on another planet, human-like intelligence is a very specific adaption to the very specific pressures that the ancestors of \emph{Homo sapiens} experienced.  So many adaptions occurred on the way that it is unlikely that sequence of influences will recur if things were slightly different \citep{Simpson64,Gould89,Mayr01}.  And in fact, most organisms get along fine without intelligence; prokaryotes form the great majority of living things \citep[see][]{Whitman98}.  There's no good reason to believe that our intelligence is anything common if our own biosphere is anything to go by \citep{Simpson64,Mayr01,Lineweaver09}.

These are, in broad strokes, \emph{a priori} arguments for and against there being extraterrestrial intelligence (ETI) in the observable Universe.\footnote{I am purposefully vague about what intelligence is, but I basically mean an organism that is biologically capable of building technology that can send signals across interstellar distances.  $\PETI$ is proportional to $f_l f_i$ in Drake's equation \citep[described in][among others]{Sagan63,Tarter01}, but also depends on the number of ``birthsites'' per planet.  A species that actually develops this technology is called a technological society in this work.  Throughout this paper, I largely ignore cultural evolution (the $f_c$ factor of the Drake equation), but in principle it could be the limiting factor in whether SETI will find anything \citep{Ashkenazi95,Davies10}.  Even then, one could use a log log prior for $f_c$ since there are a finite number of distinct societies according to the Bekenstein Bound.}  It is clear that there is an astronomical number of rocky, terrestrial planets in our past light cone where life and intelligence could develop, roughly $\sim 10^{21}$ \citep{Zackrisson16}.  The number of planets is known to around an order of magnitude, but it is unfortunately useless unless we know the odds that intelligence actually does arise on a planet --- for all we know, it could be $10^{-100}$ or smaller, in which case we are effectively alone.  On the other hand, there is no well-motivated estimate for that probability.  Even granting that \emph{Homo sapiens} is unlikely to recur, there could be many ways for intelligence to arise \citep[as in][]{Gould87,Cirkovic14}, and a planet has millions and millions of chances to find even one.  Our uniqueness in Earth's biosphere might indicate that probability of intelligence evolving is much less than 1 \citep[e.g.,][]{Mayr01,Lineweaver09}.  But our uniqueness on Earth, by itself, is about equally compatible with the probability being $0.01$, $10^{-10}$, $10^{-20}$, $10^{-100}$, and $10^{-10^{9}}$, values which would have very different implications for how populated the Universe is.

Ideally, the matter could be settled empirically, which is the aim of the Search for Extraterrestrial Intelligence (SETI; \citealt{Cocconi59,Tarter01}).  SETI has sought evidence for extraterrestrials through many programs and an increasing number of methods, from the traditional surveys for radio broadcasts \citep[e.g.,][]{Tarter85,Blair92,Horowitz93,Anderson02,Gray02,Siemion10}, to searches for laser light \citep{Shvartsman93,Reines02,Howard04,Hanna09,Borra12}, high energy radiation (\citealt{Harris02}; see also \citealt{Learned94,Corbet97,Lacki15}), extraterrestrial technology in the Solar System \citep{Freitas83,Steel95}, artificial ``megastructures'' the sizes of planets \citep{Arnold05,Wright16,Boyajian16} or star systems \citep{Slysh85,Timofeev00,Jugaku04,Carrigan09,Villarroel16}, and the engineering of entire galaxies \citep{Kardashev64,Annis99,Wright14-Results,Griffith15,Zackrisson15,Lacki16}.  But so far, no alien societies have been found yet, and there is no consensus about what that means \citep{Brin83,Cirkovic09-Fermi}.  Do extraterrestrials exist but remain too quiet to be observable yet \citep[e.g.,][]{Freitas85,Scheffer94,HaqqMisra09}, or would they rapidly grow until they become obvious even across cosmic distances \citep[as in][]{Hart75,Tipler80,Wright14-SF}?  

\subsection{The Anthropic and Copernican Principles}
\label{sec:Anthropic-Copernican}
In the absence of hard evidence, the debate has occasionally turned to philosophical arguments.  It is indisputable that life with human-like intelligence exists in the form of humanity.  As with the SETI null results, the interpretation of even this trivial positive result is disputed.  The debate frequently centers around two principles: the Anthropic Principle and the Copernican Principle.

The Anthropic Principle essentially says that our existence or situation is somehow inevitable, regardless of how special or improbable we are \citep{Carter74}.  The most commonly invoked version is the Weak Anthropic Principle, which applies if the Universe is very large.  The Weak Anthropic Principle can be formulated in terms of observations, as a statement about inference: we cannot deduce the probability of our evolution just from our existence.  Our situation could be very special (but not unique) and still be consistent with observation.  In statistical terms, as long as the probability of intelligent life evolving is nonzero, the likelihood of it existing in an infinite Universe is 1.  The inferential Weak Anthropic Principle is also interpreted as a statement about selection bias \citep{Carter74,Carter83}.

The Weak Anthropic Principle can also be stated in in terms of theoretical predictions, as a statement about causality: in a sufficiently big Universe, the appearance of humanity somewhere is to be expected, even if the probability of that happening in a particular location is very low but non-zero.  The conditions necessary for our evolution may just be one of the special, rare events that occasionally happen in an infinite Universe.  

There are stronger versions of the Anthropic Principle, which apply not just to the contingent circumstances within the Universe, but to the fundamental laws of physics themselves \citep{Carter74,Barrow83}.  The most extreme formulations argue that conscious observers or humanity play some crucial role in the functioning of the Universe.  These versions imply that the Universe is actually compelled to produce us, and it would be logically impossible for the Universe to exist without us \citep{Barrow83}.  This is in contrast with the causal Weak Anthropic Principle, in which the probability of our not existing in an infinite Universe has measure $0$; the impossibility in the Weak case is merely statistical rather than logical.  As such, stronger Anthropic Principles say that humanity is truly special in terms of function or role, not just in terms of being rare, and they are far more controversial.

In tension with the Anthropic arguments is the Copernican Principle.  The fundamental argument of the Copernican Principle is that we are not in a special location of the Universe, as demonstrated by centuries of astronomical observation \citep[e.g.,][]{Sagan94}.  More generally, we should assume that our circumstances are not special (as in its controversial application by \citealt{Gott93}).  As far as we can observe, the Earth, Solar System, and Galaxy are fairly typical in astrophysical terms; other bodies like them should be common throughout the Universe.  After all, they are the result of natural processes that could occur anywhere.  The Copernican Principle is related to the Principle of Mediocrity: in the absence of information, we are more likely to find ourselves in a typical situation than a rare one \citep[e.g.,][]{Brin83,Vilenkin95}.

Even if the evolution of intelligent life is extremely rare, we can still hold to a weaker argument that I will call the Weak Copernican Principle.  The Weak Copernican Principle states that our evolution is the result of physical processes that have a non-zero probability of occurring independently elsewhere in the Universe.  If one accepts a naturalistic view of evolution, this may seem trivial, but it is not.  It is logically possible for a class of events to have probability measure $0$: for example, one could flip a fair coin infinitely many times and get only heads. 

The Weak Copernican Principle is also technically distinct from the causal Weak Anthropic Principle, even if they both imply our existence is expected in a large enough Universe.  The Weak Copernican Principle implies the evolution of intelligence at a suitably distant location is independent of our evolution.  In contrast, one could imagine that the evolution of intelligence on one planet somehow makes it impossible for it to evolve anywhere else in the Universe.  For example, the Universe might be a computer simulation which is designed to randomly place life on one and only one planet.  This could still be consistent with the Weak Anthropic Principle, as long as the probability of life appearing the first time is large enough.  On the other hand, the Weak Copernican Principle doesn't say that our evolution is inevitable; it still applies if the chance that life evolves around a star is $10^{-100}$ and there are only $10$ stars in the Universe.

In some sense, it is clear that both the Anthropic and Copernican arguments are partly true.  The Weak Anthropic Principle is true in that most of the Universe's volume is not filled with intelligent life forms, even if every planet is habitable.  For that matter, the other planets in our Solar System and the Sun are unfit for human habitation; this was not clear a few centuries ago \citep{Crowe99}.  The Copernican Principle is true in that there are planets besides the Earth, and solar systems besides our own.  Just a few decades ago, it was an open question if exoplanets existed at all or if the Solar System was the result of an improbable stellar event \citep{Dick98}, and the existence of other planets was also unclear a few centuries ago.  Now the problem is to figure out how to extrapolate these principles beyond the evolution of stars and planets to the evolution of life and intelligence.

\subsection{The question of priors}
\label{sec:Priors}

The philosophical debate about the existence of aliens can be understood as a debate about priors.  In Bayesian probability theory, a prior is a subjective judgment about how much one believes in a hypothesis before an observational test is done.  Given a continuously varying parameter $\alpha$, the prior $dP_{\rm prior}/d\alpha$ takes the form of a probability distribution function (PDF) over the allowed values of $\alpha$ \citep{Trotta08}.

When new evidence arrives, Bayes' theorem describes how the prior can be transformed into a posterior describing subjective levels of belief after considering evidence from an observation \citep{Trotta08}:
\begin{multline}
\label{eqn:Bayes}
P_{\rm posterior} ({\rm Hypothesis} | {\rm Observation})  = \\
{\cal L} ({\rm Hypothesis} | {\rm Observation}) \frac{P_{\rm prior} ({\rm Hypothesis})}{P ({\rm Observation})}.
\end{multline}
Bayes' theorem requires the likelihood ${\cal L} ({\rm Hypothesis} | {\rm Observation}) = P({\rm Observation} | {\rm Hypothesis})$, which is the probability that one would make a given observation if the hypothesis is true.  The likelihood can frequently be estimated theoretically for a well-characterized model and a well-understood experiment.  In addition, Bayes' theorem requires an evidence factor $P({\rm Observation})$, which is a normalizing factor.  It basically is the probability that one would make an observation according to a prior, including the cases where the hypothesis is true and where the hypothesis is false.  For the continuous parameter $\alpha$, Bayes' theorem is phrased as
\begin{equation}
\label{eqn:BayesPDF}
\frac{dP_{\rm posterior}}{d\alpha} = \frac{{\cal L} (\alpha | {\rm Observation}) \displaystyle \frac{dP_{\rm prior}}{d\alpha}}{\displaystyle \int {\cal L} (\alpha | {\rm Observation}) \frac{dP_{\rm prior}}{d\alpha} d\alpha}.
\end{equation}

Although the choice of prior is subjective, the Principle of Mediocrity is a general guiding principle.  It says that, in the absence of evidence, we should assume that no particular value is special, and therefore we should favor no value over another \citep{Trotta08}.  Otherwise, if all of the prior weight is concentrated into a few hypotheses, we effectively assume whichever hypothesis we wish to prove.  Then even if evidence strongly points towards an alternate hypothesis, we essentially ignore it and cling to the old theory (for example, the ``Presumptuous Philosopher'' thought experiment recounted in \citealt{Bostrom03}).  For a continuous parameter, the prior should not be too strongly weighted towards one value, which is fulfilled if it is flat.  Note that a flat prior for a parameter $\alpha$ does not remain flat if the variable $\alpha$ is transformed, such as if we then consider $\ln \alpha$.  If $\alpha$ might have values that vary over orders of magnitude, a logarithmic prior that is uniform in $\ln \alpha$ (flat log prior) seems like a reasonable choice, since it has no scale. \citep[e.g.,][]{Trotta08,Spiegel12,Tegmark14}.

Different conclusions are reached if different facts of our evolution are emphasized as representative.  The timescales for our evolution is a common source of speculative reasoning.  The idea is that a habitable planet has some unchanging chance of producing intelligent life in a unit of time, $\Gamma_{\rm ETI}$.  This is appropriate if the appearance of intelligence depends on a process that is independent of history, like an evolutionary process generating certain key traits through a random walk \citep{Carter83}.

Life appeared quite early in our planet's history, which would not be typical if life arose through such random processes ($\Gamma_{\rm life} \ll 10^{-10}\ \yr$).  \citet{Lineweaver02} interprets this observation as evidence that life arises quickly on planets and is common \citep[as in][]{Ward04}, but others argue that intelligence can only arise if life appears early and this atypicality could be an anthropic bias \citep{Hanson98}.  The choice of a prior on $\Gamma_{\rm life}$ also affects whether meaningful constraints are then set on $\Gamma_{\rm life}$ \citep{Spiegel12}.  \citet{Behroozi15} applies an analogous argument on cosmological scales, arguing for a large $\Gamma_{\rm ETI}$ large because the Earth formed before most of the Universe's virialized gas had a chance to collapse into planets.

On the other hand, \citet{Carter83} noted that the timescale for humanity's evolution is close to the habitable lifespan of the Earth.  He essentially interprets this observation according to a flat log prior in the evolutionary timescale.  Of all the many orders of magnitude this timescale could have been, it is unlikely to have matched the Earth's lifespan so closely, so he interprets the coincidence as the result of anthropic selection; the expected timescale is much longer, and intelligent life is very rare.  Furthermore, using a simple model of evolution, he argues that the timing is related to the number of unlikely steps that occurred along the way of our evolution (\citealt{Carter83}; see also \citealt{Hanson98,Carter08,Davies10}).  

But this argument too has been disputed; \citet{Cirkovic09-Reset} argues that astrophysical extinction events like gamma ray bursts can slow down the actual time it takes for intelligent life to evolve even if the unimpeded timescale is fairly short.  It's also possible that a critical evolutionary step is directly tied to the sun's properties \citep{Chyba05}.  \citet{Livio99} suggests the critical step is the development of an ozone layer, which is related to the photodissociation of atmospheric water vapor into oxygen; this process is only efficient for blue stars no more long-lived than the Sun.

It's also possible that the evolution timescale is not the relevant factor, because the evolution of life is constrained by earlier arbitrary events.  For example, it is unlikely that a life form will greatly change the genetic code mapping amino acids to DNA nucleotide sequences; the cost is too great as it risks turning all of the genes into gibberish \citep{Crick68}.  But this code was set very early in the development of life.  If the appearance of intelligence depended on having a particular genetic code \citep[compare with][]{ConwayMorris03}, then whether it evolves on a particular planet could have little to do with the time available.  

By itself, this wouldn't explain the coincidence between the Sun's lifetime and the time for humanity to evolve, but one could imagine that events very early in life's evolution launches it on a nearly fixed course that predetermines whether and how long until intelligence appears.  If there's a small chance that this trajectory leads to intelligence appearing in 10 Gyr, a much smaller chance that intelligence appears in 1 Gyr, a much smaller chance that it appears in 100 Myr, and so on, most intelligence would appear near the end of their planet's lifespan, without depending on the presence of discrete evolutionary barriers along the way.

In a recent book, \citet{Tegmark14} presented a relatively simple argument that suggests that we are alone.  The probability that intelligent life $\PETI$ arises on a given planet is unknown, even at an order of magnitude level, so we can adopt a prior that is uniform in $\log_{10} \PETI$.  If $\log_{10} \PETI$ is between $-21$ and $0$, then we are not alone.  But we have no reason to set $-21$ as a lower limit; given our ignorance, it could easily extend to $-100$ or further.  Because of the huge range in $\log_{10} \PETI$ allowed by our ignorance, relatively little weight is left to be spread over the range of $-21 < \log_{10} \PETI < 0$, so this reasonable prior indicates that we are likely alone in the observable Universe \citep{Tegmark14}.\footnote{This telling is slightly altered from its presentation in \citet{Tegmark14}.  The first difference is that the book focuses on a flat log prior in the distance to the nearest alien society, although a flat log prior in $\PETI$ is given as the motivation.  The fundamental quantity is the probability that intelligence evolves on a world, and the number of worlds tracks comoving volume.  The logarithm of the comoving distance is not proportional to the logarithm of the comoving distance if space is slightly curved.  Second, \citet{Tegmark14} applies the Fermi Paradox \citep[described in][]{Hart75,Tipler80} to rule out $-10 \la \log_{10} \PETI < 0$, closing the window for which aliens could exist in the observable Universe.  But systematic uncertainties can blunt null results, and the prior weight in this window could be so low for the log prior that applying the Fermi Paradox has insignificant value.}

But this argument has a problem --- the lower bound is left undefined, which can lead to strange results.  Fundamental physics implies that there are at most $\sim e^{3 \times 10^{122}}$ configurations for the observable Universe \citep{Egan10}, so we might take $\log_{10} \PETI = -10^{122}$ as a lower bound.  Unfortunately, this prior then makes it essentially impossible to convince its holder that aliens exist, since the prior probability for their presence in the observable Universe is $\sim 10^{-122}$.  Besides requiring an extremely high level of statistical confidence before concluding a positive result is correct, it is essentially impossible to rule out the possibility of systematic errors to that degree.  Even for a well characterized experiment, you could always decide that the evidence is always fraudulent, or that you are hallucinating.  While such possibilities are unlikely, can you really be sure that they are more unlikely than $1$ part in $10^{122}$?

\subsection{Introducing the log log prior}
\label{sec:IntroLogLog}
A fundamental disagreement in estimates of $\PETI$ is the number of conditions that are required for intelligent life to evolve.  Suppose there are $N$ conditions that must be fulfilled for aliens to appear, and for simplicity, each is independent of each other and the probability of each holding is $1/2$.  Then $\PETI = 2^{-N}$: so if $N = 1$, $\PETI = 1/2$; if $N = 10$, $\PETI = 1/1,024$; if $N = 100$, $\PETI \approx 10^{-30}$ and so on.  If $N$ is uncertain at the order of magnitude level, then even the order of magnitude of $\PETI$ is also uncertain at the order of magnitude level.  This accounts for the vast disagreements about $\PETI$: an optimist who believes that only a few conditions are relevant can end up thinking that $\PETI \approx 1$, while a pessimist that believes that thousands of conditions need to be fulfilled can find combinatorially small estimates of $\PETI \ll 10^{-1,000}$.

The notion of an uninformative prior thus suggests that we should use a prior for the number of conditions that is constant in $\log N$ \citep[e.g.,][]{Trotta08}.  This translates to a prior that is constant in $\log |\log \PETI|$.  The advantage of this prior is that it can handle scenarios where $N$ is allowed to range up to $\sim 3 \times 10^{122}$, the entropy of the Universe (with $\PETI \approx e^{-3 \times 10^{122}}$), while remaining responsive to any future evidence that aliens exist.  

\subsection{Outline and conventions}
The loose motivation for the log log prior is developed further in more quantitative terms in Section~\ref{sec:Basis}.  The concepts of entropy, information, and state space dimensionality play key roles.  I discuss some problems that arise when trying to formulate a log log prior and apply it.  I also provide a simple model of a SETI experiment to demonstrate the prior's response.

The Bayesian credibility that ETIs exist in our past light cone is calculated in Section~\ref{sec:Results}.  I use various estimates of the entropy of biological systems and their environments to establish a lower limit to $\PETI$ for a planet.  I also describe what happens if we consider smaller birthsites, to allow for the possibility that intelligence evolves off of planets or can evolve many times on a planet.

In Section~\ref{sec:SETIReach}, I evaluate SETI surveys according to how much of the prior's weight they might constrain.  Then I discuss some additional problems and implications of the log log prior in Section~\ref{sec:Discussion}: (1) In a small Universe, one can construct a joint prior on the Universe's size and $\PETI$, complicating the weighting.  (2) The log log prior suggests that the diversity of intelligent species is beyond astronomically vast.  (3) The small probabilities considered for $\PETI$ raise the issue if there's a similarly small probability that intelligent life is starfaring, which would neutralize the Fermi Paradox.  (4) A log log prior might be useful for estimating credibility in the rates of any complex phenomenon, including life itself.  I conclude the paper with a summary (Section~\ref{sec:Summary}).

I use the values of the fundamental constants and cosmological parameters listed in Table~\ref{table:Constants} throughout this paper.

\begin{deluxetable}{lll}
\tablecolumns{3}
\tablecaption{Constants used in this paper\label{table:Constants}}
\tablehead{\colhead{Name} & \colhead{Value} & \colhead{Description}}
\startdata
$c$                      & $2.998 \times 10^{10}\ \cm\ \sec^{-1}$                 & Speed of light\\
$h$                      & $6.626 \times 10^{-27}\ \erg\ \sec$                    & Planck's constant\\
$G$                      & \parbox[c]{3cm}{$6.674 \times 10^{-8}$\\ \hphantom{1.5cm}$\times \dyn\ \cm^2\ \gram^{-2}$} & Newton's constant\\
$k_B$                    & $1.381 \times 10^{-16}\ \erg\ \Kelv^{-1}$              & Boltzmann's constant\\
$N_A$                    & $6.022 \times 10^{23}$                                 & Avogadro's number\\
$\amu$                   & $1.661 \times 10^{-24}\ \gram$                         & Atomic mass unit\\
$\Msun$                  & $1.989 \times 10^{33}\ \gram$                          & Solar mass\\
$\Mpc$                   & $3.0857 \times 10^{24}\ \cm$                           & Megaparsec\\
$H_0$                    & $67.74\ \km\ \sec^{-1}\ \Mpc^{-1}$                     & Hubble's constant\\
$\Omega_b$               & $0.04866$                                              & Cosmic baryon density\\
$\Omega_r$               & $5.385 \times 10^{-5}$                                 & Cosmic photon density\\
$\Omega_m$               & $0.3089$                                               & Cosmic matter density\\
$\Omega_{\Lambda}$       & $0.6911$                                               & Dark energy density
\enddata  
\tablecomments{The values of the fundamental constants and units are taken from \citet{Olive15}.  I use the $H_0$, $\Omega_b$, and $\Omega_m$ from \citet{Ade15-Params} (the ``TT, TE, EE+lowP+lensing+ext'' column of Table 4).  The value of $\Omega_r$ is calculated under the assumption that only the Cosmic Microwave Background contributes to the cosmic radiation density, and that it has a temperature $2.725\ \Kelv$
\citep{Fixsen09}.  I assume $\Lambda$CDM cosmology with $\Omega_{\Lambda} = 1 - \Omega_m - \Omega_r$.}
\end{deluxetable}

This paper discusses both Bayesian probability, our confidence in a hypothesis, and frequentist probability, an inherent property of stochastic processes in the Universe.  Although both can appear together in Bayes' equation, they are very different in meaning.  To help distinguish them, I will use the symbol $P$ for Bayesian probabilities and ${\cal P}$ for frequentist probabilities.  The most common of each symbol is $P_{\rm crowded}$, the Bayesian credibility assigned to the hypothesis that there are aliens in our past light cone, and $\PETI$, the frequentist fraction of birthsites that evolve intelligent life.

This paper also makes combinatorial arguments about the multitudes of ways of combining a number of objects or traits.  I use $N$ or $\ell$ to enumerate physical things or properties which are actually present in the Universe, like the number of planets in our past light cone or the number of amino acids in a protein.  For the number of possible combinations of these objects, almost all of which will never be realized within the observable Universe, I use ${\cal N}$.  Finally, I use $S$ for physical, measurable entropy and ${\cal S}$ for unitless Boltzmann-like entropies.

\section{A quantitative formulation of the log log prior}
\label{sec:Basis}

\subsection{Finding aliens in state space}
\label{sec:StateSpace}
Intelligent life is generated when certain circumstances apply at a given birthsite.  A birthsite is an independent opportunity for ETI to evolve presented by a complex system that serves as a habitat.\footnote{Throughout this paper, I effectively consider humanity an ETI when describing the evolution of intelligence.}  The probability that a birthsite leads to intelligent life is $\PETI$.  For example, a habitat may be a planet, and a birthsite may just be the planet's existence or specific events in the planet's history, like a speciation.  In the latter case, $\PETI$ may be multiplied by the rate that birthsites occur to find $\Gamma_{\rm ETI}$, how often a habitat produces ETI.

Habitats are very complicated.  The number of possible states of a system grows exponentially with the number of independent parameters needed to describe it.  The immense variety is often conceptualized as a state space $\Omega$, with each independent condition corresponding to an entire dimension of the space.  Each possible state is a point in the space; because of considerations from fundamental physics, I shall assume that the spaces are discrete and finite \citep{Bekenstein81,Bousso02}, with ${\cal N}_{\rm all}$ points in total.  Each combination is a microstate of a system, and the volume of a subset is given by the number of points inside it.  There are many possible state spaces that could describe the evolution of a habitat: the Hilbert space of quantum wavefunctions of the Universe, a planet, or an organism; sequence spaces enumerating every possible genome (the ``Library of Mendel'' of \citealt{Dennett95})\footnote{This is a reference to the fictional Library of Babel, a physical space of all possible texts with a certain length of certain letters, space, and punctuation marks.  Most of the texts are gibberish, and this library is chaotic in its organization \citep{Borges62}.  Some debates about evolution and probability echo the problems inherent in the Library of Babel --- are most proteins in protein space non-functional and how does evolution ``find'' them \citep[e.g.,][]{MaynardSmith70,Dill99,Dryden08}?  Likewise, the microstates in $\Omega_{\rm gen}$ may be rare enclaves of coherence, chaotically scattered all over the abyssal reaches of state space.  The fact that most books have nearly indistinguishable copies \citep[see also][]{Dennett95} is related to the implied diversity of intelligent species I discuss in Section~\ref{sec:Diversity}.}, protein \citep{MaynardSmith70}, or combination of alleles in an organism \citep{Kauffman95}; and morphology spaces describing the basic shapes of proteins \citep{Dill99} or organisms \citep{ConwayMorris03,Cirkovic14}.  Each microstate has some probability assigned to it. 

Only some states of a habitat host ETIs.  Judging from the lack of technological societies during most of Earth's history, only a small region $\Omega_{\rm ETI}$ of a habitat's state space includes these states.  Many more microstates of a habitat could lead to the generation of intelligent life during the habitat's lifetime.  These are included in a region $\Omega_{\rm gen}$.  One can consider the union of ${\cal N}_{\rm gen}$ microstates contained in $\Omega_{\rm gen}$ to be a kind of macrostate.  Note, however, that this macrostate does not need to be connected in state space, and it might be much coarser or finer than thermodynamic macrostates.  One can define a Boltzmann-like entropy for $\Omega_{\rm gen}$:
\begin{equation}
{\cal S}_{\rm gen} = \ln {\cal N}_{\rm gen},
\end{equation}
which can range between $0$ and ${\cal S}_{\rm all} = \ln {\cal N}_{\rm all}$.  If each microstate is equally likely, then the probability that a birthsite will generate intelligence is the probability that it starts out in $\Omega_{\rm gen}$:
\begin{equation}
\label{eqn:PETIasNegentropy}
\PETI = \frac{{\cal N}_{\rm gen}}{{\cal N}_{\rm all}} = e^{{\cal S}_{\rm gen} - {\cal S}_{\rm all}}.
\end{equation}
The fundamental quantity here is the difference in entropy.  

If, however, microstates have differing probabilities, then the probability $\PETI$ of starting out in $\Omega_{\rm gen}$ cannot be calculated simply by counting states and its accompanying Boltzmann entropy.  To complicate matters further, ``entropy'' usually refers to Gibbs (Shannon) entropy for physical systems or information sources \citep{Pierce60}.  This is smaller than the Boltzmann entropy and quantifies the mean amount of information one needs to determine a system's microstate.   Gibbs entropy makes no reference to macrostates \citep{Carroll10} and is not directly related to $\PETI$ either.  To simplify matters, I assume that the Boltzmann entropy provides a useful lower bound estimate $\PMin \equiv e^{-{\cal S}_{\rm all}}$ for $\PETI$.  It provides a natural cutoff for the log log prior.

Almost all state spaces that can be imagined to describe biological or ecological systems have a vast number of dimensions.  Wandering into the state neighborhood of the present Earth is essentially impossible in the exponentially larger spaces.  Indeed, the sheer size of state space is so great for some biological processes, like protein folding, that it sometimes seems puzzling that life can survive at all (some proposed solutions include \citealt{MaynardSmith70,Kauffman95,Dill99}).  Since it is exponentially unlikely a planet would resemble Earth in all details that might affect the evolution of humanity's technical abilities, maybe $\PETI$ is exponentially suppressed.  That is, $\Omega_{\rm ETI}$ might be too small of a target to ever hit on a random trajectory through state space (\citealt{Cirkovic14} describes \citealt{Simpson64}'s argument in this way).  

But these arguments include a number of assumptions that have been disputed.  First, humanity is not necessarily the only possible kind of intelligent life.  There could be a nearly endless variety of species that can develop interstellar communication, in which case $\Omega_{\rm ETI}$ and $\Omega_{\rm gen}$ are far larger than the small neighborhood around the Earth's current state \citep{Gould87,Cirkovic14}.  Along these lines, many of the details that describe a birthsite could be completely irrelevant --- this is clear when we consider the Hilbert space of quantum wavefunctions for the entire Universe.  Each irrelevant detail expands $\Omega_{\rm gen}$ exponentially.  Second, it's possible that the microstates in $\Omega_{\rm gen}$ are vastly more likely than typical microstates.  Complex systems frequently include great attractor basins that channel virtually all trajectories towards them \citep{Kauffman95}.  As long as this evolution completes while the habitat survives, $\Omega_{\rm gen}$ could include basins that span $\Omega$.  Finally, many of the microstates might actually be impossible \citep{ConwayMorris03}.  For example, not all genetic sequences correspond to viable organisms \citep{Dennett95}.  The evolution of the habitat could be constrained strongly to remain in an Earth-like region \citep{Bieri64,ConwayMorris03}.  The state space would be ripped by great holes where nothing could wander; less a fitness landscape and more a fitness tunnel.

With the number of relevant parameters itself uncertain, even to order of magnitude, I suggest that we adopt a flat log log (double log) prior for $\PETI$:
\begin{equation}
\label{eqn:LogLogPriorApprox}
\frac{dP_{\rm prior}}{d\ln |\ln \PETI|} \sim \frac{1}{\ln \ln \PMin} \approx \frac{1}{\ln {\cal S}_{\rm all}}.
\end{equation}
This prior has several interpretations.  Even if there are not easily observed independent parameters that determine whether intelligence evolves, equation~\ref{eqn:LogLogPriorApprox} can describe uncertainty in the number of ``extra'' \emph{dimensions} in the state space.  As seen in equation~\ref{eqn:PETIasNegentropy}, the log log prior corresponds to a flat log prior in the \emph{entropy difference} of $\Omega_{\rm gen}$.  The log log prior also describes a flat log prior in the amount of \emph{information} one needs to know about a birthsite before confidently concluding that it will generate an ETI.  And, in fact, all of these are uncertain to order of magnitude.

Why use a flat log entropy difference prior instead of a flat log entropy prior, though?  A conceptual problem with a log entropy prior is that one can always add irrelevant details to the description of a system, inflating its information content.  The actual value of $\PETI$ does not depend on these nuisance parameters, but the state space nonetheless expands exponentially.  And while $\Omega_{\rm gen}$ also grows exponentially, this is not reflected in a naive log entropy prior.  In fact, a log-entropy prior is less responsive than a simple log-probability prior: if $dP_{\rm prior}/d{\cal S}_{\rm gen} \sim 1/{\cal S}_{\rm all}$, then $dP_{\rm prior}/d|\ln \PETI| = 1/[({\cal S}_{\rm all} + \ln \PETI) \ln {\cal S}_{\rm all}] < 1/\ln {\cal S}_{\rm all}$.  For a concrete example of the difference between a log entropy difference prior and a log entropy prior, imagine $-10,000 \le \ln \PETI \le -1$.  The log entropy difference prior, which is what I use in this paper, places equal weight on the possibilities that $-10,000 \le \ln \PETI \le -1,000$, $-1,000 \le \ln \PETI \le -100$, $-100 \le \ln \PETI \le -10$, and $-10 \le \ln \PETI \le -1$.  In contrast, the log entropy prior places equal weight on the possibilities that $-10,000 \le \ln \PETI \le -9,999$, $-9,999 \le \ln \PETI \le -9,990$, $-9,990 \le \ln \PETI \le -9,900$, $-9,900 \le \ln \PETI \le -9,000$, and $-9,000 \le \ln \PETI \le -1$.  Clearly, the log entropy prior inappropriately favors incredibly small chances for ETIs arising.

The log log prior makes sense if all the microstates have equal probability, or if $\Omega_{\rm gen}$ is disproportionately likely, but what if $\Omega_{\rm gen}$ is some kind of repulsor state, disproportionately unlikely?  Then maybe $\PETI$ could be arbitrarily small, even zero in defiance of the Weak Copernican Principle, leading to the same kinds of problems that haunted the log prior.  I do not believe this is a realistic possibility, at least if thermodynamic entropies are used for $\PMin$.  A fundamental assumption of statistical mechanics is that the thermodynamic microstates of closed, equilibrium systems are equiprobable \citep{Nash06}.  We might therefore expect technological societies, including ours, to appear by random thermal fluctuations in any sufficiently large heat bath --- perhaps the ocean of a lifeless planet, the photon background of the early Universe, or the horizon of a black hole.  The Universe itself has an event horizon that appears as a heat bath.  In the distant future it may evolve into an equilibrium of some sort, perhaps with occasional macroscopic fluctuations and equiprobable microstates \citep{Albrecht15}.  Then, the entropy of the observable Universe bounds the probability of $\Omega_{\rm gen}$.  Using the cosmic entropy still results in a responsive prior because of the weak dependence on ${\cal S}_{\rm all}$ in equation~\ref{eqn:LogLogPriorApprox}.  

Additionally, the inhabitants of these fluctuation societies would be Boltzmann brains, with memories and ``knowledge'' that do not actually correspond to the actual Universe (the implications for cosmology are discussed in \citealt{Albrecht04} and many others).  Unless our experiences are almost certainly phantasms, our evolution as we remember it must be far more likely than thermal fluctuations.  This implies that $\ln \PETI$ is far bigger than $-{\cal S}_{\rm all}$ for a heat bath big enough to create a Boltzmann brain.

The log log prior, as given in equation~\ref{eqn:LogLogPriorApprox}, is timeless, in that it makes no reference to the time it takes intelligence to evolve in a habitat.  If we consider planets as habitats, it is like saying that either planets do not evolve ETIs at all, or they evolve it in 4.5 billion years.  This might make sense if we assume that planets start in one side of state space and launch their biospheres on a ballistic trajectory through state space; after all, ETIs are unlikely to be the first life forms on a planet.  But it might be the case that $\PETI$ depends on how long a habitat is hospitable, which would definitely be the case if the trajectory jumped randomly throughout state space.  In that case, the number of \emph{birthsites} should scale with time --- perhaps each individual organism is a birthsite.  In reality, there may be a whole multitude of ways durations affect $\PETI$, with astronomical conditions being nearly unchanging while biological conditions are capable of changing quickly.  For simplicity, I mainly ignore this time evolution, but I do present an upper bound on the number of temporal birthsites in Section~\ref{sec:SmallHabitats}.  The log log prior is only weakly sensitive to the number of birthsites.

By itself, the log log prior does not take into account the observation that we exist.  If the Universe is small, having relatively few birthsites, and if our existence is due to random chance, then one could update the prior with the Anthropic observation.  Although I consider the possibility of a small Universe in Section~\ref{sec:SmallUniverse}, for the rest of the paper, I assume the Universe is essentially infinite, in accordance with many currently proposed cosmologies. Then the likelihood of our existence is $1$, as noted by the causal Weak Anthropic Principle, and the resulting ``posterior'' is the same as the original prior.  

I use natural logarithms for the log log prior in equation~\ref{eqn:LogLogPriorApprox}, but this choice is somewhat arbitrary.  Unlike the log prior, the choice of logarithm base $b$ does affect the integrated probability that $\PETI$ lies in some range.  Conceptually, the base of logarithm describes the inverse probability that a parameter has the right value for an ETI to appear.  It can also describe the width of a state space along one dimension.  For a genome sequence space, there are $4$ possible nucleotide bases at each location, so $b = 4$ might be more appropriate; for a protein sequence space, there are $20$ possible amino acids at each location, so $b = 20$ might be more appropriate.  But the dependence on $b$ is very weak, and the possible variety of alien biologies makes even these suppositions uncertain, so I just use $b = e$ throughout the paper.

Finally, equation~\ref{eqn:LogLogPriorApprox} does not handle well cases where $\PETI \approx 1$.  Strictly speaking, a pure log log prior places infinite weight on $\PETI \to 1$, since $|\ln 0| = \infty$.  For simplicity, I will define a parameter
\begin{equation}
\Pi \equiv \ln (1 - \ln \PETI)
\end{equation}
that has a finite value even when $\PETI = 1$.  Throughout this paper, I will then use a log-$\Pi$ prior
\begin{equation}
\label{eqn:LogLogPrior}
\frac{dP_{\rm prior}}{d\Pi} = \frac{1}{\ln (1 - \ln \PMin)} = \frac{1}{1 + \ln {\cal S}_{\rm all}} \equiv \frac{1}{\Pi_{\rm max}},
\end{equation}
which is essentially identical to equation~\ref{eqn:LogLogPriorApprox} for small $\PETI$, but leaves some moderate weight on the possibility that $\PETI \approx 1$.  This quick fix hides conceptual issues about what a birthsite is, though.  A potential birthsite we consider could actually produce a great number of intelligent species: this can easily happen if we consider too large of a site, like an entire galaxy supercluster.  If the limiting factor in the appearance of ETIs is the origin of life, it's also entirely possible that life arises many times on a planet.  Then we may need to either consider much smaller birthsites or place more weight on $\PETI \approx 1$.  The problem is especially insidious for the origin of life case since only one biosphere may emerge from all of the independent kinds of life, or so it may appear to distant astronomers \citep[c.f.,][]{Davies09}.  If we are unsure whether the deciding factor for $\PETI$ is the origin of life or the evolution of intelligence, there may not even be a clear definition of birthsite to use.

\subsection{The probability that we are ``alone''}
\label{sec:CrowdedDef}
If there are an infinite (or very large finite) number of birthsites, then the Weak Copernican Principle virtually guarantees that ETIs exist elsewhere.  In that sense, of course we are not alone \citep{Wesson90}.

But it does not do us much good if our nearest neighbors are outside the observable Universe.  The only ones we can learn about must be within our past light cone.  In order to be visible, the ETIs must also have arisen after the Big Bang, at a reasonably low redshift (say, $\la 10^9$).  In bouncing or some inflationary cosmologies, there may be ETIs in a previous universe, but these are presumably hidden; in fact, a version of the Fermi Paradox implies that evidence of technological societies can only survive for a finite time \citep{Tipler82}.  There are then only a finite number $N_{\rm LC}$ of birthsites within this volume besides Earth; if terrestrial planets are assumed to be the birthsites, then $N_{\rm LC} \approx 4.9 \times 10^{20}$ \citep{Zackrisson16}.

I shall say that we live in a \emph{crowded} Universe if there are other intelligent life forms within our past light cone sometime between $z = 10^3$ and $z = 0$.  In contrast, I will say that we are \emph{isolated} if there are no other intelligent life forms in this region of spacetime.  As long as the birthsites evolve basically independently of each other, and ignoring the time it takes for ETIs to evolve, the probability that we are isolated for a particular $\PETI$ can be found using the binomial distribution:
\begin{equation}
\label{eqn:PIsolated}
{\cal P} ({\rm isolated} | \PETI) = (1 - \PETI)^{N_{\rm LC}} \approx e^{-N_{\rm LC} \PETI}.
\end{equation}
The prior probability that we are isolated,
\begin{align}
\nonumber P_{\rm isolated} & = \int_{0}^{\Pi_{\rm max}} {\cal P} ({\rm isolated} | \PETI) \frac{dP_{\rm prior} (\PETI)}{d\Pi} d\Pi \\
                           & \approx \frac{1}{\Pi_{\rm max}} \int_{0}^{\Pi_{\rm max}} e^{\displaystyle -N_{\rm LC} e^{\displaystyle 1 - e^{\Pi}}} d\Pi,
\end{align}
does not have an obvious closed form expression, but it is can be approximated to a precision of a few percent as
\begin{equation}
\label{eqn:PIsolatedApprox}
P_{\rm isolated} \approx \frac{\ln(1 + \ln N_{\rm LC})}{\ln(1 - \ln \PMin)}
\end{equation}
thanks to the triple exponential.\footnote{Roughly speaking, $y \equiv P ({\rm isolated} | \PETI) \approx \exp(-N_{\rm LC} \PETI)$ changes from $0$ to $1$ over a range of $\sim |d\Pi/dy|_{y = 0.5} = 3 / (1.4 + \ln N_{\rm LC})$ in $\Pi$.}  I use this approximation throughout this paper.  The probability that we live in a crowded Universe is 
\begin{equation}
P_{\rm crowded} \equiv 1 - P_{\rm isolated} \approx 1 - \frac{\ln(1 + \ln N_{\rm LC})}{\ln(1 - \ln \PMin)}.
\end{equation}

\subsection{A quantitative demonstration}
\label{sec:PosteriorExample}
\begin{figure*}
\centerline{\includegraphics[width=9cm]{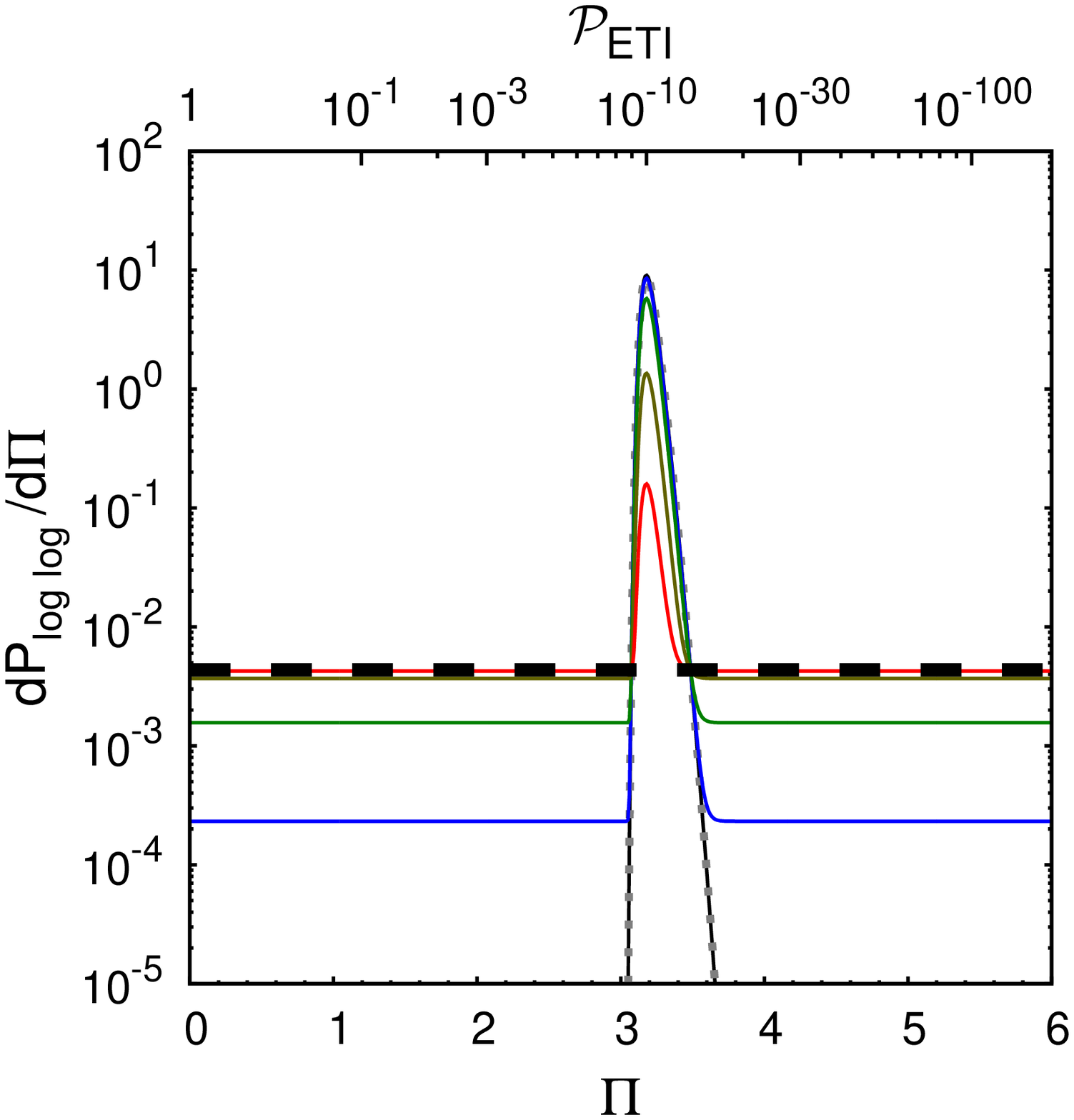}\includegraphics[width=9cm]{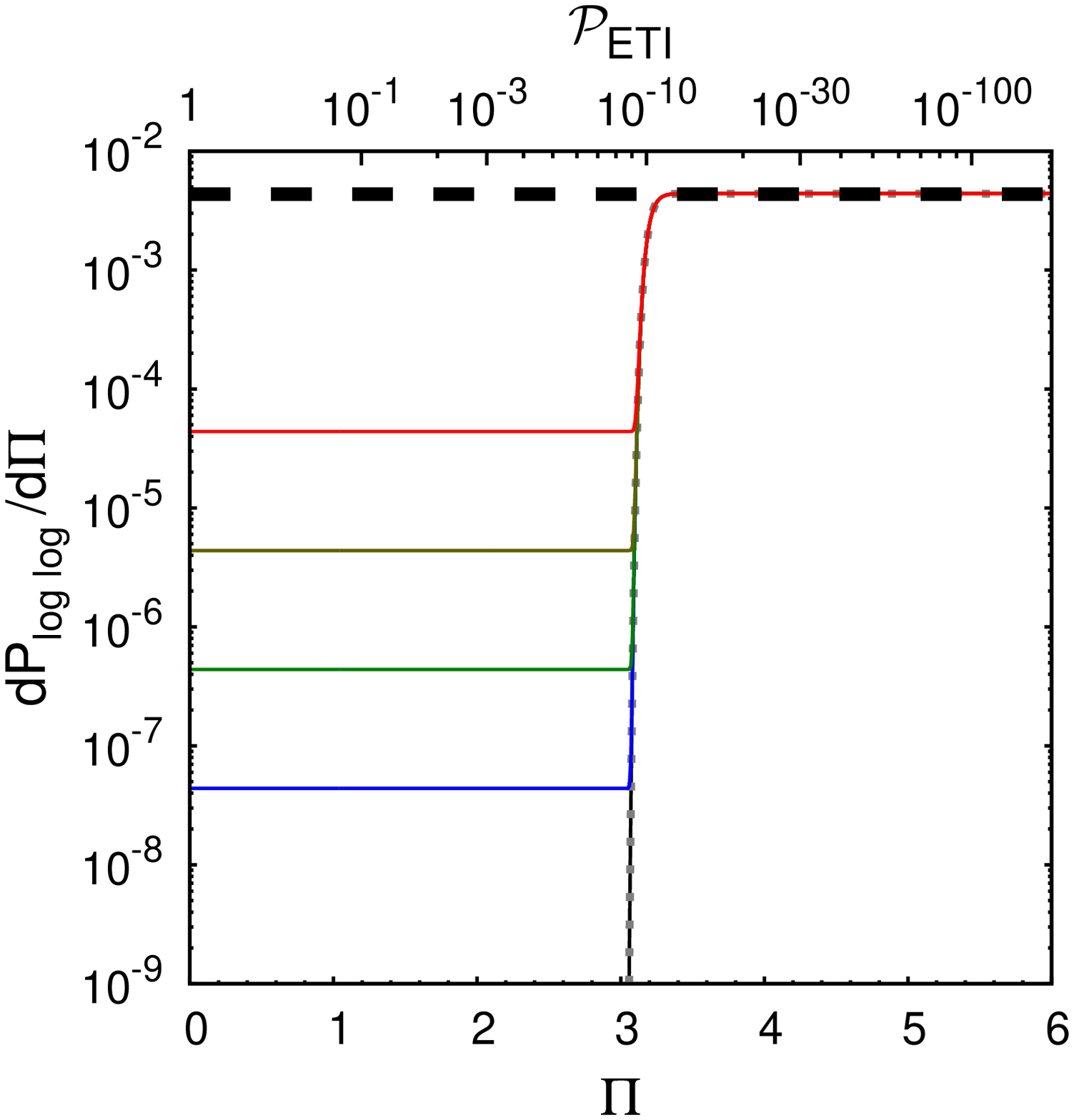}}
\figcaption{How the log log prior responds to a single detection of an extraterrestrial society (left) and a null result (right).  The thick black dashes show the log log prior itself.  If there are no systematic errors ($\varepsilon = 0$), the posterior has the form shown by the solid black line.  Increasing systematic errors shift weight towards $\PETI$ values that contradict the observations.  Results are shown for $\varepsilon$ of $10^{-10}$ (grey dotted), $10^{-5}$ (blue solid), $10^{-4}$ (green solid), $10^{-3}$ (gold solid), and $10^{-2}$ (red solid). \label{fig:LogLogPosterior}}
\end{figure*}

Here is an example demonstrating how the log log prior is responsive when the log prior is not because of systematic errors.

\begin{figure*}
\centerline{\includegraphics[width=9cm]{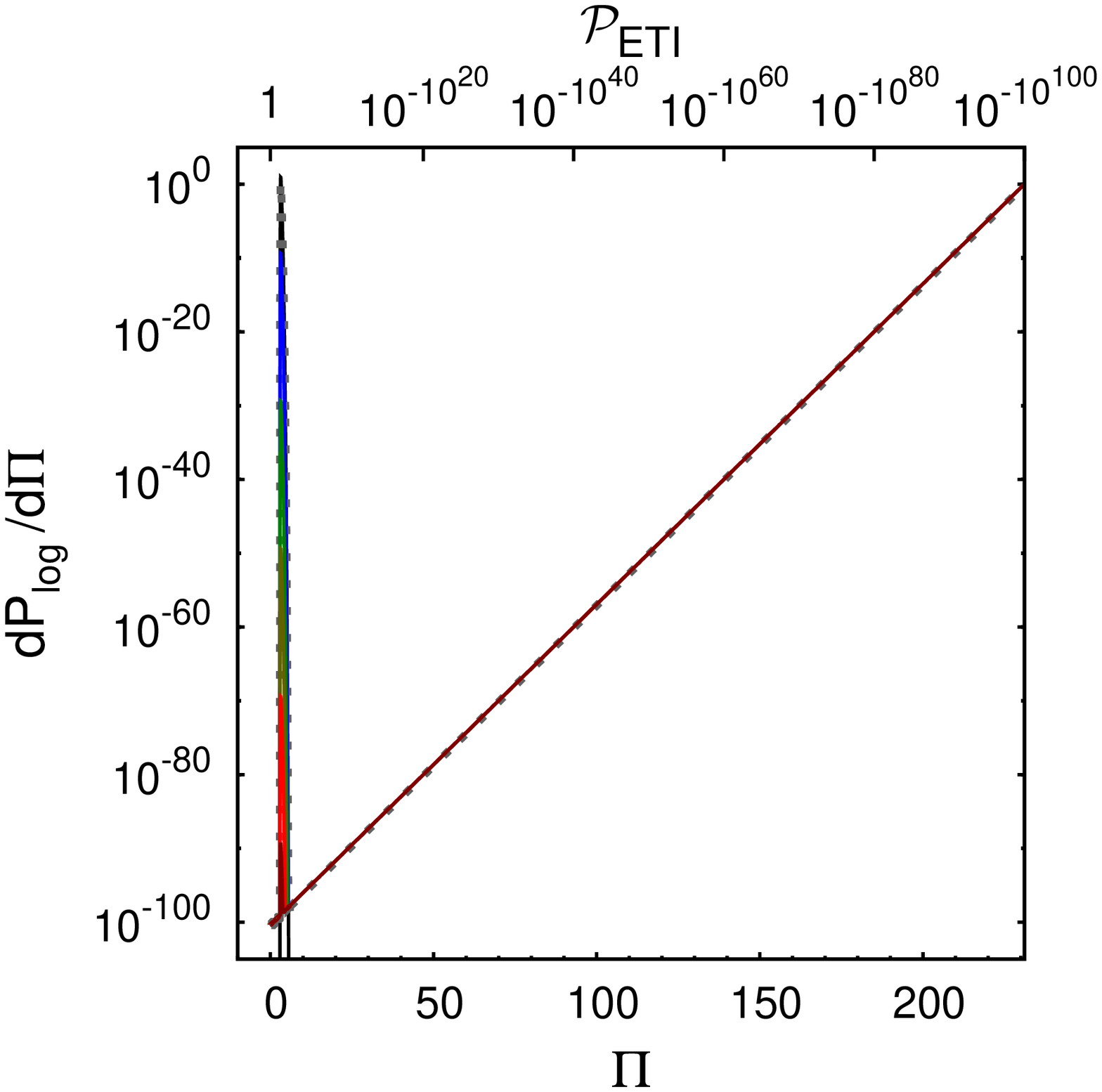}\includegraphics[width=9cm]{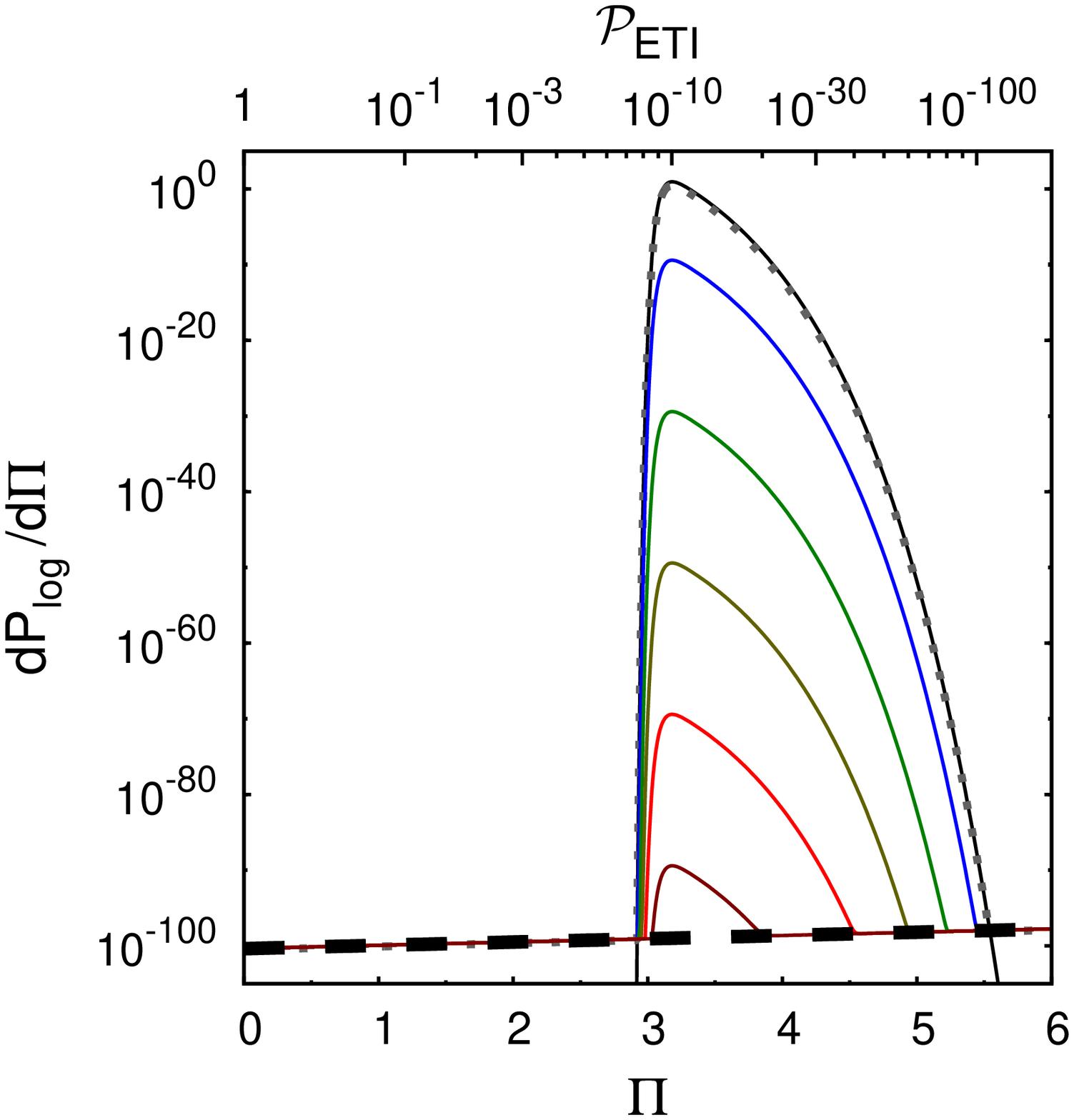}}
\figcaption{The weak response of the flat log prior to the discovery of an extraterrestrial society, showing the entire range of $\Pi$ values (left) and a close-up of $\Pi \le 6$ (right).  The thick black dashes show the prior, and the thin solid black line is the posterior if $\varepsilon = 0$.  Also shown are the posteriors for $\varepsilon = 10^{-100}$ (grey dashes), $10^{-90}$ (blue solid), $10^{-70}$ (green solid), $10^{-50}$ (gold solid), $10^{-30}$ (bright red solid), and $10^{-10}$ (dark red solid). \label{fig:LogPosterior}}
\end{figure*}

Suppose we believe that the minimum chance life can arise on a planet is $10^{-10^{100}}$.  According to the log log prior, $dP_{\rm log log}/d\Pi = 1/\ln(1 + 10^{100} \ln 10) = 1/231$ (the thick, dashed, black lines in Figure~\ref{fig:LogLogPosterior}).  According to a flat log-prior, $dP_{\rm log}/d\ln \PETI = 1/(10^{100} \times \ln 10)$, so in terms of $\Pi$, $dP_{\rm log}/d\Pi = e^{\Pi} / (10^{100} \times \ln 10)$.  As seen in the left panel of Figure~\ref{fig:LogPosterior}, the log prior's weight (thick, dashed, black lines) varies by $\sim 100$ orders of magnitude over the range of possible $\PETI$.  Compared to the log log prior, it concentrates almost all of this weight near the minimum possible $\PETI$.  For all possible values where the Universe is likely to be crowded, $dP_{\rm log}/d\Pi \approx 10^{-100}$ (right panel of Figure~\ref{fig:LogPosterior}).

Now imagine a survey that observes $N_{\rm survey} = 10^{10}$ planets and is able to determine if they ever possessed intelligent life.  (Note that, in reality, most societies might be long dead by the time the planets are observed.)  The survey is reported to have found $N_{\rm report}$ planets that have hosted ETIs.  Due to systematic errors, $N_{\rm report}$ need not be $N_{\rm host}$, the actual number of planets observed that have hosted aliens.  Instead, whatever $N_{\rm host}$ actually is, a glitch causes $N_{\rm report}$ to be reported with a probability of $\varepsilon$, while the correct $N_{\rm host}$ is reported with probability $1 - \varepsilon$.  This glitch can give a false positive, a false negative, or even accidentally leave $N_{\rm report}$ unchanged.  The likelihood of the reported observation is
\begin{multline}
\label{eqn:ToyLikelihood}
{\cal L}(N_{\rm report}|\PETI) = \varepsilon \\ + (1 - \varepsilon){\cal L}_{\rm binomial} (N_{\rm report} | N_{\rm survey}, \PETI),
\end{multline}
where ${\cal L}_{\rm binomial} (N_{\rm report}|N_{\rm survey}, \PETI)$ is the likelihood that $N_{\rm report}$ planets of $N_{\rm survey}$ host ETIs given some $\PETI$ value according to the binomial distribution.  The binomial distribution likelihood is given by
\begin{multline}
\ln {\cal L}_{\rm binomial} (N_{\rm report}| N_{\rm survey}, \PETI) = \ln \binom{N_{\rm survey}}{N_{\rm report}} \\
+ N_{\rm report} \ln \PETI + (N_{\rm survey} - N_{\rm report}) \ln(1 - \PETI).
\end{multline}
The posterior PDFs for $\PETI$ then follow from Bayes' Theorem (equation~\ref{eqn:BayesPDF}). 

The posteriors that result from using the log log prior when $N_{\rm report} = 1$ and $0$ are shown on the left and the right of Figure~\ref{fig:LogLogPosterior}.  Generally, for the $N_{\rm report} = 1$ case, the posterior has a spike near $\PETI \approx 10^{-10}$, whereas there is a steep drop at this point for the $N_{\rm report} = 0$ case.  When there are no systematic errors, with $\varepsilon = 0$, these are the only features (solid black lines).  Increasing systematic errors (grey through red lines) add a floor of posterior weight.  These errors suppress the probability spike when the positive detection is reported, because of the normalization term in Bayes' Theorem.  

\begin{figure*}
\centerline{\includegraphics[width=9cm]{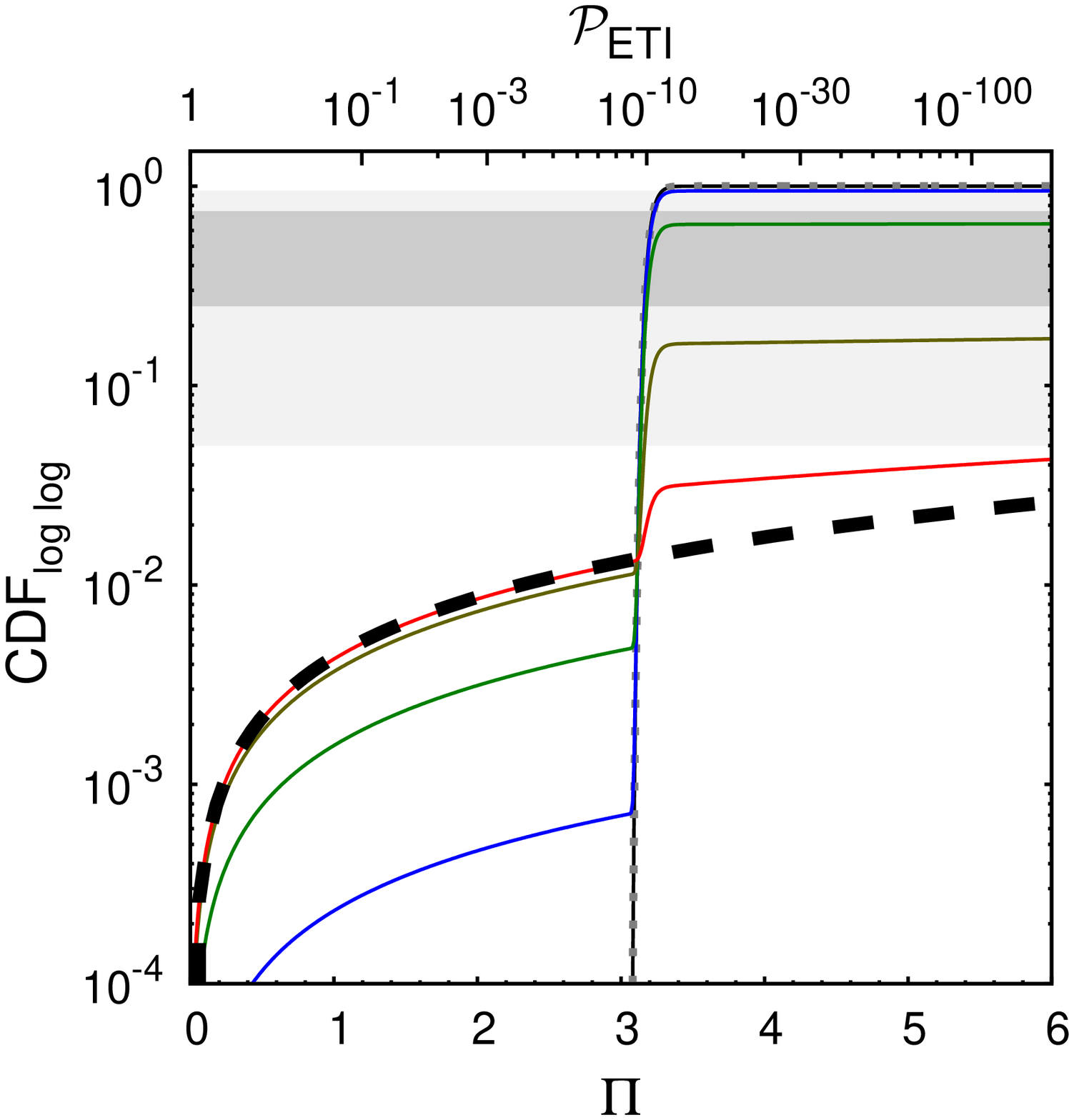}\includegraphics[width=9cm]{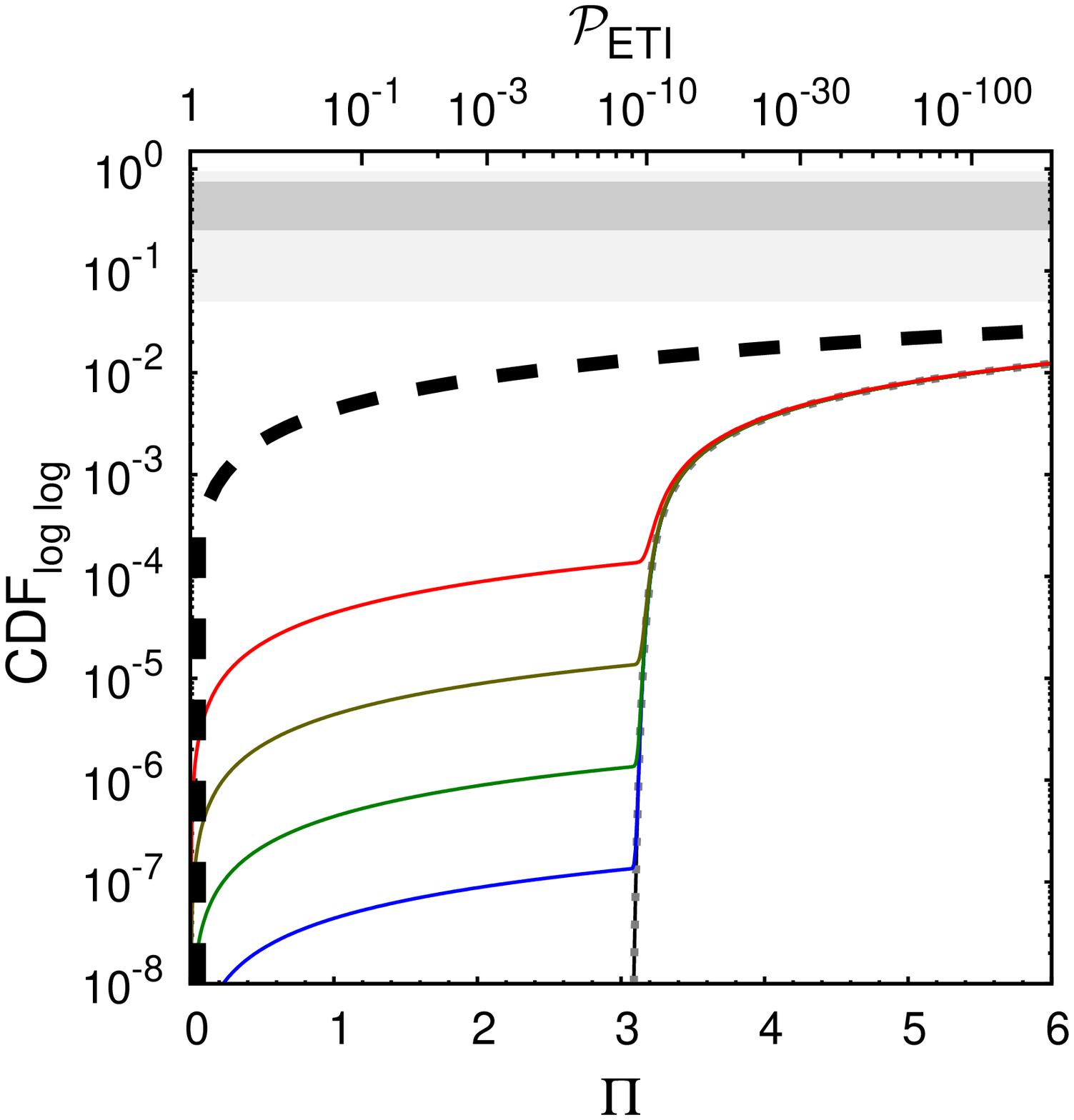}}
\figcaption{The posterior CDF using the log log prior for a single detection (left) and no detections (right).  CDF values between $0.05$ and $0.95$ are lightly shaded grey, and those between $0.25$ and $0.75$ are shaded darker grey.  The CDFs' line styles match their PDFs' styles in Figure~\ref{fig:LogLogPosterior}.  \label{fig:Credibility}}
\end{figure*}

Would these results convince us that $\PETI$ was near $10^{-10}$ if $N_{\rm report} = 1$?  This is determined by the posterior cumulative distribution function (CDF),
\begin{equation}
{\rm CDF} (\Pi^{\prime}) = \int_{0}^{\Pi^{\prime}} \frac{dP (\Pi)}{d\Pi} d\Pi,
\end{equation}
which determines how much of the posterior weight is below a $\Pi$ value.  These functions are shown for $N_{\rm report} = 1$ in Figure~\ref{fig:Credibility} (left panel).  The shaded regions represent credibility intervals containing 50\% and 90\% of the posterior weight.  If $\varepsilon$ is small enough, the CDFs reach $\sim 1$ within the shown range, indicating that we would conclude that a detection was correct.  If $\varepsilon$ is large, $\ga 10^{-4}$, however, the spike in the posterior PDF merely adds a small step to the CDF that does not breach the credibility intervals; our opinion on $\PETI$ is not changed much from the original log log prior.  The cumulative increase in the CDF over the ``spike'' is nearly
\begin{multline}
\Delta {\rm CDF} \approx (1 - \varepsilon) \\
\times \frac{\displaystyle \int_0^{\Pi_{\rm max}}  {\cal L}_{\rm binomial} [N_{\rm report} | N_{\rm survey}, \PETI(\Pi)] \frac{dP_{\rm prior}}{d\Pi} d\Pi}{\displaystyle \int_0^{\Pi_{\rm max}} {\cal L} [N_{\rm report} | N_{\rm survey}, \PETI(\Pi)] \frac{dP_{\rm prior}}{d\Pi} d\Pi},
\end{multline}
which is about $[1 + \varepsilon (1 - \varepsilon)^{-1} \Pi_{\rm max} N_{\rm found} [1 - \ln (N_{\rm found} / N_{\rm survey})]]^{-1}$ if $1 \la N_{\rm found} \ll N_{\rm survey}$.  Because $N_{\rm report} \ll N_{\rm survey}$, the survey actually could rule out high values of $\PETI$: this is seen as the steep cliffs in the CDF when $\varepsilon$ is small.  If $\varepsilon$ is large, though, the survey does not rule out large $\PETI$, since these could be false negatives.  This is the gist of arguments against the Fermi Paradox, which apparently rules out other technological societies in the Milky Way's history, but depends on uncertain assumptions about alien behavior --- we might assign some moderately high probability $\varepsilon$ to the possibility that they all stay on their home planets, or that their presence would be undetectable, blunting the paradox.  Another example of this is seen in the right panel, which shows the CDFs for $N_{\rm report} = 0$.  

\begin{figure}
\centerline{\includegraphics[width=9cm]{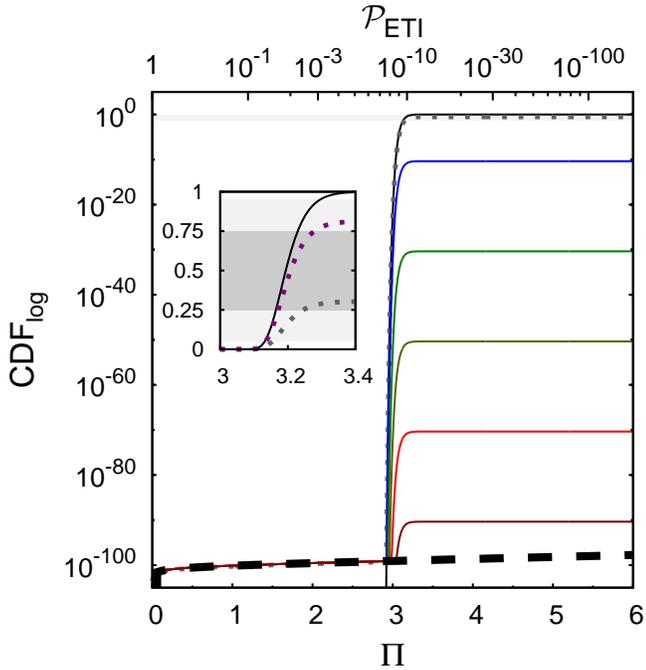}}
\figcaption{The posterior CDF using the log prior for a single detection.  The shading is the same as in Figure~\ref{fig:Credibility}, and the CDFs' line styles match their PDFs' styles in Figure~\ref{fig:LogPosterior}.  The purple dashed line in the inset is the CDF for $\varepsilon = 10^{-101}$.\label{fig:LogCredibility}}
\end{figure}

If there are no systematic errors, the log prior responds to the evidence well (solid black line in Figure~\ref{fig:LogPosterior}), showing a similarly sharp spike near $\PETI \approx 10^{-10}$ for $N_{\rm report} = 1$, but systematic errors wash away these features rapidly.  When the systematic errors are as small as $\varepsilon = 10^{-100}$ (dotted grey line), the spike seems basically unaffected, but the prior weights for other $\PETI$ are not exponentially suppressed.  This indicates that some posterior weight has already been lost from the spike.  In fact, when the CDF is calculated, it climbs only to a value of $0.3$ in this region (Figure~\ref{fig:LogCredibility}) --- we would conclude that $\PETI$ is most likely far below $10^{-10}$.  Decreasing $\varepsilon$ to $10^{-101}$ actually makes the detection credible (dotted magenta line, inset), if not secure.  But no real survey could rule out systematic errors to this degree, if only because of the possibility of fraud or outright hallucination.  Increasing $\varepsilon$ to only $10^{-90}$ already suppresses the spike's amplitude by a factor of $10^{10}$ (blue lines).  The CDF now plateaus at an insignificant value of $10^{-10}$.  While in a relative sense, this is still much greater than the prior CDF value of $\sim 10^{-100}$, it still amounts to disregarding the results in an absolute sense.  The stiffness of the log prior only grows with $\varepsilon$.  When $\varepsilon = 10^{-10}$, which might be a very conservative estimate for a real study, the CDF plateau has subsided to a mere $10^{-90}$.  As with the log log prior, the errors also blunt null results.  The CDF for large $\PETI$ falls proportionally to $\varepsilon \ga 10^{-100}$, though it hardly matters since the CDF was only $\sim 10^{-100}$ to start with.  In this scenario, the null result is entirely redundant; one is far more committed to the prior conclusion that there are no aliens in our past light cone than to any data.

While this demonstration uses an extremely small $\PMin$, it shows that the log log prior responds to both positive and negative results, avoiding a potential pathology of the log prior.  As noted above, actual surveys would need to account for their efficiency at detecting any ETIs who have ever lived.  In addition, the use of a single $\varepsilon$ is a very simplistic model.  More realistically, one might use some function to assign the probability that $N_{\rm host}$ host planets will be reported as $N_{\rm report}$ detections.

\section{Estimates of $\PCrowded$}
\label{sec:Results}
The entropy of finite systems, and so the log log prior has a reasonable cutoff for $\PMin$.  The ultimate upper bound on entropy is set by the cosmological constant; there are only finitely many possible states for an observable Universe.  It is unlikely that every last particle in every last galaxy needs to be precisely arranged for intelligence to evolve on a planet, though, motivating more stringent bounds on the entropy.  The entropy of a biosphere is limited by the amount of mass near a planet's surface, and it may be sufficient for a single intelligent organism to appear, in which case the entropy is bounded by the mass of an organism.  Finally, we can dispense with thermodynamic entropy altogether, since it mostly measures variations on the molecular level, and consider more abstract biological properties.

By default, I will assume that ``birthsite'' refers to a terrestrial planet.  From the Earth's example, a biosphere grows until it permeates the surface and ocean of a planet, filling essentially all habitable volume.  If the evolution of intelligence is constrained by global properties of the biosphere --- like which kinds of life prevail --- then a planetary biosphere is an appropriate choice for a birthsite.  But if the evolution of intelligence is constrained by a few local events, we should consider smaller, more numerous birthsites.  These sites can be individual sites for the origin of an ancestral protocell, or individual speciation events of intelligent species.  I set some upper limits on the number of possible birthing events in Section~\ref{sec:SmallHabitats}.

\begin{deluxetable*}{lcccccc}
\tablewidth{18cm}
\tablecaption{Log log prior estimates for $P_{\rm crowded}$\label{table:PCrowdedEstimates}}
\tablehead{\colhead{Entropy bound} & \colhead{${\cal S}$} & \multicolumn{4}{c}{Birthsite types} & \colhead{Section} \\ & & \colhead{FGKTP} & \colhead{TP} & \colhead{C} & \colhead{Max}}
\startdata
Number of birthsites &                     & $2 \times 10^{19}$ & $4.9 \times 10^{20}$ & $4 \times 10^{32}$ & $4 \times 10^{119}$\\
\hline
Cosmic             & $2.9 \times 10^{122}$ & 0.0135 & 0.0138 & 0.0154 & 0.0199 & \ref{sec:CosmicEntropy}\\
Biosphere          & $10^{48}$             & 0.0345 & 0.0351 & 0.0392 & 0.0509 & \ref{sec:ChemicalEntropy}\\
Organism           & $10^{29}$             & 0.0572 & 0.0582 & 0.0649 & 0.0842 & \ref{sec:ChemicalEntropy}\\
Protocell          & $10^{12}$             & 0.138  & 0.141  & 0.157  & 0.203  & \ref{sec:ChemicalEntropy}\\
Genome             & $4 \times 10^9$       & 0.173  & 0.176  & 0.196  & 0.254  & \ref{sec:GenomeEntropy}\\
Proteome           & $3.3 \times 10^7$     & 0.220  & 0.224  & 0.250  & 0.325  & \ref{sec:ProteomeEntropy}\\
Protein shape      & $1.5 \times 10^5$     & 0.320  & 0.326  & 0.363  & 0.472  & \ref{sec:ProteomeEntropy}
\enddata
\tablecomments{The birthsite types considered are terrestrial planets around F, G, and K dwarfs (FGKTPs), terrestrial planets (TPs), comets (C), and the maximum number from the \citet{Margolus98} bound (Max).}
\end{deluxetable*}

\subsection{Cosmological bounds on entropy: A lower bound on $\PCrowded$}
\label{sec:CosmicEntropy}

The Weak Copernican Principle follows from considerations of fundamental physics that limit the entropy of the observable Universe.  These limits require no assumptions about biology, and presumably could be developed by alien societies living in environments very different from the Earth.  The Bekenstein bound proposes that a weakly gravitating, isolated system has a maximum entropy of 
\begin{equation}
S_{\rm max} = \frac{2 \pi k_B E R}{\hbar c},
\end{equation}
where $E$ is the mass-energy of the system and $R$ is the radius of a sphere that can fully contain it \citep{Bekenstein81}.  The covariant version of the bound is more general and limits the entropy along null surfaces extending from the boundary of any region that fulfills some general conditions \citep{Bousso02}.  The observable Universe as a whole can be bounded by its cosmic particle horizon (CPH) and cosmic event horizon (CEH), which can serve as these null surfaces.  

The particle horizon surrounds every location in space that has ever been in the past light cone of the Earth.  The proper distance of the particle horizon at a time $t$ is 
\begin{equation}
R_{\rm CPH} = c a(t) \int_{0}^{t} \frac{dt^{\prime}}{a(t^{\prime})},
\end{equation}
where $a(t^{\prime})$ is the scale factor of the Universe at a time $t^{\prime}$ \citep[e.g.,][]{Davis04}.  Then the covariant Bekenstein bound limits the entropy along the past light cone the CPH to less than \citep{Bousso02}
\begin{equation}
S_{\rm CPH} \le \frac{k_B c^3}{G \hbar} \pi R_{\rm CPH}^2.
\end{equation}

According to the standard $\Lambda$CDM cosmology, the observable Universe also has an event horizon that surrounds every location in spacetime that can send a signal that will ever reach the Earth.  The proper distance of the cosmic event horizon at a time $t$ is \citep[as in][]{Davis04}:
\begin{equation}
R_{\rm CEH} = c a(t) \int_{t}^{\infty} \frac{dt^{\prime}}{a(t^{\prime})}.
\end{equation}
Event horizons, including the CEH have a maximal entropy, which is \citep{Gibbons77,Egan10}
\begin{equation}
S_{\rm CEH} = \frac{k_B c^3}{G \hbar} \pi R_{\rm CEH}^2.
\end{equation}
The CEH entropy, $2.9 \times 10^{122} k_B$, dwarfs the entropy of everything else contained within the observable Universe \citep[as noted by][]{Egan10}.

As long as dark energy is stable, the entropy of the observable Universe can never be much greater than the present $S_{\rm CEH}$.  Therefore, we can use $S_{\rm CEH} / k_B$ for a maximum value of the dimensionless entropy, ${\cal S}_{\rm cosmic}$.  Then, according to the log log prior, the probability that we are not isolated is $1.4\%$ (Table~\ref{table:PCrowdedEstimates}).  The odds highly favor our being isolated, but not to the point that it is an absurd hypothesis.

\begin{figure}
\centerline{\includegraphics[width=9cm]{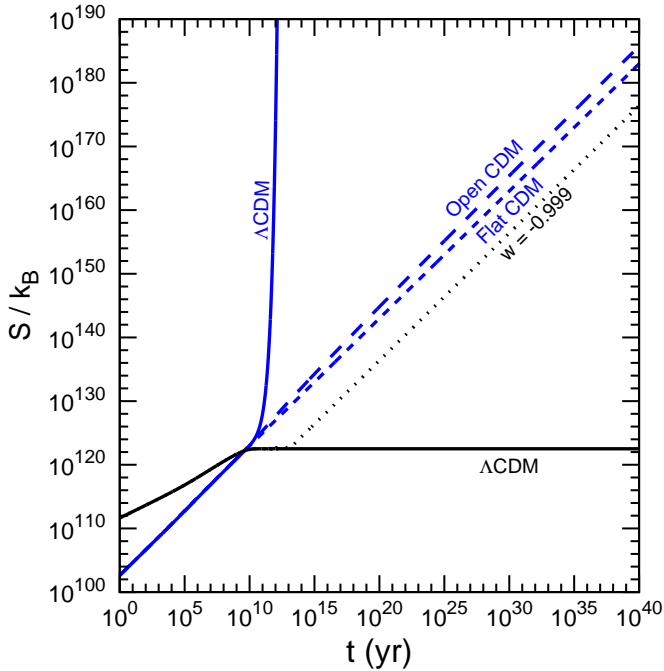}}
\figcaption{Time evolution of the maximum entropy contained in the Universe's particle (blue) and event (black) horizons, according to the Bekenstein bound.  The cosmologies considered are the standard $\Lambda$CDM cosmology (solid), a quintessence cosmology with $w = -0.999$ (dotted), a flat CDM cosmology with $\Omega_m = 1 - \Omega_r$ (short-dashed), and an open CDM cosmology with $\Omega_{\Lambda} = 0$ and $\Omega_m$ equal to its value in our Universe (medium dashed).  $H_0$ has the same value in all cases. \label{fig:HorizonEntropy}}
\end{figure}

Since the log log prior could have been developed before there was compelling evidence for the cosmological constant, or by aliens living in the distant past or future, the log log prior is guaranteed to have a reasonable cutoff only if $\ln {\cal S}_{\rm cosmic}$ doesn't vary too much with the cosmology.  The evolution of $S_{\rm CEH}$ for the $\Lambda$CDM cosmology is plotted in Figure~\ref{fig:HorizonEntropy} as the solid black line.  Its value is stable into the indefinite future, and $\ln S_{\rm CEH}$ is only $\sim 10\%$ smaller $1$ year after the Big Bang.  The particle horizon (solid blue line) could have been used to set more constraining limits on $\ln {\cal S}_{\rm cosmic}$ in the distant past, but will not be an effective limit in the distant future.  If the smaller of both horizons is chosen, the horizon entropy evolution only affects $\ln {\cal S}_{\rm cosmic}$ by $\sim 20\%$. 

If dark energy is not a cosmological constant, the future evolution of the cosmic event horizon will be dramatically different.  Phantom dark energy ($w < -1$) results in the event horizon collapsing as the Universe's scale factor approaches infinity in a finite time.  Unless $w$ is minutely more than $-1$, the collapse affects $\ln S_{\rm CEH}$ at the factor of $2$ level only very close to the end (less than $1\ \Myr$).  It is unlikely that an ETI would evolve in this narrow window.  If dark energy is quintessence ($-1 < w < -1/3$), $R_{\rm CEH}$ stalls at its current radius for a short time, but ultimately starts growing linearly with time (dotted black line in Figure~\ref{fig:HorizonEntropy}).  The value of $\ln S_{\rm CEH}$ does not exceed twice its current value for over $10^{40}$ years, but it does grow without limit.  It is possible that all baryonic matter will have decayed by then (the effects of hypothetical proton decay are described in \citealt{Adams97} among others), so there will be no ETIs drawing wildly divergent conclusions, but the estimate is clearly not perfectly robust.

Until relatively recently, cosmologies without dark energy were plausible.  These cosmologies lack an event horizon, so the particle horizon would be the only limit to the cosmic entropy.  The dashed blue lines in Figure~\ref{fig:HorizonEntropy} show the evolution of the particle horizon entropy in open and flat cosmologies without a cosmological constant.  Its value is basically the same as in $\Lambda$CDM until the present and continues growing quadratically with time into the distant future.  Like the event horizon in quintessence cosmology, $\ln S_{\rm CPH}$ grows slowly, remaining within a factor of two for over $10^{40}$ years but never stabilizing.  The horizon entropy is therefore not a robust limit on $\gg 10^{100}$ year timescales in these cosmologies, if anybody is around then.

The CPH may also be much bigger than the radii I calculate here.  In inflationary theories, comoving regions much larger than the observable Universe were once in our past light cone and fell out of contact due to a period of exponential expansion \citep{Harrison91}.  For $\Lambda$CDM cosmology, the CEH should still bound the entropy of the Universe, but it would be an issue if dark energy decays away or if the Universe didn't have it in the first place.

\citet{Krauss07} questioned whether observers living in the distant future of our Universe would even be able to derive $\Lambda$CDM cosmology \citep[see also][]{Rothman87}.  Red dwarf stars can shine for over a trillion years, so it is conceivable that ETIs will evolve in these late times on planets around these stars (\citealt{Loeb16}; \citealt{Stevenson13} discusses the geological difficulties in maintaining habitability for that long).  Since all distant galaxies would have vanished beyond the event horizon by then, observers might conclude that there is nothing beyond their own host galaxy, which is sitting in an otherwise empty Universe.  Although they would know of no cosmic horizons, these observers could still apply the Bekenstein bound to limit the entropy in a sphere that safely contains their galaxy.  In that way, they would derive a $\ln {\cal S}_{\rm cosmic}$ that is not too different from $\ln (S_{\rm CEH} / k_B)$.

While fairly robust, the cosmological entropy bounds are greatly overpowered.  Most of the degrees of freedom in the Universe play no role in the evolution of life or intelligence, and can be neglected.  More realistic estimates of $\ln {\cal S}_{\rm cosmic}$ lead to higher estimates of $\PCrowded$.

\subsection{The thermodynamic entropies of habitats and $\PCrowded$}
\label{sec:ChemicalEntropy}
Actual organic life is made of chemical matter with far less entropy than allowed under Bekenstein's bound.  Common materials found in Earth's biosphere have an entropy of about $1\,k_B$ per baryon, as shown in Table~\ref{table:ChemicalEntropies} \citep{Lodders98}.  The actual specific entropy is a few times larger for lightweight gases like molecular hydrogen, and smaller for complex molecules.  Since living things are mostly made of and live in liquid water, with an entropy of $0.47\,k_B$ per baryon, I use $1\,k_B$ per baryon as an upper estimate for their entropy.\footnote{This is smaller than the Bekenstein bound by a factor of $R / (1\ \fm)$; the Compton wavelength of a nucleon is $1\ \fm$.}  The maximum dimensionless entropy of an organism or ecosystem of mass $M$ is then approximately
\begin{equation}
{\cal S}_{\rm max} \approx \frac{M}{\amu}.
\end{equation}

\begin{deluxetable}{lccccc}
\tablecaption{Molar entropies of selected materials\label{table:ChemicalEntropies}}
\tablehead{\colhead{Material} & \colhead{Formula} & \colhead{$m$} & \colhead{Phase} & \colhead{$S^0$} & \colhead{$s_{b}$} \\
 & & \colhead{($\amu$)} & & \colhead{$\left(\frac{\Joule}{\mole\,\Kelv}\right)$} & }
\startdata
Hydrogen             & $\HChem_2$                      & 2.02   & (g)  & 130.68 & 7.78\\
Helium               & $\HeChem$                       & 4.00   & (g)  & 126.15 & 3.79\\
Nitrogen             & $\NChem_2$                      & 14.00  & (g)  & 191.61 & 1.65\\
Oxygen               & $\OChem_2$                      & 16.00  & (g)  & 205.15 & 1.54\\
Methane              & $\CChem \HChem_4$               & 16.04  & (g)  & 186.26 & 1.40\\
Water                & $\HChem_2 \OChem$               & 18.02  & (l)  & 69.95  & 0.47\\
                     &                                 &        & (g)  & 188.84 & 1.26\\
Sodium               & $\NaChem^+$                     & 22.99  & (a)  & 58.45  & 0.31\\								
Ethane               & $\CChem_2 \HChem_6$             & 30.07  & (g)  & 229.60 & 0.92\\
Chlorine             & $\ClChem^-$                     & 35.45  & (a)  & 56.60  & 0.19\\
Argon                & $\ArChem$                       & 39.95  & (g)  & 154.85 & 0.47\\
\parbox[c]{0.5cm}{Carbon\\dioxide} & $\CChem \OChem_2$               & 44.01  & (g)  & 213.79 & 0.58\\
Halite               & $\NaChem \ClChem$               & 58.44  & (s)  & 72.1   & 0.15\\
\parbox[c]{0.5cm}{Calcium\\carbonate} & $\CaChem \CChem \OChem_3$       & 100.09 & (s)  & 91.7   & 0.11
\enddata  
\tablecomments{The molecular mass is $m$ and the molar entropy of the material is $S^0$.  The entropy per baryon is $s_{b} \equiv S^0 / (N_A m k_B)$.  Values are for standard temperature and pressure, from Table 1.21 of \citet{Lodders98}.  The phases are (g) for gases, (l) for liquids, (s) for solids, and (a) for solutions in water.}
\end{deluxetable}

I consider three possible systems that might need to evolve in order for an alien society to exist: a planet's biosphere, an individual organism, and a single protocell.  

Biological species evolve in response to their planetary environment.  They also depend on their environments for sustenance.  Technological societies capable of interstellar communication require access to great amounts of materials and energy.  The maximum number of external factors in the planetary environment that could contribute to the evolution of an alien society is given by the chemical entropy of their host biosphere.  On the Earth, most of the biosphere lives in the hydrosphere, which has a mass of $1.7 \times 10^{24}\ \gram$ \citep{Lodders98}.  The entropy of the hydrosphere and its environment is at most $4 \times 10^{47} k_B$, which I round up to give a fiducial value of ${\cal S}_{\rm biosphere} = 10^{48}$.  I then find $\PCrowded$ is $3.5\%$ (Table~\ref{table:PCrowdedEstimates}).

Of course, the evolution of an organism does not depend on every degree of freedom in the biosphere, most of which describe minute arrangements of individual, distant molecules.  The number of possible organisms is limited by the entropy of an individual, suggesting this provides a more realistic lower limit on the probability that a planet will produce a given organism.  Since a human body has a mass of $\sim 100\ \kg$, with an entropy of $\sim 6 \times 10^{28}\ k_B$, I adopt a fiducial value of ${\cal S}_{\rm organism} = 10^{29}$ for individual organisms.  This entropy bound gives a $\PCrowded$ of $5.8\%$ (Table~\ref{table:PCrowdedEstimates}).  

Finally, it is possible that the rate limiting step for the evolution of ETIs is the origin of life itself.  The origin of life has been hypothesized to be an extremely rare event requiring a precise combination of molecules \citep[e.g.,][]{ConwayMorris03,Yockey00}.  I assume that the worst case is that an entire protocell must be generated from a thermal fluctuation.  A protocell is presumably no more massive than a modern prokaryote, $\sim 5 \times 10^{11}\ \amu$, from the mean carbon mass of a prokaryote multiplied by $10$ to include its water \citep{Whitman98}.  Again I round this up to the next power of $10$ and arrive at a maximum protocell entropy of ${\cal S}_{\rm protocell} = 10^{12}$.  The resulting log log prior estimate for the probability that we are not isolated is $14\%$ (Table~\ref{table:PCrowdedEstimates}).

It is unlikely that the probability an organism evolves on a planet is as small as implied by its chemical entropy.  Thermal fluctuations in the planetary environment are expected to produce organisms at least that often, subject to the availability of materials and internal energy.  $\PETI$ is as low as these $\PMin$ if we appeared as thermal fluctuations; in effect, we would be Boltzmann brains.  While that's possible under the laws of physics, the memories of Boltzmann brains are unreliable since they are also the result of the random fluctuations rather than actual life of the organism.  All of our knowledge, including that leading to the conclusion that we arose from random fluctuations, would then be completely unrelated to the external world, so adopting this conclusion is self-defeating \citep[as in cosmology,][]{Albrecht04}.  On a philosophical basis, we probably have to assume the likelihood we exist is much greater.  The Boltzmann brain problem does not apply to the protocell entropy estimate, however.  It does not matter if the internal state of a protocell reflects its history, as long as its descendents include sentient beings with memories that are reliable.

Since the number of possible living beings is limited by the entropy of an individual organism, is there any point to using the biosphere and cosmic entropy estimates?  While they almost certainly overestimate the number of relevant factors, they are in a sense more robust, since organisms are not closed thermal reservoirs.  Not all microstates are equally likely, since an organism can self-correct and radiate away ``errors'' into the surrounding environment \citep[as in][]{Lloyd00}.  For example, one could imagine the natural state of an ecosystem is an ocean filled with just one species of cells that have no mutations because they use error correction mechanisms.  In order for multicellular organisms to appear, some external factor would have to disrupt this ecosystem.  Or consider our own planet: in most microstates, there is no widespread life.  Yet, life will continue to thrive on the Earth despite the smaller number of microstates where this happens, barring some catastrophe inflicted from the outside, because the Earth is not a closed system.  In addition, microstates can be blocked because of energetic barriers that prevent certain chemical reactions.  

Life also needs certain elements in order to exist.  Conserved quantities like electric charge cannot be altered by thermal fluctuations.  The amount of each element in a biosphere is also effectively conserved.  A pure water ocean that contained no carbon could not give rise to life.  Although it is hypothetically possible to form necessary elements through pycnonuclear reactions, these rates are very slow \citep{Adams97}.  The limited lifespan of a planet without the necessary elements for life would have a much smaller $\PETI$ than implied by ${\cal S}_{\rm organism}$.  Finally, if chemicals are dilutely spread throughout an ocean, they must be concentrated into a small region in order to form into an organism.  

At the very least, one would probably have to consider whether the astrophysical environment of a planet is compatible with life.  Of course, most of the microscopic details of the sun are irrelevant for life, and the amount of extra entropy introduced by fine tuning the star's bulk properties is insignificant compared to the chemical entropies.  After all, life must be robust to survive, and not easily killed because of some minor thermal fluctuation in the sun.

While these problems are unlikely to occur in practice, they provide a philosophical basis for using the larger entropy estimates as very conservative limits on ${\cal S}_{\rm all}$.

\subsection{The entropy of the human genome: A best estimate for $\PCrowded$}
\label{sec:GenomeEntropy}
A common line of argument against the existence of aliens is that it is unlikely that \emph{Homo sapiens} would evolve on a different planet \citep{Simpson64,Mayr01,Lineweaver09}.  The probability that this would happen is bounded by the number of possible species of DNA-based organisms, assuming DNA-based organisms are fairly common.  A species can be defined by its genome, and the amount of information defines a genetic entropy.  The genetic entropy ${\cal S}_{\rm genome}$ provides the most reasonable bound on ${\cal S}_{\rm all}$, as long as life frequently arises on planets (more often than $1$ in $e^{{\cal S}_{\rm genome}}$).

The genome of a DNA based organism consists of a sequence of $\ell_{\rm bp}$ base pairs, each of which is either adenine, cytosine, guanine, or thymine (uracil when transcribed into RNA) on one strand of the DNA.  The other strand of the DNA then usually has thymine, guanine, cytosine, or adenine, respectively.  The human genome has $3 \times 10^9$ base pairs \citep{Venter01,Abdellah04}.  The DNA is interpreted according to a genetic code, with each three letter combination of base pairs (codon) corresponding to an amino acid.  These amino acids are assembled into proteins according to the sequence of codons in the genome \citep{Crick68}.  Earth life mostly uses $20$ amino acids, and two codons are reserved to signal the starting point and ending point of a gene.  However, many other amino acids exist; the genetic code that arises on different planets could map any codon to any amino acid, leading to vastly different organisms from the same DNA sequence.  

The genome entropy of \emph{Homo sapiens} is given by
\begin{multline}
{\cal S}_{\rm genome} = \ln {\cal N}_{\rm sequence} + \ln {\cal N}_{\rm codes} + |\ln {\cal P}_{\rm length}| \\
+ \ln {\cal N}_{\rm expression} - \ln {\cal N}_{\rm individuals},
\end{multline}
with ${\cal N}_{\rm sequence}$ being the number of possible DNA sequences with a length $\ell_{\rm bp}$, ${\cal N}_{\rm codes}$ is the number of possible genetic codes, ${\cal P}_{\rm length}$ is the probability that an organism has a genome with length $\ell_{\rm bp}$, ${\cal N}_{\rm expression}$ is the number of possible ways to express genes in all the cells of the body, and ${\cal N}_{\rm individual}$ is the number of possible DNA sequences that are part of a typical species.  Almost all of the entropy derives from the information in the DNA sequences themselves.  If each base pair is chosen independently, the number of possible  sequences is given by
\begin{equation}
\ln {\cal N}_{\rm sequence} = \ell_{\rm bp} \ln 4 = 4 \times 10^9,
\end{equation}
which is $4 \times 10^9$ for \emph{Homo sapiens}.  If there are $N_{\rm amino}$ amino acids that can be assigned randomly and independently to triplet base pair codons, the number of genetic codes is given by
\begin{equation}
\ln {\cal N}_{\rm codes} = 4^3 \ln N_{\rm amino}.
\end{equation}
Even if $N_{\rm amino}$ is $\sim 1,000$, and extraterrestrial genetic codes have $>4$ nucleotides \citep{Baross07} and $>3$ base pairs per codon \citep[e.g.,][]{Anderson04}, the genetic code entropy can be ignored.  The length of the genome of a human-like organism is bounded to $\ll 10^{13}$ simply by the number of baryons in a typical cell \citep[from the cell counts in][]{Bianconi13}.  As a worst case, if each possible length between $1$ and $10^{13}$ is equally likely, ${\cal P}_{\rm length} = 10^{-13}$.  Therefore, the length of the genome makes an insignificant contribution to ${\cal S}_{\rm genome}$.

The behavior of a cell can vary greatly depending on how the genes encoded in the DNA are expressed.  Different combinations of gene expression lead to different types of body cells.  Suppose an organism contains $N_{\rm celltypes}$ types of cells, and its genome includes $N_{\rm genes}$ genes that can be in $N_{\rm switch}$ states.  Then the number of possible combinations is given by
\begin{equation}
\ln {\cal N}_{\rm expression} = N_{\rm celltypes} N_{\rm genes} \ln N_{\rm switch}.
\end{equation}
The human body starts from $N_{\rm celltypes} = 1$ type of cell, the zygote, and grows into an adult with at least $411$ distinct cell types \citep{Vickaryous06}.  The genome codes $N_{\rm genes} \approx 2.2 \times 10^4$ genes \citep{Abdellah04}.  A value of $N_{\rm switch} = 2$ is enough for an enormous diversity of cell expressions, as noted by \citet{Kauffman95}.  I will use $\ln N_{\rm switch} = 1$ as a generic value, but even if $N_{\rm switch} = 1,000$, $\ln N_{\rm switch}$ would increase by only a factor of $\sim 7$.  These values imply that $\ln {\cal N}_{\rm expression} \approx 2 \times 10^4 \text{--} 9 \times 10^6$, a $\sim 0.2\%$ correction on the genetic entropy that I neglect.

To estimate the number of possible individuals in a species, I assume that a fraction $\varpi$ of the base pairs in an individual's genome can differ from some baseline sequence.  This fraction measures the genetic variability in a species.  I ignore any considerations of where those variations may occur in the genome; an individual is part of a species as long as there are $\le \varpi \ell_{\rm bp}$ differences from the baseline sequence.  Each mutation can be to one of $3$ base pairs.  Thus, the number of possible individuals is 
\begin{equation}
{\cal N}_{\rm individual} = \sum_{i = 0}^{\varpi \ell_{\rm bp}} 3^{i} \binom{\ell_{\rm bp}}{i} < \varpi \ell_{\rm bp} 3^{\varpi \ell_{\rm bp}} \binom{\ell_{\rm bp}}{\varpi \ell_{\rm bp}}.
\end{equation}

Humans today have relatively little genetic diversity, with $\varpi \la 0.1\%$, but this is a lower limit on the possible diversity within \emph{Homo sapiens} \citep{Li91,Sachidanandam01}.  An upper limit on $\varpi$ comes from the divergence of the human genome from the genomes of chimpanzees and bonobos, which is $\sim 1.2 \text{--} 1.3\%$ \citep{Mikkelsen05,Prufer12}.  I adopt a value of $1\%$, so that $\ln {\cal N}_{\rm individual} \la 0.067 \times (3 \times 10^9) = 2 \times 10^8$.  This is a minor correction to $\ln {\cal N}_{\rm sequence}$ that I ignore.

The human body is home to trillions of microbes \citep{Sender16}, each potentially having its own genome.  Some of these microbes may be vital to the body's functions \citep{Turnbaugh07}.  I ignore their contribution, though, since working out which ones are needed and the distinctiveness of their genomes is far beyond the scope of this work.

If the entropy is capped at ${\cal S}_{\rm genome} = 4 \times 10^9$, the log log prior implies that there is an $18\%$ chance that we are not isolated.  This is my best estimate --- more likely than not, we are alone in the observable Universe, but the possibility that aliens exist within our past light cone is a reasonable one.

\subsection{The proteome entropy and generous estimates for $\PCrowded$}
\label{sec:ProteomeEntropy}
The genes coded in DNA represent proteins, which actually are responsible for most biological functions.  The great majority of the DNA is non-coding \citep{Abdellah04}, however, and so might not affect the organism's traits.  Instead, the phenotype of an organism may be dependent on its proteome, the collection of proteins encoded in its genes and the ways those proteins are expressed in its cells \citep{Wilkins96}.  I make higher estimates of $\PCrowded$ from the proteome entropy.

We can calculate the proteome entropy as
\begin{equation}
{\cal S}_{\rm proteome} = N_{\rm coded} \ln {\cal N}_{\rm proteins} + \ln {\cal N}_{\rm expression},
\end{equation} 
where ${\cal N}_{\rm proteins}$ is the number of different proteins possible, $N_{\rm coded} \ga N_{\rm genes}$ is the number of proteins coded into the genome, and ${\cal N}_{\rm expression}$ is the number of gene expression patterns.  The proteins, which are first synthesized as chains of amino acids, can be enumerated:
\begin{equation}
\ln {\cal N}_{\rm proteins} = \ell_{\rm protein} \ln N_{\rm amino} \approx 500 \ln 20.
\end{equation}
About $3.4 \times 10^7$ base pairs in the human genome code protein sequences, for a mean gene length of $\sim 1,500$ base pairs, or a mean protein length of $\sim 500$ amino acids \citep{Abdellah04}.  With $22,000$ genes, and assuming $1$ protein per gene (although this ratio is probably larger in practice; \citealt{Wilkins96}), there are up to $\exp(3.3 \times 10^7)$ possible combinations of proteins possible for a proteome the size of ours.  I previously found $2 \times 10^4 \text{--} 9 \times 10^6$ for $\ln {\cal N}_{\rm expression}$, so the proteome entropy is dominated by the first term.

The proteome entropy of ${\cal S}_{\rm proteome} = 3.3 \times 10^7$ is about $1\%$ of ${\cal S}_{\rm genome}$, as might be expected since $\sim 1\%$ of the human genome codes proteins \citep{Abdellah04}.  Using this value slightly raises $P_{\rm crowded}$ to $22\%$, leaving the basic conclusion of the previous section unchanged.

Even more abstractly, it's possible that only the basic shape of a protein matters for its function, not the actual sequence of amino acids in the proteins \citep{Dill99,Dryden08}.  The actual number of possible shapes ${\cal N}_{\rm shape}$ seems to be quite low: \citet{Lau90} proposed a reduced sequence space with $\sim 10^{20}$ possible varieties, and \citet{Dill99} argued that there are perhaps $\sim 10^{10}$ functionally distinct structures.  The actual number of basic shapes might be a few thousand for a protein domain, where most proteins are built out of just a few of these domain units \citep{Rose06}.  I will consider an extremely low value of ${\cal N}_{\rm shape} = 1,000$.  In addition, patterns of gene expression still add $\ln {\cal N}_{\rm expression}$.  Following this line of thought,
\begin{align}
\nonumber {\cal S}_{\rm shape} & = N_{\rm coded} \ln {\cal N}_{\rm shape} + N_{\rm celltypes} N_{\rm genes} \ln N_{\rm switch} \\
                               & \approx 1.5 \times 10^5.
\end{align}
This most liberal estimate leads to $P_{\rm crowded}$ values of $33\%$.  The protein shape entropy should be regarded as highly speculative, since it ignores all actual chemistry.  

In principle, there could be a phenotype entropy describing the combinations of possible body traits while ignoring everything on the biochemical level.  If the phenotype entropy is small enough, the Universe being crowded is favored by the prior; ${\cal S}_{\rm phenotype} = 100$ implies $\PCrowded \approx 84\%$, for example.  Using it requires a way to enumerate ``body traits'', which is beyond the scope of this work.  The trait list would also have to be fine-grained enough to not simply assume that intelligence is common \citep[c.f. the ``skeleton space'' in][]{ConwayMorris03}: equation~\ref{eqn:PETIasNegentropy} assumes the worst case is that each combination of traits is equiprobable, which is clearly not the case if the only trait we consider is ``has a big brain''.  Also, a phenotype entropy would apply only if complex multicellular organisms evolve frequently, or if the state space includes a lot of parameters describing individual cell morphologies, to account for the probability that single-celled organisms do not evolve into complex multicellular life forms.  Lastly, for phenotype entropy to be useful, the evolution of intelligence must not depend on any of the details of the organism's biochemistry.

\subsection{What if birthsites are much smaller than planets?}
\label{sec:SmallHabitats}
Both $N_{\rm LC}$ and ${\cal S}_{\rm all}$ affect the calculated $P_{\rm crowded}$.  Although I have been assuming that the birthsites for ETIs are the entire histories of whole planets, that may not be appropriate.

Within the Solar System, the moons Europa, Titan, and Enceladus are fairly widely considered to be possible habitats for life \citep{Chyba05,Lammer09,Lunine09}.  In addition, liquid water oceans plausibly existed (or still exist) in the larger icy moons and Kuiper Belt objects \citep{Hussmann06}.  Habitable regions may have existed in large carbonaceous asteroids early in the Solar System's history, making them a conceivable, but remote habitat for life \citep{Abramov11}.  \citet{Dyson03} has even proposed that non-Earthlike life could evolve in the Kuiper Belt.  All of these worlds in principle could raise the number of world habitats in the Solar System to dozens, although they may have too little free energy or insufficient materials to support complex life \citep{Lammer09}.

A very generous estimate could be that every body in the Solar System that is at least $\sim 1\ \km$ wide is a possible birthsite.  By number, most of these bodies are comets in the Oort Cloud, where there are perhaps $\sim 10^{13}$ of them \citep[e.g.,][]{Weissman96}.  To extrapolate this estimate to other star systems, I assume that the number of comet birthsites scales with the mass of the host star: a stellar population of mass $M_{\star}$ hosts $N_{\rm LC} = 10^{13} (M_{\star} / \Msun)$ birthsites.

As shown in Table~\ref{table:PCrowdedEstimates}, including all comets as birthsites raises $P_{\rm crowded}$ somewhat, to $1.5\%$ to $36\%$.  This slight increase comes from the fact that $N_{\rm LC}$ increases from $\sim 10^{21} \approx 10^{10^{1.3}}$ to $\sim 10^{33} \approx 10^{10^{1.5}}$.  But this apparent optimism comes with a price: about $3/4$ of the weight in $P_{\rm crowded}$ now comes from scenarios where more than one ETI arises per star system.  About half of the weight is for $\PETI \ga 10^{-4}$, for which our Solar System would have hosted $\ga 10^8$ species of intelligent life.  This is in dire contradiction with the null results of SETI and the lack of evidence for widespread interplanetary migration.  The prior probability that the Universe is crowded but the Solar System is not, with $\la 1$ ETI arising per $\Msun$ of stars, is reduced to a few percent.  

If the main bottleneck for the appearance of intelligence is the origin of life, the number of birthsites might increase vastly further still.  It could be that every pool, every rock, every bubble provides an independent opportunity for the origin of life.  When the chance of aliens evolving is just the chance of any life appearing at all, every new chance for abiogenesis must be considered a birthsite.

Additional birthsites may also be located at different times in the Solar System's evolution, not just different places.  As the Sun expands into a red giant, many of the outer icy worlds will become warm enough for surface liquid water oceans \citep{Lopez05}.  Titan, especially, may evolve from a world that could host hypothetical methane dwelling life-forms to one hosting completely independent water dwelling life-forms \citep{Lorenz97}.  A world can also host many birthsites, with multiple chances to play out life's evolution, if mass extinctions devastate it frequently and if the bottlenecks occur late in evolution \citep{Cirkovic09-Reset}.  Early in the Earth's history, life may have arisen many times after being repeatedly destroyed by impacts \citep{Sleep89}.  If the development of ETIs requires a particularly robust kind of life prevailing in the biosphere, then Earth may have had several opportunities to develop it.  

The highest estimate for the number of birthsites in the Universe is provided by the \citet{Margolus98} (ML98) bound, a limit on how fast the quantum state of a system can change.  This is a necessary requirement for any kind of dynamics, including biological processes.  We can divide the Universe at any given moment into subsystems which may serve as birthplaces for life: these can be as large as the entire observable Universe or as small as individual baryons.  According to the ML98 bound, the rate of dynamics is limited to $\Gamma_{\rm M} \le 2 \Delta E / h$, where $\Delta E$ is the average difference between the internal energy of the system and its ground state energy \citep{Margolus98}.  Then the number of birthsites in a region of spacetime is found by integrating each system's $\Gamma_{\rm M}$ over its volume and adding them together.  Since $\Gamma_{\rm M}$ is linear in energy, the total number of birthsites is directly proportional to the amount of internal energy in the region.  For cold systems like planets and stars with masses of $M$, $\Delta E < M c^2$.  Within the past light cone, the baryonic birthsite number is below
\begin{equation}
N_{\rm LC}^{\rm max} = \int_0^{10^3} \frac{8\pi}{3}\frac{D_C^3 (z) \rho_b^{\rm com}(z) c^2}{h} \left(\frac{dt}{dz}\right) dz = 4.3 \times 10^{119},
\end{equation}
where $D_C (z)$ is the comoving distance of an object observed at redshift $z$ along the past light cone and $\rho_b^{\rm com} = \Omega_b \times 3 H_0^2 / (8 \pi G) \times (\Omega_{\Lambda} + \Omega_m (1 + z)^3 + \Omega_r (1 + z)^4)$ is the comoving baryonic density \citep{Hogg99}.  

The values for $P_{\rm crowded}$ now range from $2\%$ to $47\%$ (Table~\ref{table:PCrowdedEstimates}), about twice the values expected for terrestrial planet birthsites.  The relatively limited effect is because $10^{120}$ is still only $\sim 10^{10^2}$, compared to the $\PMin$ values of $10^{-10^5}$ or less; $\ln \ln 10^{120}$ isn't much bigger than $\ln \ln 10^{21}$.

If birthsites did not interfere with each other, then virtually all of the weight in $P_{\rm crowded}$ is for scenarios where there is more than one intelligent species per terrestrial planet, as with the cometary birthsite case.  In fact, of the crowded Universe scenarios, about half of the weight is for $\PETI \ga 10^{-10}$, naively implying $\sim 10^{90}$ ETIs per planet.  The prior probability that the Universe is crowded but not every planet is packed with intelligences is just $\sim 0.1\%$.  But interaction between birthsites can reduce the number of intelligent species that actually appear on a planet.  Even if life originates a great number of times on a world, only one biosphere may result, either through merger and symbiosis or elimination and competition.

One can also consider more restrictive definitions of birthsites.  \citet{Zackrisson16} estimates that most terrestrial planets orbit red dwarfs.  The habitable lifespan of such planets might greatly exceed the age of the Universe \citep{Stevenson13}, and their $\PETI$ values may include intelligences that won't evolve for tens of billions of years.  If one only considers those orbiting F, G, or K dwarfs to be potential habitats, then the number of birthsites falls to $N_{\rm LC} = 2 \times 10^{19}$.  While the resulting $P_{\rm crowded}$ values are slightly smaller, the decrease is of order $\sim 0.3\%$ (total) or $\sim 3\%$ (relative) (Table~\ref{table:PCrowdedEstimates}).

\subsection{A note on small probabilities}
The probabilities considered for $\PMin$ defy our normal intuitions, being vastly smaller than the probabilities of many situations that seem to violate common sense.  Rare astronomical phenomena, like nearby gamma-ray bursts, are not so rare compared to these kinds of odds.  Likewise, whatever the odds are for panspermia, the transfer of life between planets, they are probably high enough that its entropy cost is relatively small \citep[e.g.,][]{Napier04,Worth13}.  A transfer of life from Mars to Earth already is a widely considered possibility \citep[e.g.,][]{Davies03}, but transfers between the Earth and the outer Solar System, across interstellar space, or passing back and forth between the inner planets a hundred times would not dent the protocellular entropy.  

By itself, this does not mean panspermia is likely --- if life is equally likely to arise and thrive on the Earth as the other worlds, then it probably started here.  But if other worlds were much more conducive to forming life for some reason, these greater probabilities could easily offset the low probability that life would survive a transfer between a distant world and the Earth.  For example, \citet{Lunine92} hypothesized that Triton had a thick hydrogen atmosphere shortly after its capture by Neptune, a bit like the simulated atmosphere in the Urey-Miller experiment \citep[see][]{Chyba05}.  If that chemical environment increased the probability of a limiting step from $10^{-200}$ on Earth to $10^{-100}$, for example, a Triton origin for life would not be so implausible.  

Or we can go much further.  One issue that makes chemical evolution difficult is that the synthesis of different chemicals favors different environments, and the reaction products tend to interfere with one another \citep{ConwayMorris03,Baross07}.  That difficulty might be evaded by placing the production sites far away from each other, even on different planets \citep[c.f.,][quoting a thought experiment by Robert Shapiro]{ConwayMorris03}.  Maybe on early Earth there was a pool that formed ribose, and it just so happened that a meteorite from a planet in a different star system delivered uracil to the pool, and \emph{coincidentally} a meteorite from a guanine planet landed at the same time in the same pool, and \emph{coincidentally} a meteorite from an adenine planet landed there too, and \emph{coincidentally} that's when a meteorite from a cytosine planet landed there, allowing these ingredients to mix into RNA.  It sounds silly, but if there are such planets, this scenario is still probably more likely than a protocell arising through thermal fluctuations.  As the probability falls, the number of bizarre paths to life increases, and it becomes essentially impossible to prove that there's only one thermodynamically unlikely way for life to arise.

\section{The reach and grasp of SETI surveys}
\label{sec:SETIReach}
The log log prior can inform judgments about which SETI surveys are effective.  While in principle we cannot conclusively say that we're isolated until we thoroughly check every habitat in our past light cone, we can be fairly convinced even if we check a fairly small fraction of them, simply because it's unlikely that $\PETI$ happens to lie very near $N_{\rm LC}^{-1}$.  Instead, we simply have to constrain most of the prior's weight before our opinion changes much --- that is when the posterior will start to diverge from the prior.

SETI surveys observe $N_{\rm targets}$ targets and look for some trace of a technological society.  However, it's important to distinguish between the \emph{reach} of a survey and its \emph{grasp}.  In this discussion, reach measures how many targets it observes, regardless of the survey efficiency.  I define the weighted reach as the weight assigned to hypotheses where $\PETI^{-1} \le N_{\rm targets}$.  For how much prior weight would the survey find a society if all societies that ever evolved left a detectable trace?  This can be estimated as
\begin{equation}
\label{eqn:WeightedReach}
{\cal R} = \frac{\ln (1 + \ln N_{\rm targets})}{\ln (1 + \ln N_{\rm LC})}.
\end{equation}
The maximum possible ${\cal R}$ is 1.

Grasp, as I define it, is how much of the prior weight could be ruled out if nothing is found:
\begin{equation}
{\cal G} = \int_0^{\Pi_{\rm LC}} {\cal L} (\PETI | {\rm discovery}) \frac{dP_{\rm prior}}{d\Pi} d\Pi
\end{equation}
The posterior $\PCrowded^{\rm posterior}$ is then $(1 - {\cal G}) \PCrowded^{\rm prior}$ after a null result.  The likelihood here is the estimated probability that at least one society will be found for a given value of $\PETI$.  It is different from reach because not all ETIs necessarily leave an observable trace.  First, we usually are uncertain about whether ETIs leave a given trace at all.  This uncertainty is systematic.  For example, whether or not aliens can build megastructures is uncertain; if megastructures are completely impossible, we will never find any no matter how many targets we search or how big $\PETI$ is.  In analogy with Section~\ref{sec:PosteriorExample}, let $\varepsilon_{\rm trace}$ be the subjective probability that aliens never leave a trace visible to the survey. 

Second, there is a statistical factor $\eta$ that measures the detection efficiency of a survey.  This factor might be parameterized as $\eta = \eta_{\rm visible} \eta_{\rm time}$, where $\eta_{\rm visible}$ is the actual fraction of societies that ever leave a given trace and $\eta_{\rm time}$ is the fraction of the traces ever produced that would be visible now.  The former factor could actually be greater than $1$ if societies replicate by interstellar travel, although it increases the survey's grasp only if the progenitor society would not have been visible in the original survey \citep{Brin83}.  The latter factor is roughly the ratio of the trace's lifespan and the Universe's age.  An analogy can be made with Drake's equation: $\eta_{\rm visible}$ corresponds to $f_c$, the probability that an intelligent species forms a communicative society, and $\eta_{\rm time}$ corresponds to the lifetime of the society.  Note that a trace's lifetime may not equal the society's lifetime --- either because the artifacts outlast the society \citep{Corbet97,Carrigan12}, or because the society is visible only for a short phase of its existence \citep{Sagan73}.

After accounting for these factors,
\begin{equation}
{\cal G} = (1 - \varepsilon_{\rm trace}) \left[1 - \int_0^{\Pi_{\rm LC}} {\cal L} (\PETI | {\rm null~result}) \frac{dP_{\rm prior}}{d\Pi} d\Pi\right],
\end{equation}
remembering that $\varepsilon_{\rm trace} = 0$ means complete certainty that some intelligent lifeforms leave an observable trace and $\varepsilon_{\rm trace} = 1$ means the possibility is too far-fetched to consider.  Plugging in the log log prior,
\begin{equation}
\label{eqn:GraspApprox}
{\cal G} \approx (1 - \varepsilon_{\rm trace}) \frac{\ln [1 + \ln (\eta N_{\rm targets})]}{\ln (1 + \ln N_{\rm LC})},
\end{equation}
if $\eta N_{\rm targets} \ga 1$.  Note that if $\eta N_{\rm targets} \gg 1$, ${\cal G}$ depends very weakly on $\eta$ --- systematic uncertainties are far more effective at reducing the grasp.  It's possible to construct more advanced models, where the $\varepsilon_{\rm trace}$ factor is replaced by some prior over possible values of $\eta$.  

The grasp requires subjective assessments that are beyond the scope of the paper, but the weighted reaches of SETI surveys is easy to calculate.  These are listed in Table~\ref{table:SurveyReaches}.  According to the log log prior, the first few orders of magnitude provide more weight than the rest.  With the fiducial number of terrestrial planet birthsites, ${\cal R} = 0.5$ is achieved in a survey of $700$ ($1,700$) stars (F, G, or K dwarfs) expected to host $400$ ($300$) terrestrial planets.  Thus, according to the log log prior, if the Universe is crowded, we expect the nearest remains of intelligent lifeforms to be in the Milky Way (${\cal R} = 0.8$).  Additionally, an examination of just one world beyond Earth effectively has a weighted reach of order $\ln[1 - \ln (1/2)] = 0.1$, as long as it is a potential habitat.

\begin{deluxetable*}{lcccccccc}
\tablecaption{Weighted reaches of some SETI surveys\label{table:SurveyReaches}}
\tablehead{\colhead{Survey} & \colhead{Method} & \multicolumn{2}{c}{Number of stars} & \multicolumn{4}{c}{${\cal R}$} & \colhead{Notes} \\ & & \colhead{FGK} & \colhead{Any} & \colhead{FGKTPs} & \colhead{TPs} & \colhead{C} & \colhead{Max} & }
\startdata
Solar System                        & Inspection   & $1$                  & $1$                  & 0.194 & 0.191 & 0.792 & 0.964 & (a)\\
META (Type 0 beacons)               & Radio        & $15$                 & $119$                & 0.176 & 0.420 & 0.817 & 0.968 & (b)\\
GBT Kepler field                    & Radio        & $86$                 & $86$                 & 0.445 & 0.437 & 0.822 & 0.968 & (c)\\ 
STACEE                              & Optical      & $187$                & $187$                & 0.393 & 0.442 & 0.827 & 0.969 & (d)\\
HRMS / Project Phoenix              & Radio        & $1,000$              & $1,000$              & 0.476 & 0.510 & 0.838 & 0.970 & (e)\\
Earth 2000                          & Optical      & $6,176$              & $6,176$              & 0.535 & 0.560 & 0.847 & 0.971 & (f)\\
HabCat                              & Radio        & $17,129$             & $17,129$             & 0.576 & 0.595 & 0.854 & 0.972 & (g)\\
Kepler                              & Artifact     & $140,000$            & $150,000$            & 0.631 & 0.645 & 0.866 & 0.974 & (h)\\
\citet{Carrigan09}                  & Artifact     & $1 \times 10^6$      & $8 \times 10^6$      & 0.673 & 0.717 & 0.882 & 0.976 & (i)\\
META (Type 1 beacons)               & Radio        & $1 \times 10^7$      & $8 \times 10^7$      & 0.716 & 0.752 & 0.893 & 0.978 & (b)\\
Milky Way                           & Isolation    & $1.8 \times 10^{10}$ & $1.5 \times 10^{11}$ & 0.818 & 0.837 & 0.928 & 0.983 & (j)\\
\citet{Annis99}                     & Artifact     & $3 \times 10^{12}$   & $2 \times 10^{13}$   & 0.871 & 0.884 & 0.947 & 0.987 & (k)\\
META (Type 2 beacons)               & Radio        & $4 \times 10^{12}$   & $3 \times 10^{13}$   & 0.875 & 0.887 & 0.949 & 0.987 & (b)\\
\citet{Zackrisson15}                & Artifact     & $3 \times 10^{13}$   & $2.0 \times 10^{14}$ & 0.892 & 0.902 & 0.956 & 0.988 & (l)\\
\GHAT                               & Artifact     & $1.8 \times 10^{15}$ & $1.5 \times 10^{16}$ & 0.927 & 0.933 & 0.971 & 0.991 & (m)\\
Galaxy Zoo                          & Artifact     & $1.8 \times 10^{16}$ & $1.5 \times 10^{17}$ & 0.944 & 0.949 & 0.979 & 0.993 & (n)\\
\citet{Lacki16}                     & Artifact     & $6.6 \times 10^{16}$ & $5.3 \times 10^{17}$ & 0.953 & 0.957 & 0.983 & 0.994 & (o)\\
\citet{Olson15}                     & Isolation    & $1.3 \times 10^{18}$ & $1.0 \times 10^{19}$ & 0.973 & 0.975 & 0.993 & 0.995 & (p)
\enddata
\tablecomments{The intra-Galactic surveys typically report the number of stars they observe, whereas the extragalactic surveys report the number of galaxies examined or are distance limited.  To convert into the number of birthsites, I use the present-day stellar mass function from  Table 1 of \citet{Chabrier03}.  I find there are $2.5$ M dwarfs (masses between $0.08$ and $0.6\ \Msun$) per $\Msun$ of stellar population, and their mean mass is $0.22\ \Msun$; there are $0.37$ F, G, or K (FGK) dwarfs (stars with masses between $0.6$ and $1.2\ \Msun$, not adjusting for those off the main sequence) per $\Msun$, and their mean mass is $0.83\ \Msun$.  When an extragalactic distance is given, I use a mean stellar density of $2.3 \times 10^8\ \Msun\ \Mpc^{-3}$ (comoving) from \citet{Baldry12}.  When only the number of galaxies observed is given, I assume that all of the galaxies have the stellar mass of the Milky Way, $5 \times 10^{10}\ \Msun$ \citep[as quoted in][]{Zackrisson16}.  Then, according to \citet{Zackrisson16}, there are $0.064$ terrestrial planets per $\Msun$ of F, G, or K dwarfs (FGKTPs), and $1.5$ terrestrial planets per $\Msun$ of F, G, K, or M dwarf (TPs) for this IMF.\\
The number of cometary (C) birthsites is $10^{13}$ per $\Msun$ of stellar population mass.  I compute the number of maximal (Max) birthsites directly from the stellar population masses, not attempting to include non-stellar mass or exclude non-planetary mass, and I use stellar ages of $12$ Gyr (except for the Solar System).}
\tablenotetext{a}{I count Mercury, Venus, and Mars as the terrestrial planets, excluding Earth because of anthropic bias.  For the number of maximal birthsites, I use an age of $4.56$ Gyr.}
\tablenotetext{b}{META was an all-sky survey for narrowband radio emission \citep{Horowitz93}.  The ``types'' of beacons refer to \citet{Kardashev64}'s scale.  These reaches apply for radio beacons that radiate continuously and isotropically.  For Type 0 and Type I beacons, \citet{Horowitz93} estimates the number of ``Sun-like'' stars that could have been detected.  I take this to mean the number of F, G, and K dwarfs, and I convert it into a total stellar mass.  For Type II beacons, I extrapolate from the given range of $22$ Mpc.}
\tablenotetext{c}{\citet{Siemion13} used the Green Bank Telescope to look for artificial radio emission from $86$ Kepler field stars known to have exoplanets.  I assume each of these stars hosts $1$ terrestrial planet.}
\tablenotetext{d}{STACEE, a Cherenkov gamma-ray telescope, observed Solar-like stars in the HabCat catalog while waiting for optimal times to observe its main target.  Because it detects gamma rays through the Cherenkov light their particle showers produced, it is sensitive to optical flashes characteristic of laser light \citep{Hanna09}.}
\tablenotetext{e}{HRMS was a canceled microwave SETI project that aimed to observe $1,000$ Solar-like stars.  Project Phoenix, a private successor project, used a catalog of $2,000$ stars and observed about $200$ per year \citep{Turnbull03}.}
\tablenotetext{f}{Earth 2000 was a dedicated optical SETI project that targeted Solar-like stars \citep{Howard04}.}
\tablenotetext{g}{The HabCat is a selection of Solar-like stars that are intended to be examined by radio SETI surveys \citep{Turnbull03}.}
\tablenotetext{h}{\emph{Kepler} searches for natural exoplanets by detecting flux variations when they eclipse their host star, but it could detect transiting artifical structures too \citep{Arnold05}. Of the 150,000 highest priority stars observed, about 140,000 are F, G, K, or A dwarfs \citep{Batalha10}.  KIC 8462852 undergoes anomalous eclipses, demonstrating that unusual transits could be noticed in the \emph{Kepler} photometry (\citealt{Boyajian16}; an artificial explanation was noted in \citealt{Wright16}).}
\tablenotetext{i}{\citet{Carrigan09} examined sources in the IRAS catalogs to see if they were Dyson spheres within the Milky Way.  The survey should have been sensitive to a volume including $10^6$ Solar-like stars, from which I extrapolate the stellar population mass.}
\tablenotetext{j}{The Fermi Paradox is frequently phrased as a limit on the number of societies that have arisen in our Galaxy from the lack of starfarers in the Solar System.}
\tablenotetext{k}{\citet{Annis99} verified that $137$ galaxies lied on the Tully-Fisher relation.  If their stellar populations were mostly cloaked in Dyson spheres, they would be optically faint for their stellar mass.}
\tablenotetext{l}{\citet{Zackrisson15} looked for outliers on the Tully-Fisher relation among $1,359$ galaxies.}
\tablenotetext{m}{The \GHAT~project examined extended sources in the WISE catalog for mid-infrared thermal emission from galaxies with stars cloaked in Dyson spheres.  \citet{Griffith15} found no galaxies with $\ge 85\%$ of their starlight captured by Dyson spheres, with a projected reach of $10^5$ galaxies.}
\tablenotetext{n}{The Galaxy Zoo project is not associated with SETI, but in principle it could search for galaxies with anomalous morphologies that might result from cosmic engineering.  Volunteers have inspected about $10^6$ galaxies \citep{Lintott11}.}
\tablenotetext{o}{\citet{Lacki16} searched the \emph{Planck} Catalog of Compact Sources 2 for microwave thermal emission from Type III societies that have shrouded entire galaxies behind cold screens.}
\tablenotetext{p}{\citet{Olson15} sets limits on the number of Type III--IV societies expanding through the cosmos from their non-presence in the Milky Way.  As an example, I use a comoving volume calculated from the mean limit on $R_1$, for the non-catastrophic evolution model with an expansion speed of $0.3 c$.}
\end{deluxetable*}

\vphantom{|}

\vphantom{|}

Radio and optical surveys for societies like our own have significant reaches.  The stars in the HabCat list of SETI Institute targets, for example, have a weighted reach of $\sim 0.6$ \citep{Turnbull03}.  Furthermore, these modes of communication are feasible even with our own level of technology, at least for reaching the nearest stars \citep[e.g.,][]{Howard04,Loeb07}.  The problem with these methods is their grasp.  It is possible that most societies self-destruct while young, or that they are interested in communicating with societies like our own for a very short time window \citep{Sagan63,Sagan73,Bates78}.  Since our own society is so young, we cannot rule out that $\eta_{\rm time} \approx (100\ \yr)/(10\ \Gyr) \approx 10^{-8}$, for example.  This concern has been voiced for decades in the context of the lifespan in Drake's equation \citep{Sagan63}.  Even a facility as powerful as the Square Kilometer Array may not be sensitive enough to detect radio transmissions from societies like our own \citep{Forgan11}.

Surveys for alien megastructures and societies that rate Type II or III on \citet{Kardashev64}'s scale have very high reach.  The recent \GHAT~survey has a reach of $\sim 0.9$ \citep{Griffith15}.  Similar reaches are achieved by surveys for Type II radio beacons \citep{Horowitz93} and very cold waste heat from Type III societies \citep{Lacki16}, and from the observation that intergalactic travelers are not here \citep{Olson15}.  Furthermore, Type III societies are presumably long-lived simply because it takes so much time to cross a galaxy and engineer it \citep{Kardashev85,Wright14-SF,Zackrisson15,Lacki16}.  But the grasp of such surveys is very uncertain.  Megastructures are huge extrapolations from our technology, and require astronomical investments.  Thus, there is a high systematic uncertainty to their existence, a large value of $\varepsilon_{\rm trace}$.  If we estimate $\varepsilon_{\rm trace} = 0.5$, these surveys can have less grasp than one confined to the Milky Way with $\varepsilon_{\rm trace} \approx 0$ and $\eta \approx 1$.

An ideal survey would search for traces of technology on par with our own that survive for a long time.  It need not have a maximal reach to have a competitive grasp.  Arguably, these conditions are fulfilled by a search for interstellar probes that may exist within the Solar System.  These probes may be very efficient forms of communication simply because they last so long, having a large $\eta_{\rm time}$ (\citealt{Bracewell60}; \citealt{Freitas83} estimates survival times of $\sim 10^6\ \yr$ for large probes in high Earth orbits).  They can also be an energy efficient way of sending information, since they only have to contact a neighboring planet instead of broadcasting across interstellar space \citep{Rose04}.  If one posits replicating probes, then probes may arrive from any society in the Galaxy (and maybe beyond), so that surveys for probes have ${\cal R} \ga 0.8$ \citep{Tipler80,Armstrong13}.  But even if all such probes are non-replicating \citep{Sagan83}, the surveys can be effective as long as societies send them to $\sim 10^3$ stars \citep[as in][]{Bracewell60}.  It's still not certain whether interstellar travel is feasible, but some fairly plausible projects have been proposed, including the recent Breakthrough Starshot\footnote{https://breakthroughinitiatives.org/Initiative/3}, although the proposed vessels generally do not stop in the destination system (\citealt{Crawford90}; one futuristic exception is \citet{Forward84}).  However, searches for the artifacts of interstellar travelers in the Solar System remain a relatively unexplored avenue of SETI (reported observations include \citealt{Freitas83,Steel95}; theoretical discussions include \citealt{Arkhipov95,Arkhipov96,Tough04,HaqqMisra12,Davies12}).

There may be some other traces on alien home planets that could remain visible for relatively long times.  Perhaps one could search for signs of widescale pollution \citep{Whitmire80,Lin14,Stevens15} or signs of geoengineering \citep{Cirkovic04,Lacki16}.  The feasibility of these methods has not been studied much, though.

\section{Discussion}
\label{sec:Discussion}

\subsection{What if the Universe is not infinite?}
\label{sec:SmallUniverse}
I have assumed that the Universe is essentially infinite, so that the likelihood of our existence is $1$.  However, multiverse scenarios have their own problems.  Aside from the difficulty in testing them, there is the measure problem, which is an uncertainty about how to assign probabilities in an infinite Universe.  Naive extrapolations of the current $\Lambda$CDM cosmology imply that most observers, even those with our exact memories, are Boltzmann brains produced by thermal fluctuations of the cosmological event horizon \citep[e.g.,][]{Dyson02,Albrecht04,deSimone10}.  A short-lived Universe can end before many Boltzmann brains appear, though \citep{Page08}.

\begin{figure}
\centerline{\includegraphics[width=9cm]{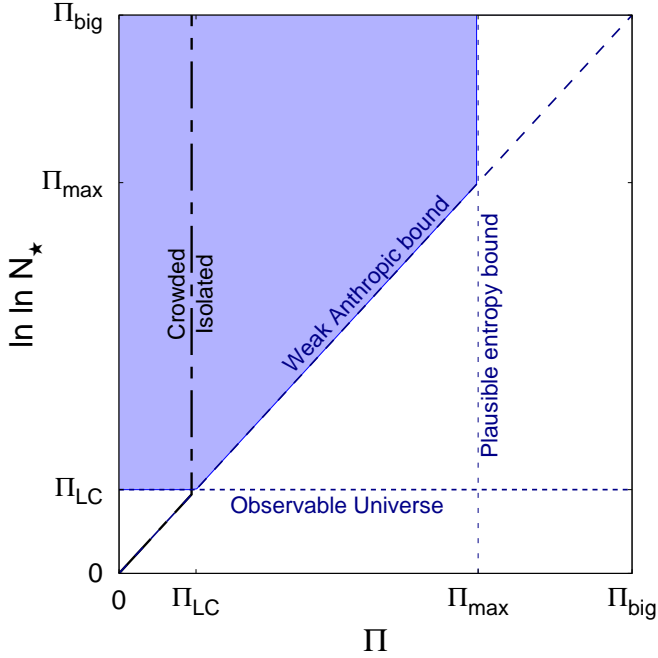}}
\figcaption{A joint prior on $\ln \ln N_{\star}$ and $\Pi$ is constrained by observations of the Universe's size, thermodynamic lower bounds on the probability of life evolving (Weak Copernican Principle), and our own existence (Weak Anthropic Principle).  We are isolated if the true values of $\PETI$ and $N_{\star}$ lie above and to the left of the heavy dash-dotted line.\label{fig:SmallUniversePrior}}
\end{figure}

If the Universe is small, though, we have to contend with at least two unknown parameters: the true number of birthsites in the universe $N_{\star}$ and $\PETI$.  We could then codify our uncertainty with a joint prior $d^2 P_{\rm prior}/(dN_{\star} d\PETI)$.  The joint prior can be integrated to find the marginal prior describing our prior belief in $\PETI$ alone:
\begin{equation}
\frac{dP_{\rm prior}}{d\PETI} = \int_0^{\infty} \frac{d^2 P_{\rm prior}}{dN_{\star} d\PETI^{\prime}}\Big|_{\PETI^{\prime} = \PETI} dN_{\star}.
\end{equation}
Likewise, the marginal prior on $N_{\star}$ is 
\begin{equation}
\frac{dP_{\rm prior}}{dN_{\star}} = \int_0^1 \frac{d^2 P_{\rm prior}}{dN_{\star}^{\prime} d\PETI}\Big|_{N_{\star}^{\prime} = N_{\star}} d\PETI.
\end{equation}

There would be a few obvious bounds on these parameters, as indicated in Figure~\ref{fig:SmallUniversePrior}.  The minimum size of the Universe is constrained by observation, and there are lower limits on $\PETI$ from bounds on the entropy of living systems.  If we presume that we are the result of a stochastic process, the inferential Weak Anthropic Principle becomes the observation that the probability of our own existence is ${\cal L} (\ge 1~{\rm society} | \PETI) \approx 1 - \exp(-\PETI N_{\star})$.  Using this likelihood in Bayes' Rule essentially removes the weight from possibilities where $\PETI N_{\star} \ll 1$, while leaving intact the weight where $\PETI N_{\star} \gg 1$.  

\begin{figure}
\centerline{\includegraphics[width=9cm]{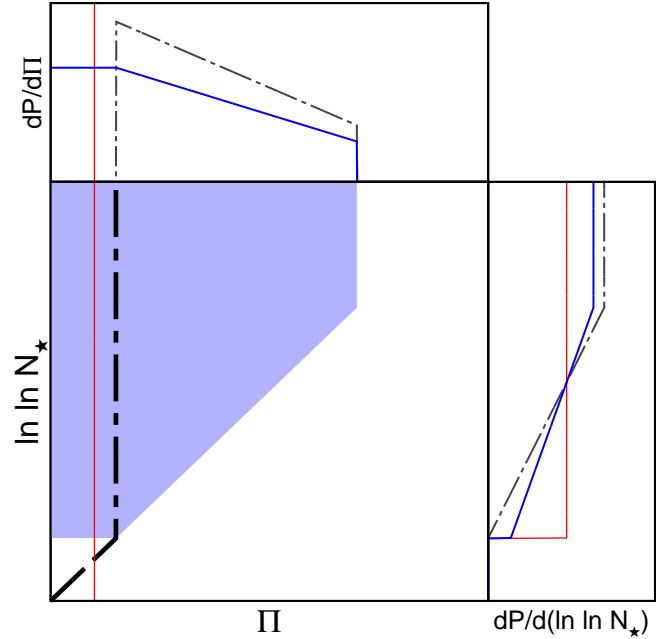}}
\figcaption{A uniform joint prior and the marginalized priors on $\Pi$ and $\ln \ln N_{\star}$ derived from it.  A joint prior with constant density favors belief in large $\PETI$ or large $N_{\star}$ after marginalization.  The red line demonstrates the effect of a detection by SETI, while the dashed grey lines demonstrate the effect of proof that we are isolated.\label{fig:UniformJointPrior}}
\end{figure}

How one would implement such a joint prior is unclear, however.  The simplest method might be to start out with a constant probability density everywhere, after transforming to the variables $\Pi$ and $\ln \ln N_{\star}$ (Figure~\ref{fig:UniformJointPrior}, left panel).  After applying the inferential Weak Anthropic Principle, the joint density remains constant for values where $\PETI N_{\star} \gg 1$.  Unfortunately, the resulting marginalized PDFs are no longer flat in $\Pi$ and $\ln \ln N_{\star}$ --- the log log prior described in this paper no longer applies to $\PETI$.  Instead, we would favor scenarios where the Universe is large.  Independently, we would favor scenarios where $\PETI$ is big.  We would not favor scenarios where both $N_{\star}$ and $\PETI$ are big simultaneously, though.  If we ever do discover aliens (red lines in Figure~\ref{fig:UniformJointPrior}), the marginalized PDF on $\ln \ln N_{\star}$ would suddenly become flat, because all values of $N_{\star}$ are compatible with abundant aliens.  Proof that we are isolated (dashed gray lines) moderately increase our belief in big Universes.

\begin{figure*}
\centerline{\includegraphics[width=9cm]{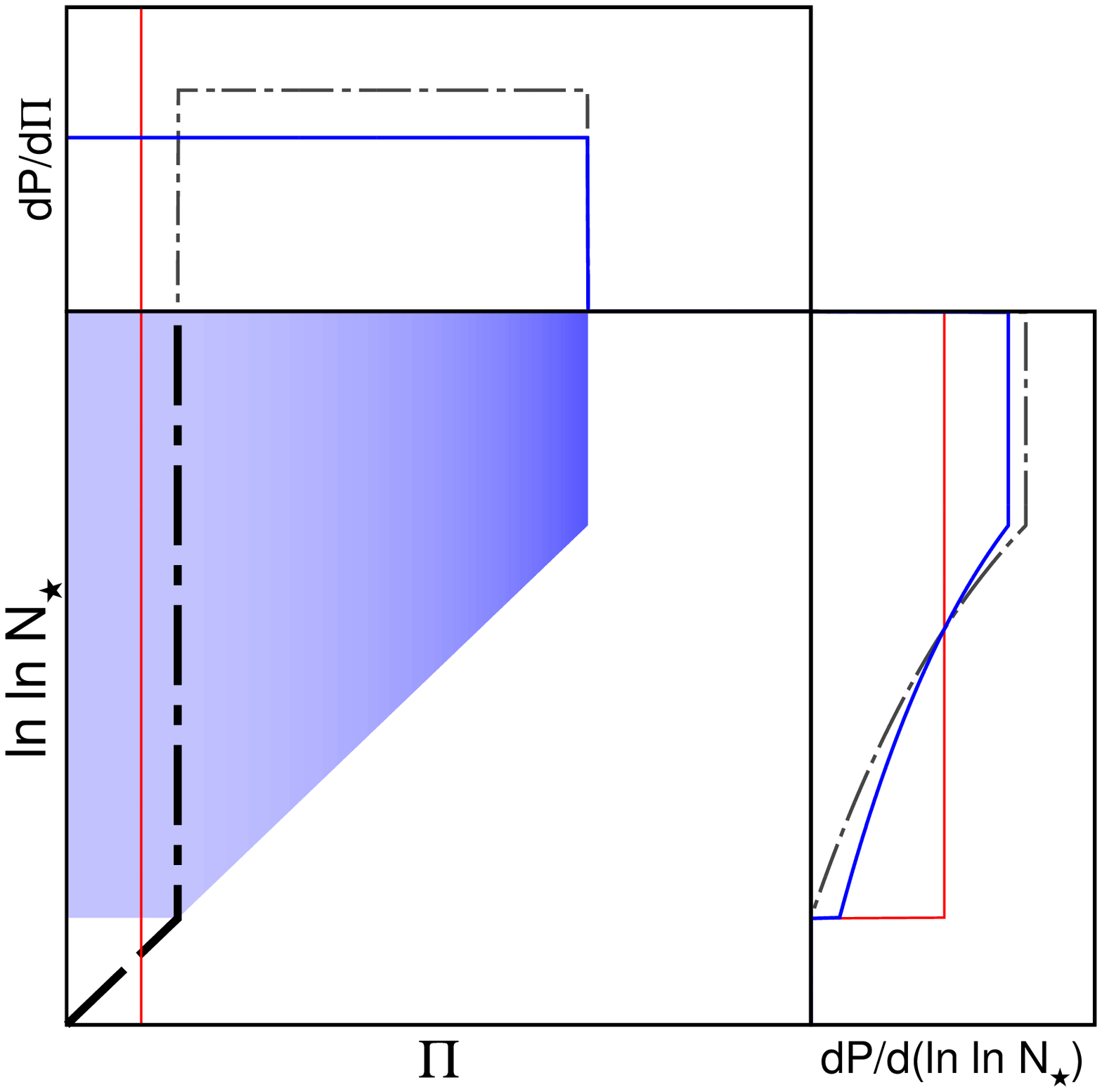}\includegraphics[width=9cm]{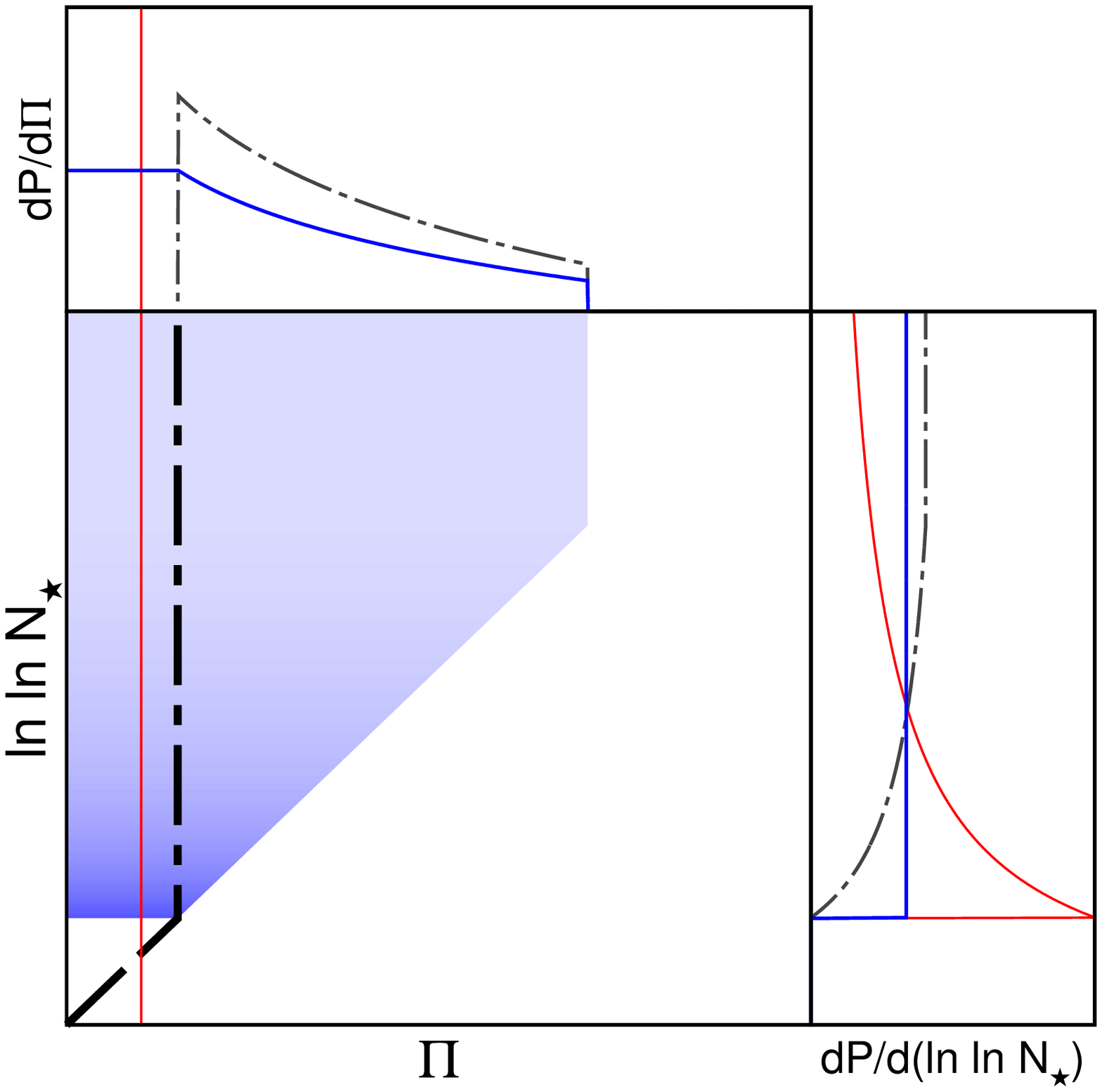}}
\figcaption{Comparison of joint priors with different weights, and their derived marginalized priors on $\Pi$ and $\ln \ln N_{\star}$.  Darker blue shading indicates a heavier joint prior density.  If the joint prior density is scaled so that the marginalized $\Pi$ prior is flat in $\Pi$ (left), then a large Universe is strongly favored.  If the joint prior density is instead scaled so that the marginalized $\ln \ln N_{\star}$ prior is uniform in $\ln \ln N_{\star}$ (right), a large $\PETI$ is strongly favored. The red line demonstrates the effect of a detection by SETI, while the dashed grey lines demonstrate the effect of proof that we are isolated.\label{fig:JointVsMarginalizedPriors}}
\end{figure*}

Another possibility would be to decree that the log log prior on $\PETI$ alone must be correct.  The joint prior would be more heavily weighted for small $\PETI$ (darker shading in Figure~\ref{fig:JointVsMarginalizedPriors}, left panel) so that the marginalized prior for $\PETI$ matches the log log prior.  This scheme heavily weights scenarios where the Universe is large, since only a large Universe is consistent with our existence if $\PETI$ is small and our evolution is random.  As with the uniform joint prior, discovering aliens flattens the marginal prior for $\ln \ln N_{\star}$, while proving our isolation increases our belief that $\ln \ln N_{\star}$ is large.   

Or we could decree that the marginal prior for $\ln \ln N_{\star}$ is constant, and place more weight in scenarios where $\ln \ln N_{\star}$ is small (Figure~\ref{fig:JointVsMarginalizedPriors}, right panel).  Now the marginalized prior on $\Pi$ is skewed to favor large $\PETI$.  SETI surveys would have interesting effects on the marginalized PDF for the Universe's size.  If we discovered aliens, the marginalized PDF for $\ln \ln N_{\star}$ would have a sharp peak near its lower limit because of the prior's weighting.  Proving we are isolated leads to a sharp fall-off near the lower limit, while slightly increasing our confidence in a big Universe.

Hence, implementing the joint prior involves somewhat arbitrary decisions about how to weight it.  One could probably consider more complicated schemes, like having a flat prior in the number of intelligent species in the Universe.  Unless the marginalized prior on $\PETI$ was specifically forced to be the log log prior, the log log prior used in the other sections no longer applies.
	
Aside from these practical difficulties, there is the philosophical issue of what counts as the Universe.  Some alternatives to inflationary cosmology are effectively multiverses in that they have an infinite number of places to live \citep{Rubenstein14}.  For example, ekpyrotic cosmologies posit that the Universe's evolution is basically cyclic, lasting an infinite time but being occasionally reset by some process \citep{Steinhardt02}.  Because birthsites can be defined temporally, the endless lifespans of the Universe in these scenarios still provide an endless number of chances for life and intelligence to evolve, as long as $\PETI$ does not change between cycles.  Even if we had proof that the Universe was finite in space and time, the many worlds interpretation of quantum mechanics still could guarantee our existence if it's true, since each branch of the Universe's wavefunction is as real as the others, and would seem real to its inhabitants \citep{Tegmark14}.  As long as we evolve on any branch of the wavefunction, the Weak Anthropic Principle applies.  The correct interpretation of quantum mechanics may never be proven experimentally, so the question of whether the Universe is small or big may always be metaphysical.

\subsection{The diversity of ETIs}
\label{sec:Diversity}
A curious aspect of the log log prior is that it suggests that there is a combinatorially high number of possible intelligent species, of order $\sim 10^{10^9}$.  This should be true as long as $\ln \PETI \ga -{\cal S}_{\rm genome}$, and if the possible genome sequences are even remotely equiprobable.  Then the number of possible intelligent species ${\cal N}_{\rm ETI}$ is roughly given by
\begin{equation}
\ln {\cal N}_{\rm ETI} \approx {\cal S}_{\rm genome} - \ln \PETI.
\end{equation}
I estimated ${\cal S}_{\rm genome} \approx 4 \times 10^9$ in Section~\ref{sec:GenomeEntropy}.  According to the log log prior, using the genome entropy bound, we generally expect $\ln \PETI$ to be several orders of magnitude below $4 \times 10^9$, so $\ln {\cal N}_{\rm ETI} \approx 4 \times 10^9$.  This conclusion can be avoided if there are a few species (presumably including \emph{Homo sapiens}) that are extreme attractors in genome space.  The odds would have to be heavily skewed in favor of these species, with them being $\sim 10^{10^9}$ times being more likely than the mean probability over genome space, to affect the estimate of $\ln {\cal N}_{\rm ETI}$, though.

If lifeforms are distinguished only by their proteomes, then we can use the proteome entropy to estimate the number of distinct types of intelligent life:
\begin{equation}
\ln {\cal N}_{\rm ETI} \approx {\cal S}_{\rm proteome} - \ln \PETI.
\end{equation}
Using the proteome entropy in the log log prior implies $\sim 10^{10^7}$ kinds of intelligent life.  While much smaller than the number of distinct species, this is still a vast number.  Most ``types'' would then consist of $\exp(\ln {\cal N}_{\rm ETI}^{\rm species} - \ln {\cal N}_{\rm ETI}^{\rm proteome}) \approx 10^{10^9}$ species of intelligent life, distinguished by their non-coding DNA and the order in which the genes are coded.

These enormous numbers are a natural result if not all of the information in the genome or proteome is relevant for the development of intelligence.  Even if intelligent species must be basically humanoid, would their development depend on the presence of hair, the number of fingers and vertebrae, having the same taste receptors as humans, much less the structure of every enzyme?  If not, then there are an exponentially large number of species possible from all the combinations of non-vital traits.  

If evolution is contingent, then the odds that intelligence evolved on Earth might be like the odds that a tornado passes through a given location on a given day.  The weather is a highly chaotic system and contingent; a single stray gust of wind just a few weeks before would completely change the weather.  It doesn't follow that every last eddy in the planet's history is necessary for there to be a tornado at that location.  If history were changed, new opportunities could arise; a breeze that in our history would have inhibited the tornado could have helped create a different one if that stray gust of wind happened.  While it's probably true, as Gould famously said, that ``\emph{Homo sapiens} is an entity, not a tendency'' \citep{Gould89}, intelligence is probably a panoply, not an entity.  Whether or not it's also a tendency is an empirical question.\footnote{Gould himself made this point in \citet{Gould87}.}

These estimates say nothing about the phenotypes of possible alien intelligences.  They could have radically different biochemistries, or they could look identical to humans while remaining a completely different species genetically. 

\subsection{Small probabilities and the Fermi Paradox}
\label{sec:HumanFuture}
The power of the Fermi Paradox is that it bends the argument from large numbers --- usually taken to be the strongest argument in favor of aliens --- against the existence of aliens.  Our being alone among the trillions of planets in the observable Universe requires an incredibly small $\PETI$.  But the existence of trillions of technological societies in the observable Universe requires that the probability that they spread into space is incredibly small.  The beyond astronomically small probabilities considered in this paper might undermine both arguments, though.  Just as a log log prior encompasses tiny $f_l f_i$, a modified version could accommodate tiny $f_c$.

One possible ``filter'' between developing technology and achieving starflight is a standoff involving nuclear weapons \citep[as in][]{Sagan83}.  Although we survived the Cold War, the Anthropic Principle reminds us that our vantage point is biased \citep{Cirkovic10} --- perhaps in virtually all histories we really did annihilate ourselves.  There were several incidents in which global nuclear war was avoided only due to the actions of a few people \citep[e.g.,][]{UoCS15}.\footnote{See also https://en.wikipedia.org/wiki/List\_of\_nuclear\_close\_calls (accessed 18 September 2016).}  Maybe those actions were themselves flukes --- a rare fluctuation in the thermal noise of someone's brain might be amplified into an otherwise unlikely stray thought that in turn stays someone's hand during a nuclear crisis.  If that was what happened, then the nuclear filter could be essentially absolute.  There are other possible filters.  In analogy with there being an unknown number $N$ of conditions necessary for intelligent life to arise, we may face an unknown number $N$ of future crises before we attain starflight and cosmic engineering.  If those crises are independent, and if the probability that a society survives each are of order $1/2$, then the odds of a society achieving starflight could easily be smaller than $10^{-21}$ if $N \ga 70$.  

A philosophical observation known as the Doomsday Argument appears to support to hypotheses that the odds are against anyone attaining starflight.  The basic idea, a kind of temporal Copernican Principle, is that it's unlikely that we are among the very first humans to have every lived, so the total number of humans who will ever live must not be many trillions \citep{Gott93}.  It is a counterweight to \citet{Hart75}'s formulation of the Fermi Paradox: an interstellar society could expand across its home galaxy at least, embracing billions of stars and many trillions of people.  \citet{Knobe06} used similar reasoning to make a Universal Doomsday Argument: because interstellar societies are so big, they would dominate the population of all sapient observers unless they were extremely rare.  Unless we are incredibly atypical, the population of the Universe cannot be concentrated into starfaring societies; thus, the fraction of technological societies that spread across interstellar space is negligible \citep{Knobe06}.

The Doomsday Argument itself is extremely contentious, though.  Obviously it is wrong for some people in the history of humanity, so it's not inconceivable that it's wrong for us \citep{Tarter07}.  Another response is the Self-Indication Assumption, which states that we should favor hypotheses that predict a larger number of observers \citep{Olum02}.  It is a bit like the causal Weak Anthropic Principle: the probability that we exist is larger if the Universe has more opportunities for us to exist.  This assumption has its own problems, as it allows essentially zero prior weight on the idea that the Universe is small \citep[e.g.,][]{Bostrom03}.  Overall, the debate around both the Doomsday Argument and the Self-Indication Assumption arises from problems similar to those of the flat log prior for $\PETI$.  Each suppresses the prior weight for some entirely reasonable sounding hypothesis --- star travel being possible, our not living in a vast multiverse, or our not being isolated --- by factors of billions at least, so that no actual evidence could ever persuade us otherwise \citep[c.f.,][]{Olum02,Bostrom03}.  

Our own future evolution, at least, has some important differences with the evolution of life and intelligence.  Most importantly, we have goals, whereas natural evolution does not.  We can also anticipate future crises.  Future crises are not necessarily independent of each other, either; many of the crises may involve fundamentally similar problems, like scarcity.  Nor is it clear that there always are bottlenecks.  For example, the nuclear standoff during the Cold War may not have been inevitable in our own history \citep{Rhodes86}, much less in alien societies.  Furthermore, like life in general, societies can be robust.  If one didn't know the history of life on Earth, one might conclude its survival for billions of years is nigh impossible given how many crises could arise.  But it has survived that long simply by being so resilient, and while this may be a fluke that we observe due to the Anthropic Principle \citep{Cirkovic10}, this is not generally thought to be the case.\footnote{If one does accept the Doomsday Argument, it also predicts that we are very unlikely to be the very last people who ever live \citep{Gott93}, suggesting that technological societies do not collapse at the slightest provocation.}

On the other hand, if one does accept the thesis behind the Fermi Paradox, the log log prior actually strengthens its power.  As noted in Section~\ref{sec:SETIReach}, the Milky Way encompasses $80\%$ of the prior's weight for crowded Universe scenarios, and about half is for $\PETI \ga 10^{-4}$.  Definitive evidence for our being alone in the Galaxy is then good evidence that we are alone in the observable Universe, with $\PCrowded$ cut to $\sim 0.2 \times 0.2 \approx 4\%$.  The null results from searches for Type III societies reduce the $\PCrowded$ estimates further, subject to systematic uncertainties about these surveys' grasp.

\subsection{The prior applied to exolife and other complex phenomena}
\label{sec:LifePrior}
The main ingredients plugged into the log log prior, the number of birthsites and the bound on entropy, are very generic.  Similar priors could be constructed for any potentially rare, complex phenomenon, including life itself.  Indeed, the origin of life has been proposed to be the limiting factor for $\PETI$ \citep{ConwayMorris03}, in which case the log log prior for intelligence is the log log prior for life.  The protocell chemical entropy provides a plausible but very conservative lower bound on the probability that life arises.  As shown in Table~\ref{table:PCrowdedEstimates}, the log log prior then implies a $15\%$ credibility that life has arisen on another planet in the Universe.  In coming decades, it may be possible to identify signs of life on the nearest exoplanets \citep{Seager14}, suggesting a test of the log log prior: we should never find any signs of independent life on any exoplanet.  This is a very weak test, though, as the credibility for lifeless neighbor planets is only $\la 85\%$.

On the other hand, favorable probabilities for life's appearance may come from applying the timing of its origin on Earth to the log log prior \citep{Lineweaver02}.  Suppose the early Earth had many birthsites, appearing at a rate of $\Gamma_{\rm birth}$.  Assuming that there was a duration $\Delta t_1$ between when life could have started and when it did \citep{Spiegel12}, the Earth had $N_{\rm birth} = \Gamma_{\rm birth} \Delta t_1$ birthsites before life got started.  The expected rate that life appears during the window is $\Gamma_{\rm life} = \PLife \Gamma_{\rm birth}$, where $\PLife$ is the probability that a birthsite generates life \citep[as in][]{Scharf16}.  For an upper limit on $\Gamma_{\rm birth}$, I take the mass of the Earth's hydrosphere and apply the ML98 bound.  The specific internal energy of liquid water at $300\ \Kelv$ and standard pressure\footnote{The difference in specific enthalpy for ice at melting and ice at absolute zero is $300\ \Joule\ \gram^{-1}$, and the specific enthalpy of melting is another $330\ \Joule\ \gram^{-1}$, all at standard pressure \citep{Feistel06}.  Heating liquid water, with a specific heat of $4.2\ \Joule\ \gram^{-1}\ \Kelv^{-1}$, from the melting point to $300\ \Kelv$ requires $110\ \Joule\ \gram^{-1}$.  I ignore the (minor) correction from enthalpy to internal energy.} is $740\ \Joule\ \gram^{-1}$, for a total $\Delta E$ of $1.3 \times 10^{34}\ \erg$, a maximum $\Gamma_{\rm birth}$ of $4 \times 10^{60}\ \sec^{-1}$, and a maximum $N_{\rm birth}$ of $1.2 \times 10^{77} (\Delta t_1 / \Gyr)$.  

\citet{Spiegel12} demonstrate how a prior is affected by the timing of life's appearance.  Roughly speaking, we can group $\Gamma_{\rm life}$ into two categories: a fast case when $\Gamma_{\rm life} \ga (\Delta t_1)^{-1}$ and a slow case $\Gamma_{\rm life} \la (\Delta t_2)^{-1}$.  In the slow case, $\Delta t_2$ describes the window that life could have appeared while still allowing intelligent life to evolve by now; because of the very weak dependence of $\Pi$ on $N_{\rm birth}$ from the double logarithm, I replace it with $\Delta t_1$ for simplicity.  Then, the posterior probability that life appears quickly is
\begin{equation}
P_{\rm fast}^{\rm posterior} \approx \left(1 + \frac{1 - P_{\rm fast}^{\rm prior}}{{\cal B} P_{\rm fast}^{\rm prior}}\right)^{-1}, 
\end{equation}
where ${\cal B}$ is the Bayes' factor\footnote{Denoted ${\cal R}$ in \citet{Spiegel12}.}, the ratio of likelihoods for the slow and fast cases.  In their ``optimistic'' case, ${\cal B} = 15$ and $\Delta t_1 = 0.2\ \Gyr$.  The log log prior, however, starts out disfavoring the fast case, which has a prior probability $\ln[1 + \ln N_{\rm birth}]/\ln[1 - \ln \PMin]$.  For the protocell entropy $\PMin$ and the maximal birthsite rate calculated above, this is $19\%$. 

The optimistic case of \citet{Spiegel12} then gives $P_{\rm fast}^{\rm posterior} \approx 78\%$.  That value weakly supports the conclusion of \citet{Lineweaver02} that life emerges quickly on planets.  Yet the ``conservative'' cases of \citet{Spiegel12} merely increase $P_{\rm fast}$ to $20\%$.  Furthermore, $\Gamma_{\rm birth}$ is probably much lower than the maximum value.  In principle, $N_{\rm birth} \approx 1$, for which $P_{\rm fast}^{\rm prior} \approx 2\%$ and $P_{\rm fast}^{\rm posterior} \approx 20\%$ in the optimistic case.  The key improvement is that the log log prior has well-defined bounds --- if ${\cal B}$ definitely exceeds $100$ then it \emph{does} favor rapidly appearing life even for $\ln N_{\rm birth} \approx 1$, unlike the logarithmic prior, which can have an arbitrarily small normalization \citep{Spiegel12}.

One potential problem with applying the log log prior to different complex phenomena indiscriminately is that these phenomena may be dependent on one another.  The evolution of intelligent life on a planet happens only if life appears on the planet first.  One can posit a whole chain of dependent phenomena of increasing rarity: life, complex multicellular life, intelligent life, humanoid intelligences, \emph{Homo sapiens}, humans that share your exact memories, and chemically identical versions of you.  Properly constraining the probability of each step would require the construction of a joint prior on the probability of each step happening.  The marginalized prior density for each step would then no longer be a log log prior.  Generally, the Bayesian expectation for the probabilities of earlier steps would be higher than with a simple log log prior (c.f. Section~\ref{sec:SmallUniverse}).
 
A stronger test of the log log prior than exolife alone may be to check whether known astrophysical phenomena are evenly distributed in $\Pi$.  For example, if our past light cone contains $N$ kinds of stellar phenomena, about $N/2$ should be expressed in a random sample of $\sim 10^3$ stars over their lifetimes.  Distinct phenomena might be classified according to criteria proposed by \citet{Harwit81}.

\section{Summary}
\label{sec:Summary}
The log log prior is a plausible framework for evaluating evidence for or against alien societies.  It can be justified from the uncertainty in the number of constraints that need to be fulfilled for intelligence to evolve, and it can be phrased in terms of entropy differences, information, or state space dimensionality.  The main advantage of the log log prior is that it can accommodate a great range of $\PETI$, from $e^{-3 \times 10^{122}}$ to $1$.  Unlike a flat log prior, it responds to observations even in the face of possible systematic errors.  The potential for systematic errors is inevitable for any realistic experiment.  The log log prior can provide a guide for measuring the relative power of SETI surveys.

Essentially by design, the log log prior is not all that profound in its content.  It basically is just the statement that, for all we know, $\PETI^{-1}$ could be $1$, $10^{1}$, $10^{10}$, $10^{100}$, $10^{1,000}$, $10^{10,000}$, $10^{100,000}$, $10^{1,000,000}$, $10^{10,000,000}$, $10^{100,000,000}$, or $10^{1,000,000,000}$, so we might as well consider any of those values as an equally valid possibility (using $\PMin = 10^{-10^9}$ here as an example).  The calculations of $P_{\rm crowded}$ just amount to the observation that for the first three of those values, we're not isolated, while for the others, we are, so the odds that we're isolated are a few to one.  The prior can even be summarized in non-Bayesian terms: we are very uncertain about the number of factors that contribute to intelligence's evolution, so we are very, \emph{very} uncertain about the probability that it happens.  

What is new is simply the emphasis placed on each value.  A flat log prior amounts to the statement that, for all we know, $\PETI^{-1}$ could be $10^0$, $10^1$, $10^2$, and so on, through $10^{999,999,998}$, $10^{999,999,999}$, or $10^{1,000,000,000}$, so we might as well consider any of \emph{those} values as an equally valid possibility.  But this inherently implies a belief that $\log_{10} \PETI \sim 10^{-9}$ with near certainty.  It tacitly implies that we are virtually certain that we are isolated.  Or, in non-Bayesian terms, we are a bit uncertain about the number of factors that contribute to intelligence's evolution, and we are very uncertain about its probability, but we are quite sure it's small.\footnote{We could do much worse still, though, with an flat inverse prior \citep{Spiegel12}.  This would be the statement that $\PETI^{-1}$ could be $1$, $2$, $3$, ..., $10^{1,000,000,000} - 2$, $10^{1,000,000,000} - 1$, or $10^{1,000,000,000}$ so we might as well consider any of \emph{these} values as an equally valid possibility.  Now one would be nearly certain that $10^{-1,000,000,000} \le \PETI \la 10^{-999,999,998}$, far too precise to be realistic.}

It is important to remember that the prior is a \emph{Bayesian} probability, measuring our confidence about the abundance of ETIs, whereas $\PETI$ is a \emph{frequentist} probability, an intrinsic feature of birthsites.  While I estimate a $\sim 15 \text{--} 20\%$ (Bayesian) credibility that there are other intelligent species in the observable Universe, for almost all $\PETI$, the frequentist probability ${\cal P} ({\rm isolated} | \PETI)$ that we're isolated is almost always $\sim 0$ or $\sim 1$.  Conversely, the fact that this frequentist probability is probably $\sim 0$ or $\sim 1$ does not mean we should be confident about whether or not we can contact aliens.  Just because there are $\sim 10^{21}$ planets in the observable Universe, it does not follow that there must be ETIs among them, because we are not totally confident that $\PETI \ga 10^{-21}$.  And even if the evolution of intelligence depends on a vast number of contingent events, it does not follow that we must be alone, because we are not confident that there are just a few ways for intelligence to evolve. 

According to the log log prior, we should approach SETI with a degree of agnosticism about whether there are aliens in the observable Universe.  Despite that, the prior consistently leans towards mild skepticism about their presence.  I find that the most realistic bound on $\PMin$, the genome entropy, implies that $P_{\rm crowded} \approx 18\%$.  Even with the most optimistic assumptions, with the maximal number of birthsites possible and using the protein shape entropy to bound $\PMin$, $P_{\rm crowded} \approx 47\%$.  Of course, $P_{\rm isolated}$ is quite far from the traditional $95\%$ credibility threshold for a conclusion.  A positive detection would be one of the most profound discoveries ever.  This epochal potential more than offsets the relatively moderate risks for the relatively low spending on SETI.  Still, it does mean the null results from SETI are not surprising.  The ``Great Silence'' is an expected result of the log log prior.

The log log prior is not without its own issues.  Most importantly, it's not clear how to define a ``birthsite''.  Should it refer to an entire planet, or maybe something smaller, like a speciation event?  If we use a large body as a birthsite, we should account for the possibility that life or intelligence arises an unknown, possibly large number of times.  If we use common events as birthsites, which is appropriate if $\PETI$ depends on the time a habitat is hospitable, then they almost certainly interact with one another, making it difficult to calculate the number of ETIs expected.  The issue is related to the subjective decision about how to handle values of $\PETI$ near $1$, where $\ln |\ln \PETI|$ itself diverges.  I chose to use the auxiliary variable $\Pi = \ln (1 + \ln \PETI)$, but this is not the only possible choice.  The prior is also slightly affected by the choice of the base of the inner logarithm.

The other main issue is which value of $\PMin$ to use.  The cosmological entropies ideally should be absolute bounds.  Yet they are not totally robust.  The entropy within the particle or event horizons increases without bound in cosmologies without dark energy or if $w > -1$, although it takes $\sim 10^{40}$ years for this to seriously affect the estimates of $P_{\rm crowded}$.  Within the $\Lambda$CDM cosmology, observers living more than $\sim 100$ billion years in the future may not be aware there is a cosmological event horizon \citep{Krauss07}.  Furthermore, while well motivated, the entropy of the cosmological horizons and Bekenstein-like bounds in general are not empirically proven.  On smaller scales, we could use the chemical entropy of an organism or a planetary biosphere to limit $\PMin$.  However, $\PETI$ values as small as these $\PMin$ essentially imply that we are Boltzmann brains, in which case none of our reasoning can be justified.  Even the larger $\PMin$ associated with the protocell entropy is compatible with strange evolutionary histories, such as those where life is juggled across the worlds of the Solar System, or even different star systems entirely.   

Values of $\PMin$ as small as those considered in this paper implicitly require that the Universe is very large if our evolution is a stochastic process.  Multiverse theories face the measure problem \citep[e.g.,][]{Albrecht04} and may not be testable.  If one favors a small Universe, one can set up a joint prior on the size of the Universe and $\PETI$, but then the marginalized prior on $\PETI$ alone is not the log log prior anymore unless the weighting is uneven.  Additionally, the Universe would have to be small in time as well as space, disallowing cyclic cosmologies.  Furthermore, the many-worlds interpretation of quantum mechanics cannot hold if the Universe is small \citep{Tegmark14}, but we may never know whether or not it is true.  
	
As frequently acknowledged, $\PETI$ itself is not the only factor determining whether we will ever find an alien society \citep{Sagan63,Sagan73,Bates78,Forgan11}.  SETI surveys are only constraining if we look for traces that are physically possible, commonly produced by technological societies, and last a long time.  The weighted reach of current SETI surveys, the number of targets they look at, is quite good according to the log log prior (Table~\ref{table:SurveyReaches}).  Their grasps, however, the amount of prior weight they can constrain after considering the effectiveness of the survey method, are debatable.  Current methods are haunted by the potentially short lifespan of radio or optical broadcasting, or by systematic uncertainties about whether megastructures are physically and socially plausible.  ISearches for small probes in the Solar System may be an effective way to proceed, because they last so long and seem fairly feasible.  The log log prior suggests that the probes do not have to be self-replicating, sweeping through the Galaxy ravenously, for this to be an effective tracer.

\acknowledgments
I thank Juna Kollmeier for discussions.  In addition, I wish to acknowledge the use of NASA's Astrophysics Data System and arXiv.


\begin{thebibliography}{}
\twocolumngrid
\bibitem[Abdellah et al.(2004)]{Abdellah04} Abdellah, Z. et al. (International Human Genome Sequencing Consortium)\ 2004, \nat, 431, 931 

\bibitem[Abramov \& Mojzsis(2011)]{Abramov11} Abramov, O., \& Mojzsis, S.~J.\ 2011, \icarus, 213, 273 

\bibitem[Ade et al.(2015)]{Ade15-Params} Ade, P.~A.~R., Aghanim, N., et al. (Planck Collaboration)\ 2015, arXiv:1502.01589 

\bibitem[Adams \& Laughlin(1997)]{Adams97} Adams, F.~C., \& Laughlin, G.\ 1997, Reviews of Modern Physics, 69, 337, arXiv:astro-ph/9701131

\bibitem[Albrecht \& Sorbo(2004)]{Albrecht04} Albrecht, A., \& Sorbo, L.\ 2004, \prd, 70, 063528, arXiv:hep-th/0405270

\bibitem[Albrecht(2015)]{Albrecht15} Albrecht, A.\ 2015, \prd, 91, 103510, arXiv:1401.7309

\bibitem[Anderson et al.(2002)]{Anderson02} Anderson, D. P., Cobb, J., Korpela, E., Lebofsky, M., \& Werthimer, D. 2002, Communications of the ACM, 45, 56

\bibitem[Anderson et al.(2004)]{Anderson04} Anderson, J.~C., Wu, N., Santoro, S.~W., et al.\ 2004, Proceedings of the National Academy of Science, 101, 7566 

\bibitem[Annis(1999)]{Annis99} Annis, J.\ 1999, Journal of the British Interplanetary Society, 52, 33 

\bibitem[Arkhipov(1995)]{Arkhipov95} Arkhipov, A.\ 1995, Progress in the Search for Extraterrestrial Life, 74, 259 

\bibitem[Arkhipov(1996)]{Arkhipov96} Arkhipov, A.~V.\ 1996, The Observatory, 116, 175 

\bibitem[Armstrong \& Sandberg(2013)]{Armstrong13} Armstrong, S., \& Sandberg, A.\ 2013, Acta Astronautica, 89, 1 

\bibitem[Arnold(2005)]{Arnold05} Arnold, L.~F.~A.\ 2005, \apj, 627, 534, arXiv:astro-ph/0503580

\bibitem[Ashkenazi(1995)]{Ashkenazi95} Ashkenazi, M.\ 1995, Progress in the Search for Extraterrestrial Life., 74, 507 

\bibitem[Baldry et al.(2012)]{Baldry12} Baldry, I.~K., Driver, S.~P., Loveday, J., et al.\ 2012, \mnras, 421, 621, arXiv:1111.5707

\bibitem[Baross et al.(2007)]{Baross07} Baross, J. A. et al. 2007. \emph{The Limits of Organic Life in Planetary Systems}, Washington DC, National Academies Press.

\bibitem[Barrow(1983)]{Barrow83} Barrow, J.~D.\ 1983, \qjras, 24, 146 

\bibitem[Batalha et al.(2010)]{Batalha10} Batalha, N.~M., Borucki, W.~J., Koch, D.~G., et al.\ 2010, \apjl, 713, L109, arXiv:1001.0349

\bibitem[Bates(1978)]{Bates78} Bates, D.~R.\ 1978, \apss, 55, 7 

\bibitem[Behroozi \& Peeples(2015)]{Behroozi15} Behroozi, P., \& Peeples, M.~S.\ 2015, \mnras, 454, 1811, arXiv:1508.01202 

\bibitem[Bekenstein(1981)]{Bekenstein81} Bekenstein, J.~D.\ 1981, \prd, 23, 287 

\bibitem[Bianconi et al.(2013)]{Bianconi13} Bianconi, E. et al. 2013, Annals of human biology, 40, 463

\bibitem[Bieri(1964)]{Bieri64} Bieri, R. 1964, American scientist, 52, 452

\bibitem[Blair et al.(1992)]{Blair92} Blair, D.~G., Norris, R.~P., Troup, E.~R., et al.\ 1992, \mnras, 257, 105 

\bibitem[Borges(1962)]{Borges62} Borges, J. L. 1962, ``The Library of Babel'', in \emph{Ficciones}, ed. Kerrigan, A., New York, Grove Press (ebook edition)

\bibitem[Borra(2012)]{Borra12} Borra, E.~F.\ 2012, \aj, 144, 181, arXiv:1210.5986

\bibitem[Bostrom \& {\'C}irkovi{\'c}(2003)]{Bostrom03} Bostrom, N., {\'C}irkovi{\'c}, M. M. 2003, The Philosophical Quarterly, 53, 83

\bibitem[Bousso(2002)]{Bousso02} Bousso, R.\ 2002, Reviews of Modern Physics, 74, 825, arXiv:hep-th/0203101

\bibitem[Boyajian et al.(2016)]{Boyajian16} Boyajian, T.~S., LaCourse, D.~M., Rappaport, S.~A., et al.\ 2016, \mnras, 457, 3988, arXiv:1509.03622

\bibitem[Bracewell(1960)]{Bracewell60} Bracewell, R.~N.\ 1960, \nat, 186, 670 

\bibitem[Brin(1983)]{Brin83} Brin, G.~D.\ 1983, \qjras, 24, 283 

\bibitem[Carrigan(2009)]{Carrigan09} Carrigan, R.~A., Jr.\ 2009, \apj, 698, 2075, arXiv:0811.2376

\bibitem[Carrigan(2012)]{Carrigan12} Carrigan, R.~A.\ 2012, Acta Astronautica, 78, 121 

\bibitem[Carroll(2010)]{Carroll10} Carroll, S. 2010, \emph{From Eternity to Here: The Quest for the Ultimate Theory of Time}, New York, Penguin

\bibitem[Carter(1974)]{Carter74} Carter, B.\ 1974, Confrontation of Cosmological Theories with Observational Data, 63, 291 

\bibitem[Carter(1983)]{Carter83} Carter, B.\ 1983, Philosophical Transactions of the Royal Society of London Series A, 310, 347 

\bibitem[Carter(2008)]{Carter08} Carter, B.\ 2008, International Journal of Astrobiology, 7, 177, arXiv:0711.1985

\bibitem[Cassan et al.(2012)]{Cassan12} Cassan, A., Kubas, D., Beaulieu, J.-P., et al.\ 2012, \nat, 481, 167, arXiv:1202.0903 

\bibitem[Chabrier(2003)]{Chabrier03} Chabrier, G.\ 2003, \pasp, 115, 763, arXiv:astro-ph/0304382

\bibitem[Chyba \& Hand(2005)]{Chyba05} Chyba, C.~F., \& Hand, K.~P.\ 2005, \araa, 43, 31 

\bibitem[{\'C}irkovi{\'c} \& Cathcart(2004)]{Cirkovic04} {\'C}irkovi{\'c}, M.~M., \& Cathcart, R.~B.\ 2004, Journal of the British Interplanetary Society, 57, 209, arXiv:physics/0308058

\bibitem[{\'C}irkovi{\'c} \& Bradbury(2006)]{Cirkovic06} {\'C}irkovi{\'c}, M.~M., \& Bradbury, R.~J.\ 2006, New Astronomy, 11, 628, arXiv:astro-ph/0506110

\bibitem[{\'C}irkovi{\'c}(2009)]{Cirkovic09-Fermi} {\'C}irkovi{\'c}, M.~M.\ 2009, Serbian Astronomical Journal, 178, 1, arXiv:0907.3432

\bibitem[{\'C}irkovi{\'c} et al.(2009)]{Cirkovic09-Reset} {\'C}irkovi{\'c}, M.~M., Vukoti{\'c}, B., \& Dragi{\'c}evi{\'c}, I.\ 2009, Astrobiology, 9, 491, arXiv:0912.4980

\bibitem[{\'C}irkovi{\'c} et al.(2010)]{Cirkovic10} {\'C}irkovi{\'c}, M. M., Sandberg, A., \& Bostrom, N. 2010, Risk analysis, 30, 1495

\bibitem[{\'C}irkovi{\'c}(2014)]{Cirkovic14} {\'C}irkovi{\'c}, M.~M.\ 2014, Biology and Philosophy, 29, 539  

\bibitem[Cocconi \& Morrison(1959)]{Cocconi59} Cocconi, G., \& Morrison, P.\ 1959, \nat, 184, 844 

\bibitem[Conway Morris(2003)]{ConwayMorris03} Conway Morris, S. 2003, \emph{Life's Solution: Inevitable Humans in a Lonely Cosmos}, New York, Cambridge University Press (2013 ebook edition)

\bibitem[Corbet(1997)]{Corbet97} Corbet, R.~H.~D.\ 1997, Journal of the British Interplanetary Society, 50, 253, arXiv:1609.00330

\bibitem[Crawford(1990)]{Crawford90} Crawford, I.~A.\ 1990, \qjras, 31, 377 

\bibitem[Crick(1968)]{Crick68} Crick, F. H. C. 1968, Journal of molecular biology, 38, 367

\bibitem[Crowe(1999)]{Crowe99} Crowe, M. J. 1999, \emph{The Extraterrestrial Life Debate, 1750-1900}, New York, Dover Publications

\bibitem[Davies(2003)]{Davies03} Davies, P.~C.~W.\ 2003, Astrobiology, 3, 673, arXiv:astro-ph/0403049

\bibitem[Davies et al.(2009)]{Davies09} Davies, P.~C.~W., Benner, S.~A., Cleland, C.~E., et al.\ 2009, Astrobiology, 9, 241 

\bibitem[Davies(2010)]{Davies10} Davies, P. C. W. 2010, \emph{The Eerie Silence: Renewing Our Search for Alien Intelligence}, New York, Houghton Miffton Harcourt (ebook edition)

\bibitem[Davies(2012)]{Davies12} Davies, P.~C.~W.\ 2012, Acta Astronautica, 73, 250 

\bibitem[Davis \& Lineweaver(2004)]{Davis04} Davis, T.~M., \& Lineweaver, C.~H.\ 2004, \pasa, 21, 97, arXiv:astro-ph/0310808

\bibitem[de Simone et al.(2010)]{deSimone10} de Simone, A., Guth, A.~H., Linde, A., et al.\ 2010, \prd, 82, 063520, arXiv:0808.3778

\bibitem[Dennett(1995)]{Dennett95} Dennett, D. C. 1995, \emph{Darwin's Dangerous Idea: Evolution and the Meaning of Life}, New York, Simon \& Schuster Paperbacks (ebook edition)

\bibitem[Dick(1998)]{Dick98} Dick, S. J. 1998, \emph{Life on Other Worlds: The 20th-Century Extraterrestrial Life Debate}, New York, Cambridge University Press

\bibitem[Dill(1999)]{Dill99} Dill, K. A. 1999, Protein Science, 8, 1166

\bibitem[Dryden et al.(2008)]{Dryden08} Dryden, D. T., Thomson, A. R., \& White, J. H., 2008, Journal of The Royal Society Interface, 5, 953

\bibitem[Dyson et al.(2002)]{Dyson02} Dyson, L., Kleban, M., \& Susskind, L.\ 2002, Journal of High Energy Physics, 10, 011, arXiv:hep-th/0208013

\bibitem[Dyson(2003)]{Dyson03} Dyson, F.~J.\ 2003, International Journal of Astrobiology, 2, 103 

\bibitem[Egan \& Lineweaver(2010)]{Egan10} Egan, C.~A., \& Lineweaver, C.~H.\ 2010, \apj, 710, 1825, arXiv:0909.3983 

\bibitem[Emery(2006)]{Emery06} Emery, N. J. 2006, Philosophical Transactions of the Royal Society of London B: Biological Sciences, 361, 23

\bibitem[Feistel \& Wagner(2006)]{Feistel06} Feistel, R., \& Wagner, W., 2006, Journal of Physical and Chemical Reference Data, 35, 1021

\bibitem[Fixsen(2009)]{Fixsen09} Fixsen, D.~J.\ 2009, \apj, 707, 916, arXiv:0911.1955

\bibitem[Forgan \& Nichol(2011)]{Forgan11} Forgan, D.~H., \& Nichol, R.~C.\ 2011, International Journal of Astrobiology, 10, 77, arXiv:1007.0850

\bibitem[Forward(1984)]{Forward84} Forward, R.~L.\ 1984, Journal of Spacecraft and Rockets, 21, 187 

\bibitem[Frank \& Sullivan(2016)]{Frank16} Frank, A., \& Sullivan, W.~T., III 2016, Astrobiology, 16, 359, arXiv:1510.08837

\bibitem[Freitas \& Valdes(1980)]{Freitas80} Freitas, R.~A., Jr., \& Valdes, F.\ 1980, \icarus, 42, 442 

\bibitem[Freitas(1985)]{Freitas85} Freitas, R.~A., Jr.\ 1985, \icarus, 62, 518 

\bibitem[Freitas(1983)]{Freitas83} Freitas, R.~A., Jr.\ 1983, \icarus, 55, 337 

\bibitem[Gibbons \& Hawking(1977)]{Gibbons77} Gibbons, G.~W., \& Hawking, S.~W.\ 1977, \prd, 15, 2738 

\bibitem[Gonzalez et al.(2001)]{Gonzalez01} Gonzalez, G., Brownlee, D., \& Ward, P.\ 2001, \icarus, 152, 185, arXiv:astro-ph/0103165

\bibitem[Gott(1993)]{Gott93} Gott, J.~R., III 1993, \nat, 363, 315

\bibitem[Gould(1987)]{Gould87} Gould, S. J. 1987, ``SETI and the Wisdom of Casey Stengel'', in \emph{The Flamingo's Smile: Reflections in Natural History}, New York, W. W. Norton \& Company (ebook edition)

\bibitem[Gould(1989)]{Gould89} Gould, S. J. 1989, \emph{Wonderful Life: The Burgess Shale and the Nature of History}, New York, W. W. Norton \& Company

\bibitem[Gray \& Ellingsen(2002)]{Gray02} Gray, R.~H., \& Ellingsen, S.\ 2002, \apj, 578, 967 

\bibitem[Griffith et al.(2015)]{Griffith15} Griffith, R.~L., Wright, J.~T., Maldonado, J., et al.\ 2015, \apjs, 217, 25, arXiv:1504.03418 

\bibitem[Hanna et al.(2009)]{Hanna09} Hanna, D.~S., Ball, J., Covault, C.~E., et al.\ 2009, Astrobiology, 9, 345, arXiv:0904.2230

\bibitem[Hanson(1998)]{Hanson98} Hanson, R. 1998, ``Must early life be easy? The rhythm of major evolutionary transitions'', accessed 5 Sep 2016, \textless http://mason.gmu.edu/~rhanson/hardstep.pdf \textgreater

\bibitem[Haqq-Misra \& Baum(2009)]{HaqqMisra09} Haqq-Misra, J.~D., \& Baum, S.~D.\ 2009, Journal of the British Interplanetary Society, 62, 47, arXiv:0906.0568

\bibitem[Haqq-Misra \& Kopparapu(2012)]{HaqqMisra12} Haqq-Misra, J., \& Kopparapu, R.~K.\ 2012, Acta Astronautica, 72, 15, arXiv:1111.1212

\bibitem[Harris(2002)]{Harris02} Harris, M.~J.\ 2002, Journal of the British Interplanetary Society, 55, 383, arXiv:astro-ph/0112490

\bibitem[Harrison(1991)]{Harrison91} Harrison, E.\ 1991, \apj, 383, 60 

\bibitem[Hart(1975)]{Hart75} Hart, M.~H.\ 1975, \qjras, 16, 128 

\bibitem[Harwit(1981)]{Harwit81} Harwit, M. 1981, \emph{Cosmic Discovery}, Brighton: Harvester Press

\bibitem[Hochner et al.(2006)]{Hochner06} Hochner, B., Shomrat, T., \& Fiorito, G. 2006, The Biological Bulletin, 210, 308

\bibitem[Hogg(1999)]{Hogg99} Hogg, D.~W.\ 1999, arXiv:astro-ph/9905116 

\bibitem[Horowitz \& Sagan(1993)]{Horowitz93} Horowitz, P., \& Sagan, C.\ 1993, \apj, 415, 218 

\bibitem[Howard et al.(2004)]{Howard04} Howard, A.~W., Horowitz, P., Wilkinson, D.~T., et al.\ 2004, \apj, 613, 1270 

\bibitem[Hussmann et al.(2006)]{Hussmann06} Hussmann, H., Sohl, F., \& Spohn, T.\ 2006, \icarus, 185, 258 

\bibitem[Johnson et al.(2010)]{Johnson10} Johnson, J.~A., Aller, K.~M., Howard, A.~W., \& Crepp, J.~R.\ 2010, \pasp, 122, 905, arXiv:1005.3084

\bibitem[Jugaku \& Nishimura(2004)]{Jugaku04} Jugaku, J., \& Nishimura, S.\ 2004, Bioastronomy 2002: Life Among the Stars, 213, 437 

\bibitem[Kardashev(1964)]{Kardashev64} Kardashev, N.~S.\ 1964, \sovast, 8, 217 

\bibitem[Kardashev(1985)]{Kardashev85} Kardashev, N.~S.\ 1985, The Search for Extraterrestrial Life: Recent Developments, 112, 497 

\bibitem[Kauffman(1995)]{Kauffman95} Kauffman, S. 1995, \emph{At Home in the Universe: The Search for Laws of Self-Organization and Complexity}, New York, Oxford University Press (ebook edition)

\bibitem[Knobe et al.(2006)]{Knobe06} Knobe, J., Olum, K.~D., \& Vilenkin, A.\ 2003, The British journal for the philosophy of science, 57, 47, arXiv:physics/0302071 

\bibitem[Krauss \& Scherrer(2007)]{Krauss07} Krauss, L.~M., \& Scherrer, R.~J.\ 2007, General Relativity and Gravitation, 39, 1545, arXiv:0704.0221

\bibitem[Lacki(2015)]{Lacki15} Lacki, B.~C.\ 2015, arXiv:1503.01509 

\bibitem[Lacki(2016)]{Lacki16} Lacki, B.~C.\ 2016, arXiv:1604.07844 

\bibitem[Lammer et al.(2009)]{Lammer09} Lammer, H., Bredeh{\"o}ft, J.~H., Coustenis, A., et al.\ 2009, \aapr, 17, 181 

\bibitem[Lau \& Dill(1990)]{Lau90} Lau, K. F., Dill, K. A. 1990, Proceedings of the National Academy of Sciences, 87, 638

\bibitem[Learned(1994)]{Learned94} Learned, J.~G.\ 1994, Royal Society of London Philosophical Transactions Series A, 346, 99 

\bibitem[Li \& Sadler(1991)]{Li91} Li, W. H., \& Sadler, L. A. 1991, Genetics, 129, 513

\bibitem[Lin et al.(2014)]{Lin14} Lin, H.~W., Gonzalez Abad, G., \& Loeb, A.\ 2014, \apjl, 792, LL7, arXiv:1406.3025

\bibitem[Lineweaver \& Davis(2002)]{Lineweaver02} Lineweaver, C.~H., \& Davis, T.~M.\ 2002, Astrobiology, 2, 293, arXiv:astro-ph/0205014

\bibitem[Lineweaver(2009)]{Lineweaver09} Lineweaver, C. H. 2009, \emph{From fossils to astrobiology}, 353, Netherlands, Springer, arXiv:0711.1751

\bibitem[Lintott et al.(2011)]{Lintott11} Lintott, C., Schawinski, K., Bamford, S., et al.\ 2011, \mnras, 410, 166, arXiv:1007.3265

\bibitem[Livio(1999)]{Livio99} Livio, M.\ 1999, \apj, 511, 429, arXiv:astro-ph/9808237

\bibitem[Lloyd(2000)]{Lloyd00} Lloyd, S.\ 2000, \nat, 406, 1047, arXiv:quant-ph/9908043

\bibitem[Lodders \& Fegley(1998)]{Lodders98} Lodders, K., Fegley, B. 1998, \emph{The Planetary Scientist's Companion}, New York, Oxford University Press, Inc.

\bibitem[Loeb \& Zaldarriaga(2007)]{Loeb07} Loeb, A., \& Zaldarriaga, M.\ 2007, \jcap, 1, 020, arXiv:astro-ph/0610377

\bibitem[Loeb \& Turner(2012)]{Loeb12} Loeb, A., \& Turner, E.~L.\ 2012, Astrobiology, 12, 290, arXiv:1110.6181

\bibitem[Loeb et al.(2016)]{Loeb16} Loeb, A., Batista, R.~A., \& Sloan, D.\ 2016, \jcap, 8, 040, arXiv:1606.08448

\bibitem[Lopez et al.(2005)]{Lopez05} Lopez, B., Schneider, J., \& Danchi, W.~C.\ 2005, \apj, 627, 974, arXiv:astro-ph/0503520

\bibitem[Lorenz et al.(1997)]{Lorenz97} Lorenz, R.~D., Lunine, J.~I., \& McKay, C.~P.\ 1997, \grl, 24, 2905 

\bibitem[Lunine \& Nolan(1992)]{Lunine92} Lunine, J.~I., \& Nolan, M.~C.\ 1992, \icarus, 100, 221 

\bibitem[Lunine(2009)]{Lunine09} Lunine, J.~I.\ 2009, arXiv:0908.0762 

\bibitem[Margolus \& Levitin(1998)]{Margolus98} Margolus, N., \& Levitin, L.~B.\ 1998, Physica D Nonlinear Phenomena, 120, 188, arXiv:quant-ph/9710043 (ML98)

\bibitem[Marino(2002)]{Marino02} Marino, L. 2002, Brain, Behavior and Evolution, 59, 21

\bibitem[Maynard Smith(1970)]{MaynardSmith70} Maynard Smith, J.\ 1970, \nat, 225, 563 

\bibitem[Mayr(2001)]{Mayr01} Mayr, E. 2001, \emph{What Evolution Is}, New York, Basic Books

\bibitem[McKay \& Smith(2005)]{McKay05} McKay, C.~P., \& Smith, H.~D.\ 2005, \icarus, 178, 274 

\bibitem[Mikkelsen et al.(2005)]{Mikkelsen05} Mikkelsen, T. S. et al. (The Chimpanzee Sequencing and Analysis Consortium) 2005, Nature, 437, 69

\bibitem[Napier(2004)]{Napier04} Napier, W.~M.\ 2004, \mnras, 348, 46

\bibitem[Nash(2006)]{Nash06} Nash, L. K., \emph{Elements of Statistical Thermodynamics}, 2nd ed., Mineola, New York, Dover Publications (ebook edition)

\bibitem[Olive et al.(2015)]{Olive15} Olive, K. A. et al. (Particle Data Group) 2014, Chin. Phys. C, 38, 090001 and 2015 update

\bibitem[Olson(2015)]{Olson15} Olson, S.~J.\ 2015, arXiv:1507.05969

\bibitem[Olum(2002)]{Olum02} Olum, K.D., 2002, The Philosophical Quarterly, 52, 164, arXiv:gr-qc/0009081v2

\bibitem[Page(2008)]{Page08} Page, D.~N.\ 2008, \prd, 78, 063535, arXiv:hep-th/0610079

\bibitem[Papagiannis(1985)]{Papagiannis85} Papagiannis, M.~D.\ 1985, The Search for Extraterrestrial Life: Recent Developments, 112, 505	

\bibitem[Petigura et al.(2013)]{Petigura13} Petigura, E.~A., Howard, A.~W., \& Marcy, G.~W.\ 2013, Proceedings of the National Academy of Science, 110, 19273, arXiv:1311.6806

\bibitem[Pierce(1960)]{Pierce60} Pierce, J. R.\ 1960, \emph{An Introduction to Information Theory: Symbols, Signals, \& Noise}, New York, Dover Publications (ebook edition)

\bibitem[Pr\"ufer et al.(2012)]{Prufer12} Pr\"ufer, K. et al. 2012, Nature, 486, 527

\bibitem[Reines \& Marcy(2002)]{Reines02} Reines, A.~E., \& Marcy, G.~W.\ 2002, \pasp, 114, 416, arXiv:astro-ph/0112479

\bibitem[Rhodes(1986)]{Rhodes86} Rhodes, R. 1986, \emph{The Making of the Atomic Bomb}, New York, Simon \& Schuster

\bibitem[Rose \& Wright(2004)]{Rose04} Rose, C., \& Wright, G.\ 2004, \nat, 431, 47 

\bibitem[Rose et al.(2006)]{Rose06} Rose, G. D., Fleming, P. J., Banavar, J. R., \& Maritan, A. 2006, Proceedings of the National Academy of Sciences, 103, 16623

\bibitem[Rothman \& Ellis(1987)]{Rothman87} Rothman, T., \& Ellis, G.~F.~R.\ 1987, The Observatory, 107, 24 

\bibitem[Rubenstein(2014)]{Rubenstein14} Rubenstein, M. J. 2014, \emph{Worlds Without End: The Many Lives of the Multiverse}, New York, Columbia University Press

\bibitem[Sachidananam et al.(2001)]{Sachidanandam01} Sachidanandam, R. et al. 2001, Nature, 409, 928

\bibitem[Sagan(1963)]{Sagan63} Sagan, C.\ 1963, \planss, 11, 485 

\bibitem[Sagan(1973)]{Sagan73} Sagan, C.\ 1973, \icarus, 19, 350 

\bibitem[Sagan \& Newman(1983)]{Sagan83} Sagan, C., \& Newman, W.~I.\ 1983, \qjras, 24, 113

\bibitem[Sagan(1994)]{Sagan94} Sagan, C.\ 1994, \emph{Pale Blue Dot: A Vision of the Human Future in Space}, New York, Ballantine Books (ebook edition)

\bibitem[Scharf \& Cronin(2016)]{Scharf16} Scharf, C., \& Cronin, L.\ 2016, Proceedings of the National Academy of Science, 113, 8127, arXiv:1511.02549

\bibitem[Scheffer(1994)]{Scheffer94} Scheffer, L.~K.\ 1994, \qjras, 35, 157 

\bibitem[Seager(2014)]{Seager14} Seager, S.\ 2014, Proceedings of the National Academy of Science, 111, 12634 

\bibitem[Sender et al.(2016)]{Sender16} Sender, R., Fuchs, S. and Milo, R., 2016, Cell, 164, 337

\bibitem[Shvartsman et al.(1993)]{Shvartsman93} Shvartsman, V., Beskin, G., Mitronova, S., et al.\ 1993, Third Decennial US-USSR Conference on SETI, 47, 381

\bibitem[Sleep et al.(1989)]{Sleep89} Sleep, N.~H., Zahnle, K.~J., Kasting, J.~F., \& Morowitz, H.~J.\ 1989, \nat, 342, 139 

\bibitem[Slysh(1985)]{Slysh85} Slysh, V.~I.\ 1985, The Search for Extraterrestrial Life: Recent Developments, 112, 315

\bibitem[Siemion et al.(2010)]{Siemion10} Siemion, A., Von Korff, J., McMahon, P., et al.\ 2010, Acta Astronautica, 67, 1342, arXiv:0811.3046

\bibitem[Siemion et al.(2013)]{Siemion13} Siemion, A.~P.~V., Demorest, P., Korpela, E., et al.\ 2013, \apj, 767, 94, arXiv:1302.0845

\bibitem[Simpson(1964)]{Simpson64} Simpson, G. G.\ 1964, Science, 143, 769 

\bibitem[Spiegel \& Turner(2012)]{Spiegel12} Spiegel, D.~S., \& Turner, E.~L.\ 2012, Proceedings of the National Academy of Science, 109, 395, arXiv:1107.3835 

\bibitem[Steel(1995)]{Steel95} Steel, D.\ 1995, The Observatory, 115, 78 

\bibitem[Steinhardt \& Turok(2002)]{Steinhardt02} Steinhardt, P.~J., \& Turok, N.\ 2002, Science, 296, 1436, arXiv:hep-th/0111030

\bibitem[Stevens et al.(2015)]{Stevens15} Stevens, A., Forgan, D., \& O'Malley-James, J.\ 2015, arXiv:1507.08530 

\bibitem[Stevenson(2013)]{Stevenson13} Stevenson, D.~S.\ 2013, \emph{Under a Crimson Sun: Prospects for Life in a Red Dwarf System, Astronomers' Universe}, New York, Springer Science+Business Media New York 

\bibitem[Tarter(1985)]{Tarter85} Tarter, J.\ 1985, The Search for Extraterrestrial Life: Recent Developments, 112, 271

\bibitem[Tarter(2001)]{Tarter01} Tarter, J.\ 2001, \araa, 39, 511 

\bibitem[Tarter(2007)]{Tarter07} Tarter, J.~C.\ 2007, Highlights of Astronomy, 14, 14 

\bibitem[Tegmark(2014)]{Tegmark14} Tegmark, M. 2014, \emph{Our Mathematical Universe: My Quest for the Ultimate Nature of Reality}, New York, Alfred A. Knopf (ebook edition)

\bibitem[Timofeev et al.(2000)]{Timofeev00} Timofeev, M.~Y., Kardashev, N.~S., \& Promyslov, V.~G.\ 2000, Acta Astronautica, 46, 655 

\bibitem[Tipler(1980)]{Tipler80} Tipler, F.~J.\ 1980, \qjras, 21, 267 

\bibitem[Tipler(1982)]{Tipler82} Tipler, F.~J.\ 1982, The Observatory, 102, 36 

\bibitem[Tough \& Lemarchand(2004)]{Tough04} Tough, A., \& Lemarchand, G.~A.\ 2004, Bioastronomy 2002: Life Among the Stars, 213, 487 

\bibitem[Trotta(2008)]{Trotta08} Trotta, R.\ 2008, Contemporary Physics, 49, 71, arXiv:0803.4089 

\bibitem[Turnbaugh et al.(2007)]{Turnbaugh07} Turnbaugh, P.~J., Ley, R.~E., Hamady, M., et al.\ 2007, \nat, 449, 804 

\bibitem[Turnbull \& Tarter(2003)]{Turnbull03} Turnbull, M.~C., \& Tarter, J.~C.\ 2003, \apjs, 145, 181, arXiv:astro-ph/0210675

\bibitem[Union of Concerned Scientists(2015)]{UoCS15} Union of Concerned Scientists\ 2015, ``Close Calls with Nuclear Weapons'', \textless http://www.ucsusa.org/sites/default/files/attach/2015/04/\\Close\%20Calls\%20with\%20Nuclear\%20Weapons.pdf \textgreater

\bibitem[Venter et al.(2001)]{Venter01} Venter, J. C. et al. 2001, Science, 291, 1304

\bibitem[Vickaryous \& Hall(2006)]{Vickaryous06} Vickaryous, M. K. \&  Hall, B. K. 2006, Biological reviews, 81, 425

\bibitem[Vilenkin(1995)]{Vilenkin95} Vilenkin, A.\ 1995, Physical Review Letters, 74, 846, arXiv:gr-qc/9406010

\bibitem[Villarroel et al.(2016)]{Villarroel16} Villarroel, B., Imaz, I., \& Bergstedt, J.\ 2016, \aj, 152, 76, arXiv:1606.08992  

\bibitem[Ward \& Brownlee(2004)]{Ward04} Ward, P. D., Brownlee, D. 2004, \emph{Rare Earth: Why Complex Life is Uncommon in the Universe}, New York, Copernicus Books

\bibitem[Weissman(1996)]{Weissman96} Weissman, P.~R.\ 1996, Completing the Inventory of the Solar System, 107, 265 

\bibitem[Wesson(1990)]{Wesson90} Wesson, P.~S.\ 1990, \qjras, 31, 161 

\bibitem[Whitman et al.(1998)]{Whitman98} Whitman, W.~B., Coleman, D.~C., \& Wiebe, W.~J.\ 1998, Proceedings of the National Academy of Science, 95, 6578 

\bibitem[Whitmire \& Wright(1980)]{Whitmire80} Whitmire, D.~P., \& Wright, D.~P.\ 1980, \icarus, 42, 149 

\bibitem[Wilkins et al.(1996)]{Wilkins96} Wilkins, M. R., Sanchez, J. C., Gooley, A. A., Appel, R. D., Humphery-Smith, I., Hochstrasser, D. F., \& Williams, K. L. 1996, Biotechnology and genetic engineering reviews, 13, 19

\bibitem[Worth et al.(2013)]{Worth13} Worth, R.~J., Sigurdsson, S., \& House, C.~H.\ 2013, Astrobiology, 13, 1155, arXiv:1311.2558

\bibitem[Wright et al.(2014b)]{Wright14-SF} Wright, J.~T., Mullan, B., Sigurdsson, S., \& Povich, M.~S.\ 2014b, \apj, 792, 26, arXiv:1408.1133

\bibitem[Wright et al.(2014a)]{Wright14-Results} Wright, J.~T., Griffith, R.~L., Sigurdsson, S., Povich, M.~S., \& Mullan, B.\ 2014a, \apj, 792, 27, arXiv:1408.1134

\bibitem[Wright et al.(2016)]{Wright16} Wright, J.~T., Cartier, K.~M.~S., Zhao, M., Jontof-Hutter, D., \& Ford, E.~B.\ 2016, \apj, 816, 17, arXiv:1510.04606

\bibitem[Yockey(2000)]{Yockey00} Yockey, H. P. 2000, Computers \& chemistry, 24, 105

\bibitem[Zackrisson et al.(2015)]{Zackrisson15} Zackrisson, E., Calissendorff, P., Asadi, S., \& Nyholm, A.\ 2015, \apj, 810, 23, arXiv:1508.02406

\bibitem[Zackrisson et al.(2016)]{Zackrisson16} Zackrisson, E., Calissendorff, P., Gonzalez, J., et al.\ 2016, arXiv:1602.00690 

\end{thebibliography}
\end{document}